\documentclass[oneside]{book}

\usepackage{arxiv}

\usepackage[utf8]{inputenc} 
\usepackage[T1]{fontenc}    
\usepackage[bookmarks, colorlinks={true}, linkcolor=blue, citecolor=blue, pdftitle={Aproximación práctica a los métodos de selección de portafolios de inversión}, pdfauthor={Carlos Minutti Martínez}, pdfsubject={Portafolios de inversion}, pdfkeywords={portfolio selection, genetic algorithms}, pdfstartview={FitH}]{hyperref}
\usepackage{url}            
\usepackage{booktabs}       
\usepackage{amsfonts}       
\usepackage{nicefrac}       
\usepackage{microtype}      
\usepackage{cleveref}       
\usepackage{lipsum}         
\usepackage{graphicx}
\usepackage{doi}

\usepackage[spanish]{babel}
\usepackage{enumerate,multicol}
\usepackage{bm}
\usepackage{multicol}
\usepackage{amssymb}
\usepackage[numbers]{natbib}

\newcommand\abstractname{Abstract}  
\makeatletter
\if@titlepage
  \newenvironment{abstract}{%
      \titlepage
      \null\vfil
      \@beginparpenalty\@lowpenalty
      \begin{center}%
        \bfseries \abstractname
        \@endparpenalty\@M
      \end{center}}%
     {\par\vfil\null\endtitlepage}
\else
  \newenvironment{abstract}{%
      \if@twocolumn
        \section*{\abstractname}%
      \else
        \small
        \begin{center}%
          {\bfseries \abstractname\vspace{-.5em}\vspace{\z@}}%
        \end{center}%
        \quotation
      \fi}
      {\if@twocolumn\else\endquotation\fi}
\fi
\makeatother

\title{Aproximación práctica a los métodos de selección de portafolios de inversión}

\date{2024}

\newif\ifuniqueAffiliation
\uniqueAffiliationtrue

\ifuniqueAffiliation 
\author{Carlos Minutti Martínez\hspace{1mm}\href{https://orcid.org/0000-0002-4002-0773}{\includegraphics[scale=0.1]{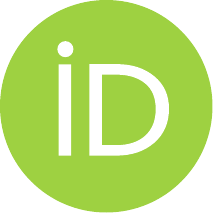}} \\[0.5cm]
	Centro de Estudios en Computación Avanzada (CECAv)\\
	Universidad Nacional Autónoma de México (UNAM)\\
	\texttt{carlos\_minutti@lapi.unam.mx} \\[0.5cm]
	Centro de Investigación e Innovación en Tecnologías \\de la Información y Comunicación  (INFOTEC) \\
	\texttt{carlos.minutti@infotec.mx}\\[0.5cm]
	\texttt{cminuttim.github.io}
}

\renewcommand{\shorttitle}{Aproximación práctica a los métodos de selección de portafolios de inversión}

\hypersetup{
pdftitle={Aproximación práctica a los métodos de selección de portafolios de inversión},
pdfsubject={Optimization, portfolio selection},
pdfauthor={Carlos.~Minutti-Martinez},
pdfkeywords={Portfolio selection, Genetic algorithms, Markowitz},
}

\begin{document}
\maketitle

\begin{abstract}
This paper explores the practical approach to portfolio selection methods for investments. The study delves into portfolio theory, discussing concepts such as expected return, variance, asset correlation, and opportunity sets. It also presents the efficient frontier and its application in the Markowitz model, which employs mean-variance optimization techniques.

An alternative approach based on the mean-semivariance model is introduced. This model accounts for the skewness and kurtosis of the asset return distribution, providing a more comprehensive view of risk and return. The study also addresses the practical implementation of these models, including the use of genetic algorithms to optimize portfolio selection.

Additionally, transaction costs and integer constraints in portfolio optimization are considered, demonstrating the applicability of the Markowitz model in real-world scenarios.

-----

Este documento explorar la aproximación práctica a los métodos de selección de portafolios para inversiones. El estudio profundiza en la teoría de los portafolios, discutiendo conceptos como el rendimiento esperado, la varianza, la correlación entre activos y los conjuntos de oportunidades. También presenta la frontera eficiente y su aplicación en el modelo de Markowitz, que utiliza técnicas de optimización media-varianza.

Se introduce un enfoque alternativo basado en el modelo media-semivarianza. Este modelo tiene en cuenta la asimetría y la curtosis de la distribución de retornos de los activos, proporcionando una visión más completa de riesgo y rendimiento. El estudio también aborda la implementación práctica de estos modelos, incluyendo el uso de algoritmos genéticos para optimizar la selección de portafolios.

Además, se consideran los costos de transacción y las restricciones enteras en la optimización del portafolio, demostrando la aplicabilidad del modelo de Markowitz en escenarios reales.
\end{abstract}



\tableofcontents \setcounter{page}{1}
\listoffigures

\chapter{Introducción}

La importancia de la Bolsa de Valores ha ido en aumento debido a que se ha constituido como un medio alternativo de financiamiento para las empresas y el gobierno, así como una alternativa de inversión. Es así que la Bolsa es considerada como el mercado financiero que mejor refleja la situación real de la economía. Es un indicador de todo lo que ocurre y afecta la economía de un país.

La subidas y bajadas en el precio de las acciones no son una casualidad, responden a situaciones específicas que marcan tendencias en sus precios. Es por esto que es posible utilizar modelos estadísticos para la selección de portafolios que cumplan con las necesidades del inversionista, quien requiere de elementos que le permitan tomar una decisión para maximizar su ganancia con un mínimo de riesgo dado. Por esto es necesario conocer los instrumentos y técnicas para invertir en la Bolsa de Valores.

Existen una serie de teorías para conformar un portafolio de inversión, cada una de estas posee sus hipótesis, ventajas y desventajas, pero en todas su objetivo es obtener un mayor rendimiento.

Los estudios estadísticos de la teoría de selección de portafolios pueden remontarse hasta Markowitz \cite{mark1}, quien trabajó el problema de selección de portafolios utilizando el criterio Media-Varianza para un sólo periodo de tiempo como criterio de selección, con lo que establecería las bases de la Teoría Moderna de Portafolios (TMP), y que se sigue usando y estudiando ({\it e.g.} \cite{andreevichproblems,erwin2023metaheuristics}).

El modelo de Markowitz tiene el propósito de determinar la combinación óptima de acciones por medio de la programación cuadrática, por lo que requiere de una gran cantidad de cálculos, sobre todo cuando el número de acciones utilizadas aumenta, ya que las combinaciones posibles crecen rápidamente, es así que su uso creció desde la década de los 80, con el desarrollo en potencia y disponibilidad de las computadoras, ayudado al uso de modelos matemáticos como este.

Debido a la problemática del poder de cómputo necesario para el modelo de Markowitz los estudios sobre el tema continuaron, y así aparecieron modelos alternativos que presentan simplificaciones al modelo original como el Modelo de Índice Simple de William Sharpe (1963). A su vez también aparecieron técnicas para la búsqueda de soluciones a ciertos modelos matemáticos, así es como en los años 70, John Henry Holland desarrolla una rama de la inteligencia artificial llamada Algoritmos Genéticos\footnote{Las simulaciones por computadora sobre la evolución comenzaron desde 1954 por Nils Aall Barricelli, en 1957 Alex Fraser publicó una serie de documentos sobre la simulación de la selección artificial, a partir de esto la simulación de la evolución se hizo más común en la década de 1960 y los métodos utilizados se describen en los libros de Fraser y Burnell (1970) y Crosby (1973). La simulación de Fraser incluye todos los elementos esenciales de los algoritmos genéticos, sin embargo, los algoritmos genéticos se popularizaron hasta el trabajo de Holland, particularmente en su libro ``Adaptation in Natural and Artificial Systems'' (1975).}, que es básicamente un método computacional, basado en probabilidad, para encontrar o aproximar soluciones a problemas de optimización.

En el presente trabajo abordamos en el primer capítulo la Teoría de Moderna de Portafolios, que nos permitirá entender cómo se administra el rendimiento y el riesgo de los mismos y nos proporcionará un criterio de selección dentro de un conjunto de opciones.

En el segundo capítulo revisamos la teoría del modelo de Markowitz, así como su aplicación en la Bolsa de Valores, donde creamos portafolios de inversión con diferentes niveles de riesgo, que nos permitirán observar el comportamiento a futuro de este modelo.

El tercer capítulo aborda un modelo con una medida alternativa de riesgo al del modelo clásico de Markowitz, y se comparan los resultados obtenidos con el modelo del segundo capítulo.

El capítulo cuarto maneja la teoría de los algoritmos genéticos y su uso en la optimización, específicamente para la inclusión de restricciones adicionales en el modelo de Markowitz, los beneficios y desventajas que conlleva, así como de su uso para problemas complejos.

Por último, en el quinto capítulo generamos un modelo agregando restricciones de costos por transacción y lotes al modelo de Markowitz, para así obtener portafolios más prácticos, que cumplen mejor las restricciones del mercado, pero para el cual será necesario utilizar los algoritmos genéticos para darle solución.

Para todos estos modelos generamos el código de programación que nos permite darles solución. Esto con la finalidad de comprender mejor cada uno de los modelos y que los resultados obtenidos en esta investigación puedan ser fácilmente reproducidos o estudiados por quien lo desee.

Con todo lo anterior, concluiremos cuál es el comportamiento de estos instrumentos en la Bolsa de Valores y determinar su eficiencia, para así ser capaces de utilizar estos modelos en una inversión real.

Este texto es dirigido a individuos con una base en estadística que buscan familiarizarse y poner en práctica alguna de los modelos que se emplean para la construcción y selección de portafolios de inversión. Será accesible tanto para aquellos que desean explorar modelos más sofisticados, como para los que estén comenzando su camino en el campo. Se presenta a un nivel adecuado para estudiantes cursando carreras relacionadas con las matemáticas o la computación.

\mainmatter

\chapter{Teoría de portafolios}

{\em 
En el mercado de valores existen muchas empresas en las que es posible invertir, cuyas cotizaciones constantemente fluctúan por diversos motivos. La teoría de portafolios trata de la selección de portafolios óptimos, es decir, aquellos que proporcionan el mayor rendimiento\footnote{Sea $P_t$ el precio de un valor en el periodo $t$, entonces su rendimiento en $t$ se define como $R_t=\frac{P_t}{P_{t-1}} -1$, es decir, la ganancia proporcional de un periodo $t$ a otro.} dado un nivel de riesgo, o los que son de mínimo riesgo dada un nivel de rendimiento.

En la teoría de portafolios suponemos que dos características fundamentales determinan
el comportamiento de un inversionista con respecto a una acción\footnote{Las acciones son las partes iguales en las que una empresa divide su capital social, por lo que cuando se adquieren acciones de una empresa, debe entenderse que se hace dueño de una parte de ésta en mayor o menor medida.} (o portafolios\footnote{Se le llama cartera o portafolios de inversión al conjunto de varios activos financieros que posee un individuo o institución, comúnmente acciones y bonos.}), primero, su rendimiento (la ganancia proporcional a su inversión) y segundo, el riesgo (las fluctuaciones del rendimiento), por lo que un inversionista tomará sus decisiones considerando estos dos parámetros.

Al no poder conocer el rendimiento futuro de las acciones, un inversionista querrá al menos pronosticar el rendimiento que estas tendrán en el próximo periodo de inversión. Para abordar este problema, Markowitz considera al rendimiento de las acciones como una variable aleatoria, por lo que podemos obtener su valor esperado y varianza (así como de una combinación lineal de las mismas). De esta manera, la varianza y el rendimiento esperado son las medidas de rentabilidad y riesgo, respectivamente, que utilizaremos.

Como es de esperarse, existe una relación intrínseca entre el rendimiento y el riesgo de 
una acción-portafolios. Consideramos que un inversionista ``razonable'' es averso al riesgo, por lo que su decisión se inclinará por aquel conjunto de acciones que proporcionen un alto rendimiento con un mínimo de riesgo.

Considerando todo lo anterior, la finalidad de la teoría de portafolios es la de proporcionar los criterios de decisión básicos, para encontrar las opciones que mejor satisfacen los objetivos del inversionista.
}

\section{Rendimiento esperado de un portafolios}

Sea la tasa de rendimiento de la acción $i$ la variable aleatoria definida como $R_{i}$, al ser un portafolio una combinación de activos financieros, la ganancia de este viene dada por la ganancia que aporta cada uno de sus elementos. De lo cual se deduce que el rendimiento de un portafolio $R_{p}$ es tal que:

\begin{equation}
R_{p} = w_{1}R_{1}+w_{2}R_{2}+\cdots+w_{n}R_{n} = \sum_{i=1}^{n}w_{i}R_{i} \label{rend}
\end{equation}
donde $n$ es el número de activos que componen el portafolio y $w_{i}$ la proporción invertida en el activo $i$, por lo que $\sum_{i=1}^{n}w_{i} = 1$.

De esta manera, también podemos considerar a un portafolio como una combinación lineal de variables aleatorias, del cual, su rendimiento es el promedio ponderado de los rendimientos de los valores individuales que lo componen.

Con el propósito de manejar una notación más práctica, podemos escribir (\ref{rend}) en notación matricial del siguiente modo:

\begin{eqnarray}
R_{p} = \sum_{i=1}^{n}w_{i}R_{i} = 	
	\left[ \begin{array}{cccc} w_{1} & w_{2} & \cdots & w_{n} \end{array} \right]
	\left[ \begin{array}{c} R_{1} \\ R_{2} \\ \vdots \\ R_{n} \end{array} \right]
	= {\bm w}'{\bm R}
\end{eqnarray}

El valor esperado para el rendimiento del portafolios será:

\begin{eqnarray}
\mu_{p} = E(R_{p}) &=& E(w_{1}R_{1}+\cdots+w_{n}R_{n}) = w_{1}E(R_{1})+\cdots+w_{n}E(R_{n}) \nonumber \\
		 		   &=&  \sum_{i=1}^{n}w_{i}E(R_{i}) = \sum_{i=1}^{n}w_{i}\mu_{i}  \label{rendE}
\end{eqnarray}

de igual forma:

\begin{eqnarray}
\mu_{p} = E(R_{p}) = E({\bm w}'{\bm R}) = {\bm w}'E({\mathbf R}) = {\bm w}'{\bm \mu}
\end{eqnarray}

donde, $\mu_{i}$ es el valor esperado para el rendimiento de la acción $i$, y ${\bm \mu}$ el vector compuesto de los $\mu_{i}$. Así, el rendimiento esperado del portafolio es el promedio ponderado de los rendimientos esperados que lo componen.

\section{Varianza de un portafolios}

Para determinar la varianza de un portafolio debemos utilizar su definición en términos del valor esperado\footnote{Si $X$ es una variable aleatoria, $V(X) = E[(X - E[X])^{2}]$.}:

\begin{eqnarray}
\sigma_{p}^{2} = V(R_{p}) &=& V\left(\sum_{i=1}^{n}w_{i} R_{i} \right) = E\left[\left(\sum_{i=1}^{n}w_{i}R_{i} -
	\sum_{i=1}^{n}w_{i}E(R_{i})\right)^{2}\right] \nonumber \\
	&=& E\left[\left(\sum_{i=1}^{n}w_{i}[R_{i} - E(R_{i})]\right)^{2}\right] \nonumber \\
	&=& E\left[ \left( \sum_{i=1}^{n}w_{i}[R_{i} - E(R_{i})] \right) \left( \sum_{j=1}^{n}w_{j}[R_{j} - E(R_{j})] \right)  \right]  \nonumber \\
	&=& E\left[\sum_{i=1}^{n}\sum_{j=1}^{n}w_{i}w_{j}[R_{i} - E(R_{i})][R_{j} - E(R_{j})]\right]  \nonumber \\
	&=& \sum_{i=1}^{n}\sum_{j=1}^{n}w_{i}w_{j}E([R_{i} - E(R_{i})][R_{j} - E(R_{j})])  \nonumber \\
	&=& \sum_{i=1}^{n}\sum_{j=1}^{n}w_{i}w_{j}Cov(R_{i},R_{j})
	= \sum_{i=1}^{n}\sum_{j=1}^{n}w_{i}w_{j}\sigma_{ij} \label{varianza}
\end{eqnarray}

donde, $\sigma_{ij}$ es la covarianza entre los rendimientos de los activos $i,j$.

Al usar la notación matricial, obtenemos una expresión que resulta muy práctica para representar el riesgo 
de un portafolio, esto es:
\begin{eqnarray}
\sigma_{p}^{2} &=& 
	\left[ \begin{array}{cccc} w_{1} & w_{2} & \cdots & w_{n} \end{array} \right]
	\left[ \begin{array}{cccc} 
		\sigma_{11} & \sigma_{12} & \cdots & \sigma_{1n} \\
		\sigma_{21} & \sigma_{22} & \cdots & \sigma_{2n} \\ 
		\vdots & \vdots & \ddots & \vdots \\ 
		\sigma_{n1} & \sigma_{n2} & \cdots & \sigma_{nn} \\ 
	\end{array} \right] \nonumber
	\left[ \begin{array}{c} w_{1} \\ w_{2} \\ \vdots \\ w_{n} \end{array} \right] 
	= {\bm w}' {\bm \Sigma} {\bm w} \nonumber
\end{eqnarray}

donde ${\bm \Sigma}$ es la matriz de covarianzas de los rendimientos que por definición es semidefinida positiva\footnote{Una matriz simétrica $\bm A$, es semidefinida positiva si $\bm x' \bm A \bm x \geq 0$ para todo vector $\bm x \neq {\bm 0}$.}.  

Se puede mostrar que ${\bm \Sigma}$ es positiva definida
\footnote{Una matriz simétrica ${\bm A}$, es definida positiva si ${\bm x}'{\bm A}{\bm x} > 0$ para todo vector $\bm x \neq {\bm 0}$.}, bajo el supuesto de que los distintos rendimientos no son todos iguales, y estos sean linealmente independientes (Green, 1986 y Mcentire, 1984).

Nótese que (\ref{varianza}) puede escribirse como: 
\begin{eqnarray}
\sigma_{p}^{2} = \sum_{i=1}^{n}\sum_{j=1}^{n}w_{i}w_{j}\sigma_{ij} 
	=\sum_{i=1}^{n}w_{i}^2\sigma_{i}^2 + \sum_{i \neq j}w_{i}w_{j}\sigma_{ij} \label{varcov}
\end{eqnarray}

donde $\sigma_{i}^2$ es la varianza del activo $i$. De esta expresión observamos que es posible separar en dos partes la varianza de un portafolio, una que depende de las varianzas de los rendimientos para
cada acción, y otra que depende de sus correlaciones, lo cual, como se verá en la siguiente sección, es un punto importante a considerar al momento de seleccionar portafolios de mínimo riesgo.

\section{Correlación entre activos}

Sabemos que el coeficiente de correlación entre los activos $i,j$ se define como:
$$ \rho_{ij}=\frac{\sigma_{ij}}{\sigma_{i}\sigma_{j}} \; , \hspace{0.7cm} \mbox{donde} \;  -1 \leq \rho_{ij} \leq 1 $$

Existen tres principales casos que nos interesa estudiar cuando hablamos del coeficiente de correlación:

\begin{itemize}
	\item[$\bullet$]
	{\em La correlación entre dos activos es uno ($\rho_{ij}=1, \;\mbox{para}\; i \neq j$)}, por lo que el rendimiento del activo $i$ posee una correlación perfectamente positiva con el activo $j$, de esta forma al aumentar uno de los rendimientos, el otro también lo hace en una determinada proporción.

	\item[$\bullet$]
	{\em La correlación entre dos activos es menos uno ($\rho_{ij}=-1, \;\mbox{para}\; i \neq j$)}, esto implica que el rendimiento del activo $i$ posee una correlación perfectamente negativa, tal que si el rendimiento
	de uno crece, el otro decrece.

	\item[$\bullet$]
	{\em La correlación entre dos activos es cero ($\rho_{ij}=0, \;\mbox{para}\; i \neq j$)}, entonces no hay correlación entre los rendimientos, por lo que no se puede deducir el comportamiento de un activo por medio del otro.
\end{itemize}

Además, podemos reescribir la varianza en (\ref{varcov}) sustituyendo la covarianza en términos de la correlación, obteniendo:
\begin{eqnarray}
\sigma_{p}^{2} = \sum_{i=1}^{n}w_{i}^2\sigma_{i}^2 + \sum_{i \neq j}w_{i}w_{j}\sigma_{ij}
	= \sum_{i=1}^{n}w_{i}^2\sigma_{i}^2 + \sum_{i \neq j}w_{i}w_{j} \rho_{ij} \sigma_{i}\sigma_{j} \label{varcorr}
\end{eqnarray}

Esta expresión es analíticamente muy útil, ya que el coeficiente de correlación varía entre $-1$ y $1$, por lo que es más fácil observar como se comporta el riesgo de un portafolios dada la correlación que existe entre los activos que lo componen.

Supongamos las varianzas fijas. Se debe observar que cuando la correlación entre los distintos activos es cero ($\rho_{ij}=0, \forall i \neq j$), 
el riesgo del portafolios es $\sum_{i=1}^{n}w_{i}^2\sigma_{i}^2$, el cual corresponde al riesgo de un portafolios no correlacionado; en este caso se elimina el segundo componente del riesgo.

Por definición las varianzas son no negativas, por lo que el primer componente del riesgo en (\ref{varcorr}) será no negativo, sin embargo, el signo para  el segundo dependerá únicamente de la correlación entre los activos, así, el riesgo disminuirá si existen correlaciones entre los distintos activos que sean negativas. Es decir, el riesgo del portafolios será menor cuando sus activos están correlacionados negativamente.

Una forma más intuitiva de apreciar esto es pensar que dado que los activos tienen correlación negativa, cuando el rendimiento de alguno baja, hay otro que sube, de esta forma las posibilidades de pérdidas son menores y por lo tanto
el riesgo del portafolios también lo es.

De todo lo anterior es evidente la importancia de la correlación para la reducción del riesgo de nuestro portafolios.

\section{Conjunto de oportunidades de un portafolios}

Para un portafolios dado\footnote{Suponiendo fijos los coeficientes de correlación, rendimientos esperados y varianzas de cada uno de los valores.}, las diferentes proporciones de inversión en cada activo generan diferentes rendimientos
esperados y varianzas, los cuales generan un conjunto de elecciones Media - Desviación estándar\footnote{Al ser la desviación estándar la raíz cuadrada de la varianza, ésta también es una medida asociada al riesgo, por lo que se puede usar cualquiera de las dos al hablar del riesgo de un portafolios.} 
$ \{ \; (\mu_{p_1},\sigma_{p_1}), \; (\mu_{p_2},\sigma_{p_2}), \; \ldots \; \}  $, al cual se le llama conjunto de oportunidades, ya que representa los posibles portafolios entre los que el inversionista puede decidir.

Al graficar este conjunto, su forma dependerá del número de activos que componen el portafolios, así como de la correlación entre estos.

Las ventas en corto\footnote{Las ventas en corto consisten en la venta de títulos que no se poseen y que es necesario comprar posteriormente para poder cumplir con la venta.} también influyen en la forma de la gráfica del conjunto de oportunidades, pero ya que en el planteamiento original se establece que no se permiten las ventas en corto, en adelante se supondrá lo mismo.

\subsection[Conjunto de oportunidades con dos acciones]{Conjunto de oportunidades de un portafolios con dos acciones}

Para lograr comprender mejor el comportamiento del conjunto de oportunidades de un portafolios, se utilizará como 
ejemplo el portafolios más pequeño posible, esto es, un portafolios de dos acciones, $a$ y $b$, donde $\sigma_{a} < \sigma_{b}$ y $E(R_{a}) < E(R_{b})$; para el cual nos interesa describir la forma del conjunto de oportunidades en tres situaciones:
\\
\\
\begin{itemize}
\item[$\bullet$] {\em \bf Cuando $\rho_{ab} = 1$} :

El rendimiento esperado y la desviación estándar del portafolios usando (\ref{rendE}) y (\ref{varcorr}) es:
\begin{eqnarray}
\mu_{p} &=& \sum_{i=1}^{n}w_{i}E(R_{i}) = w_{a}E(R_{a}) + (1-w_{a})E(R_{b}) \nonumber \\
	&=& w_{a}\left[ E(R_{a})-E(R_{b}) \right] + E(R_{b}) \label{ejemEsp} \\
\sigma_{p}^{2} &=& \sum_{i=1}^{n}w_{i}^2\sigma_{i}^2 + \sum_{i \neq j}w_{i}w_{j} \rho_{ij} \sigma_{i}\sigma_{j} \nonumber \\
	&=& w_{a}^2\sigma_{a}^2 + (1-w_{a})^2\sigma_{b}^2 + 2w_{a}(1-w_{a}) \rho_{ab} \sigma_{a}\sigma_{b} \nonumber \\
	&=& w_{a}^2\sigma_{a}^2 + 2w_{a}(1-w_{ab})\sigma_{a}\sigma_{b} + (1-w_{a})^2\sigma_{b}^2 \nonumber \\
	&=& \left( w_{a}\sigma_{a} + (1-w_{a})\sigma_{b} \right)^2 \nonumber \\
	\sigma_{p} &=& w_{a}\sigma_{a} + (1-w_{a})\sigma_{b}  \label{ejemDes}
\end{eqnarray}

Para describir el rendimiento en términos de la desviación estándar, usamos la proporción de inversión $w_{a}$, despejando de (\ref{ejemDes}):

$$ w_{a} = \frac{\sigma_{p}-\sigma_{b}}{\sigma_{a}-\sigma_{b}} $$

Sustituyendo en (\ref{ejemEsp}):
\begin{eqnarray*}
\mu_{p} &=& w_{a}\left[ E(R_{a})-E(R_{b}) \right] + E(R_{b}) \\
	&=& \frac{\sigma_{p}-\sigma_{b}}{\sigma_{a}-\sigma_{b}} \left[ E(R_{a})-E(R_{b}) \right] + E(R_{b}) \\
	&=& \sigma_{p}\frac{E(R_{a})-E(R_{b})}{\sigma_{a}-\sigma_{b}} - \sigma_{b}\frac{E(R_{a})-E(R_{b})}{\sigma_{a}-\sigma_{b}} + E(R_{b}) 
\end{eqnarray*}

Esta función corresponde a la ecuación de una recta ($\; ax+b \; $) con pendiente  $\frac{E(R_{a})-E(R_{b})}{\sigma_{a}-\sigma_{b}}$, 
así, cada incremento de $\sigma_{a}-\sigma_{b}$ en la desviación estándar incrementa el rendimiento esperado del portafolios en $E(R_{a})-E(R_{b})$.

\begin{figure}[h]
	\begin{center}
		\includegraphics[width=11cm]{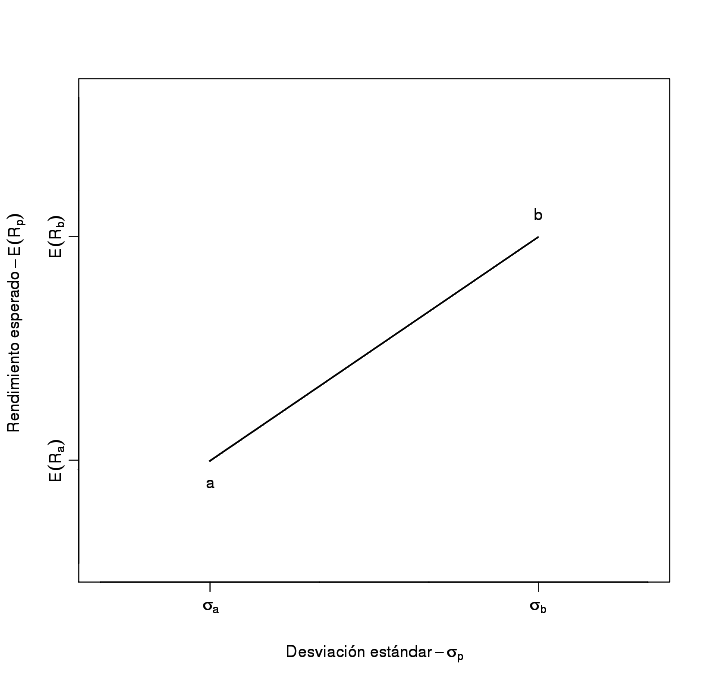} 
		\caption{Rendimiento esperado vs desviación estándar ($\rho_{ab} = 1$)} \label{grafRend1}
	\end{center}
\end{figure}

En la Figura \ref{grafRend1} podemos observar claramente este comportamiento del riesgo con relación al rendimiento de un portafolios. Además, cada uno de los extremos de la recta corresponde al portafolios formado únicamente por $a$ o $b$.
\\
\\
\item[$\bullet$] {\em \bf Cuando $\rho_{ab} = -1$} :

Nuevamente determinamos el rendimiento esperado y la desviación estándar:
\begin{eqnarray}
\mu_{p} &=& w_{a}\left[ E(R_{a})-E(R_{b}) \right] + E(R_{b}) \label{ejemEsp2} \\
\sigma_{p}^{2} &=& \sum_{i=1}^{n}w_{i}^2\sigma_{i}^2 + \sum_{i \neq j}w_{i}w_{j} \rho_{ij} \sigma_{i}\sigma_{j} \nonumber \\
	&=& w_{a}^2\sigma_{a}^2 + (1-w_{a})^2\sigma_{b}^2 + 2w_{a}(1-w_{a}) \rho_{ab} \sigma_{a}\sigma_{b} \nonumber \\
	&=& w_{a}^2\sigma_{a}^2 - 2w_{a}(1-w_{a})\sigma_{a}\sigma_{b} + (1-w_{a})^2\sigma_{b}^2 \nonumber \\
	&=& \left( w_{a}\sigma_{a} - (1-w_{a})\sigma_{b} \right)^2 \hspace{3mm}\mbox{o}\hspace{3mm} \left( -w_{a}\sigma_{a} + (1-w_{a})\sigma_{b} \right)^2 \nonumber \\
	\sigma_{p} &=& w_{a}\sigma_{a} - (1-w_{a})\sigma_{b} \hspace{4mm}\mbox{o}\hspace{4mm} -w_{a}\sigma_{a} + (1-w_{a})\sigma_{b} \label{ejemDes2}
\end{eqnarray}

Obtenemos:
$$ w_{a} = \frac{\sigma_{p}+\sigma_{b}}{\sigma_{a}+\sigma_{b}} \hspace{4mm}\mbox{o}\hspace{4mm} w_{a} = \frac{\sigma_{b}-\sigma_{p}}{\sigma_{a}+\sigma_{b}}$$

Sustituyendo en (\ref{ejemEsp2}) surgen dos rectas:
\begin{eqnarray*}
\mu_{p_1} &=& w_{a}\left[ E(R_{a})-E(R_{b}) \right] + E(R_{b}) \\
	&=& \frac{\sigma_{p}+\sigma_{b}}{\sigma_{a}+\sigma_{b}} \left[ E(R_{a})-E(R_{b}) \right] + E(R_{b}) \\
	&=& \sigma_{p}\frac{E(R_{a})-E(R_{b})}{\sigma_{a}+\sigma_{b}} + \sigma_{b}\frac{E(R_{a})-E(R_{b})}{\sigma_{a}+\sigma_{b}} + E(R_{b}) \\
\mu_{p_2} &=& w_{a}\left[ E(R_{a})-E(R_{b}) \right] + E(R_{b}) \\
	&=& \frac{\sigma_{b}-\sigma_{p}}{\sigma_{a}+\sigma_{b}} \left[ E(R_{a})-E(R_{b}) \right] + E(R_{b}) \\
	&=& -\sigma_{p}\frac{E(R_{a})-E(R_{b})}{\sigma_{a}+\sigma_{b}} + \sigma_{b}\frac{E(R_{a})-E(R_{b})}{\sigma_{a}+\sigma_{b}} + E(R_{b})	
\end{eqnarray*}

\begin{figure}[h]
\begin{center}
	\includegraphics[width=11cm]{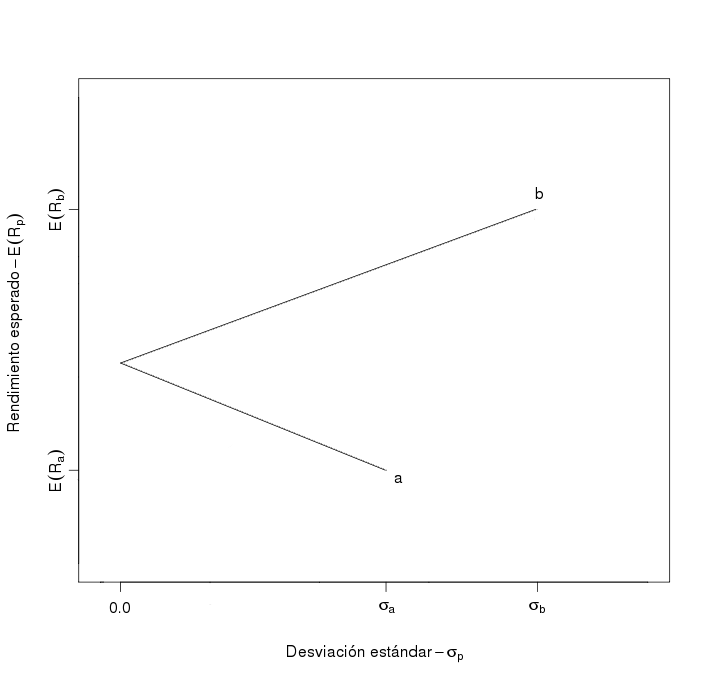} 
	\caption{Rendimiento esperado vs desviación estándar ($\rho_{ab} = -1$)} \label{grafRend2}
\end{center}
\end{figure}

En la Figura \ref{grafRend2} se describe el comportamiento del rendimiento frente a la desviación estándar cuando la correlación entre los activos es negativa. Nuevamente los portafolios formados por $a$ o $b$ se encuentran en los extremos. Observe que existe un segmento de recta donde al combinar los activos disminuye el riesgo (hasta llegar a cero) 
y aumenta el rendimiento, a este efecto se le llama diversificación, el cual se verá en detalle más adelante.
\\
\\
\item[$\bullet$] {\em \bf Cuando $\rho_{ab} = 0$} :

Al ser cero la correlación, la segunda componente del riesgo se elimina, de esta forma las ecuaciones de rendimiento
y riesgo son:
\begin{eqnarray}
\mu_{p} &=& w_{a}\left[ E(R_{a})-E(R_{b}) \right] + E(R_{b}) \label{ejemEsp3} \\
\sigma_{p}^{2} &=& \sum_{i=1}^{n}w_{i}^2\sigma_{i}^2 + \sum_{i \neq j}w_{i}w_{j} \rho_{ij} \sigma_{i}\sigma_{j} \nonumber \\
	&=& w_{a}^2\sigma_{a}^2 + (1-w_{a})^2\sigma_{b}^2 \label{ejemDes3}
\end{eqnarray}

Obtenemos:
$$ w_{a} = \frac{\sigma_{b}^2 \pm \sqrt{ \sigma_{p}^2(\sigma_{a}^2+\sigma_{b}^2)-\sigma_{a}^2\sigma_{b}^2 }}{\sigma_{a}^2+\sigma_{b}^2} $$

Sustituyendo en (\ref{ejemEsp3}):
\begin{eqnarray*}
\mu_{p} &=& w_{a}\left[ E(R_{a})-E(R_{b}) \right] + E(R_{b}) \\
	&=& \frac{\sigma_{b}^2 \pm \sqrt{ \sigma_{p}^2(\sigma_{a}^2+\sigma_{b}^2)-\sigma_{a}^2\sigma_{b}^2 }}{\sigma_{a}^2+\sigma_{b}^2} \left[ E(R_{a})-E(R_{b}) \right] + E(R_{b}) \\
	&=& \pm \sqrt{ \sigma_{p}^2(\sigma_{a}^2+\sigma_{b}^2)-\sigma_{a}^2\sigma_{b}^2 }\frac{E(R_{a})-E(R_{b})}{\sigma_{a}^2+\sigma_{b}^2} + \sigma_{b}^2\frac{E(R_{a})-E(R_{b})}{\sigma_{a}^2+\sigma_{b}^2} + E(R_{b}) \\
\end{eqnarray*}

El comportamiento del rendimiento cuando la correlación entre los dos activos es cero 
se muestra en la Figura \ref{grafRend3}. Como puede verse, éste se describe como una curva, la cual presenta una segmento donde el riesgo disminuye y el rendimiento aumenta al variar las proporciones de ambos activos. Así, existe un portafolios de mínimo riesgo, el cual, posee un rendimiento mayor al del portafolios conformado únicamente por el activo $a$.
\begin{figure}[h]
\begin{center}
	\includegraphics[width=11cm]{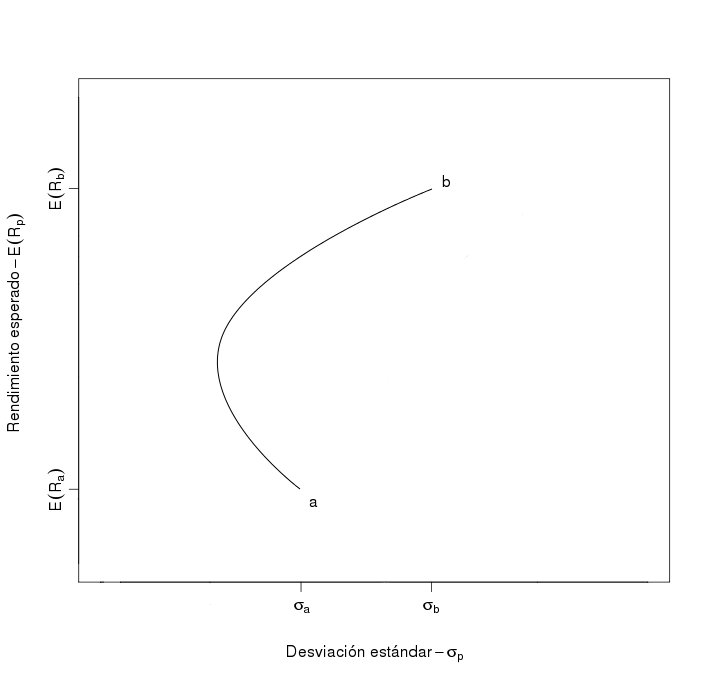} 
	\caption{Rendimiento esperado vs desviación estándar ($\rho_{ab} = 0$)} \label{grafRend3}
\end{center}
\end{figure}

\end{itemize}

\begin{figure}[h]
\begin{center}
	\includegraphics[width=11cm]{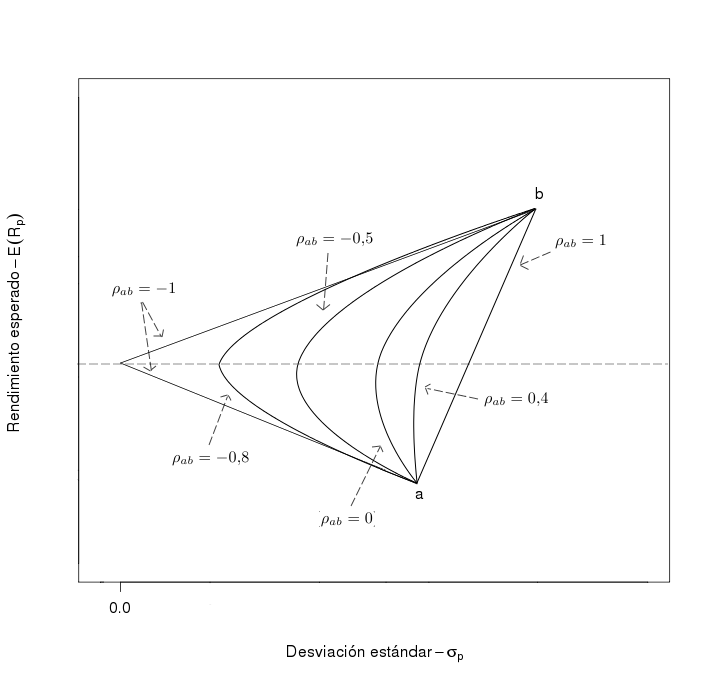} 
	\caption{Rendimiento esperado vs desviación estándar - forma general} \label{grafRend4}
\end{center}
\end{figure}

Con la información anterior podemos describir el caso general del comportamiento para el rendimiento de un portafolios de dos acciones, esto se obtiene al variar las proporciones de inversión en diferentes niveles de correlación entre sus activos; así, la gráfica de la Figura \ref{grafRend4} representa este caso general.

En la gráfica cada una de las lineas representa un conjunto de oportunidades para diferentes niveles de correlación. 

Como puede verse, en los extremos se encuentran los portafolios formados únicamente por los activos $a$ o $b$, además, cuando los rendimientos están perfectamente correlacionados ($\rho_{ab}=1$) se forma una recta, la cual implica al conjunto de portafolios de mayor riesgo. En cambio, si $-1<\rho_{ab}<1$ se forma una curva, que será más o menos pronunciada dependiendo del coeficiente de correlación, teniendo un mínimo donde los rendimientos están inversamente correlacionados ($\rho_{ab}=-1$), formando dos rectas que se intersectan en $\sigma_{p}=0$, siendo este punto el portafolios de mínimo riesgo posible.

Así, el triángulo generado en la gráfica establece lo que podríamos llamar la frontera del conjunto de oportunidades, dentro de la que estará cualquier portafolios posible.

\subsection{Conjunto de oportunidades para más de dos acciones}

En las gráficas del conjunto de oportunidades para dos acciones encontramos que la forma general de éste, es obtenida por una curva o una recta debido a que cada nivel de rendimiento del portafolios sólo puede ser generado por una  combinación de los dos activos que lo componen (asimismo la proporción de cada activo genera una varianza de portafolios, la cual puede no ser necesariamente única), sin embargo, el conjunto de oportunidades para más de dos valores es generado por un área, la razón se debe a que el rendimiento de portafolios puede ser obtenido por más de una sola combinación lineal, por ejemplo, si tenemos las acciones $a$, $b$, $c$ con rendimientos $a = 1.0$, $b = 2.0$, $c = 3.0$, entonces:

\begin{eqnarray}
\mu_{p} &=& (0.5)a + (0.0)b +(0.5)c = (0.5)(1) + (0.0)(2) +(0.5)(3) = 2.0 \nonumber \\
	&=& (0.0)a + (1.0)b +(0.0)c = (0.0)(1) + (1.0)(2) +(0.0)(3) = 2.0 \nonumber
\end{eqnarray}

\begin{figure}[h]
\begin{center}
	\includegraphics[width=11cm]{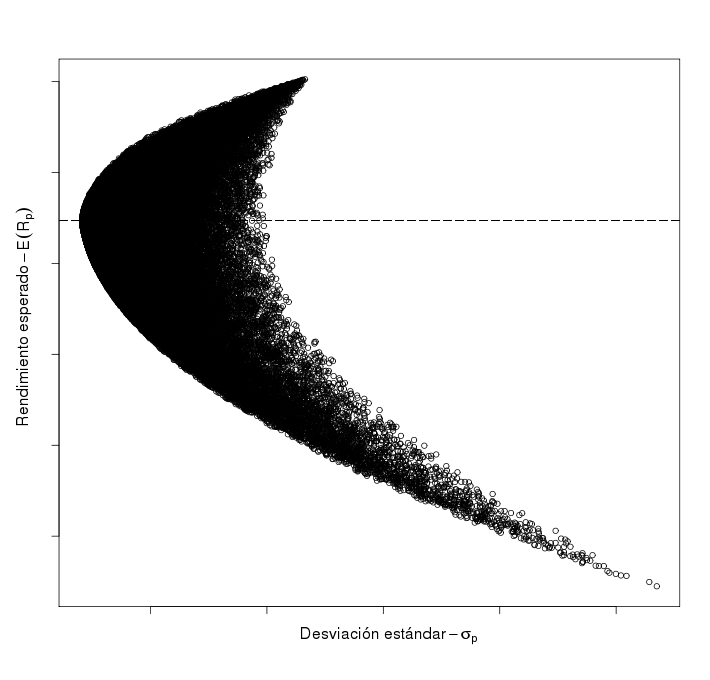} 
	\caption{Conjunto de oportunidades generado por 100,000 combinaciones para un portafolios de seis acciones} \label{front1}
\end{center}
\end{figure}

Una simulación en {\bf R}\footnote{100,000 combinaciones lineales, con proporciones de inversión aleatorias, generadas a partir de un portafolios de seis acciones (\^GSPC, DELL, AAPL, IBM, MSFT, INTC), utilizando el periodo de 2004-01-01 a 2007-01-01.} nos proporciona el conjunto de oportunidades que se muestra en la Figura \ref{front1}, así, al trazar una línea horizontal podemos observar que existen cientos de combinaciones que nos otorgan el mismo rendimiento, pero con diferentes riesgos.

El conjunto de oportunidades de un portafolios se encontrará en los límites del mínimo al máximo rendimiento esperado de cada uno de los activos que lo componen, esto es:

\begin{eqnarray}
& \mbox{Si} \quad M = \max \lbrace E(R_{i}) | i = 1, \cdots , n \rbrace \quad ; \quad m = \min \lbrace E(R_{i}) | i = 1, \cdots , n \rbrace & \label{rendminmax} \\
& \therefore E(R_{p}) \in [m, M] & \nonumber
\end{eqnarray}

La forma general del conjunto de oportunidades se deriva de una superficie sólida-curva que, a su vez, es producto de la unión de una serie de parábolas, de tal forma que nunca hay un salto o pliegue al pasar de una a otra, por lo que la gráfica del conjunto de oportunidades puede variar enormemente, pero utilizaremos la que se encuentra en la Figura \ref{front1} de manera general como un conjunto de oportunidades factible, ya que nos ayudará a marcar fácilmente varias de las características que se presentan en éste.

\section{Frontera eficiente}

\begin{figure}[h]
\begin{center}
	\includegraphics[width=11cm]{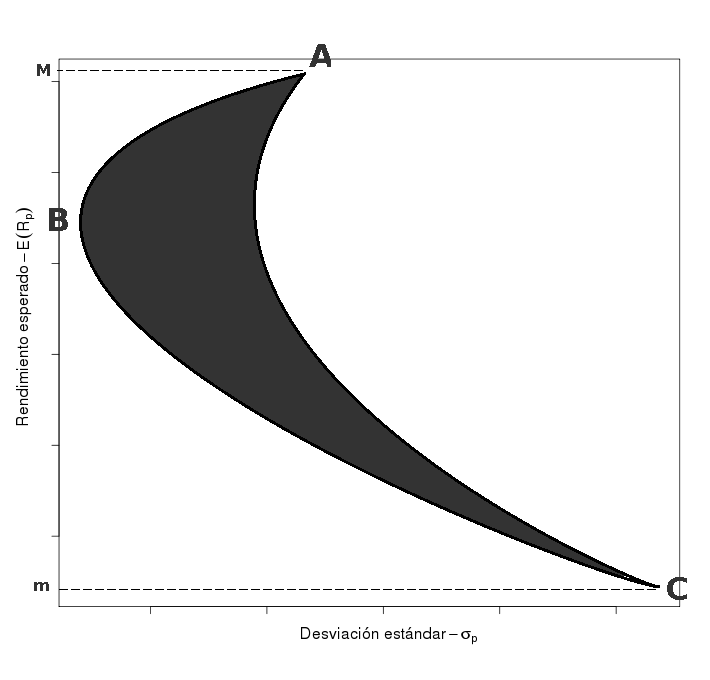} 
	\caption{Conjunto de oportunidades - forma general} \label{front2}
\end{center}
\end{figure}

En la Figura \ref{front2} se muestra un conjunto de oportunidades, puede observarse, como se mencionó anteriormente, que existen portafolios con un mismo nivel de rendimiento y diferente varianza, de esta forma la curva ABC es conocida como el {\bf Conjunto de oportunidades de varianza mínima}, ya que proporciona la varianza mínima posible para el rendimiento y por lo tanto la desviación estándar mínima.

Los inversionistas tienen diferentes actitudes frente al riesgo, siendo más o menos aversos a éste. Sin embargo, el conjunto de oportunidades de varianza mínima no necesariamente presenta los portafolios más apropiados, por esta razón se introduce el concepto de {\bf portafolios eficiente}.

Se dice que un portafolios es eficiente si la varianza asociada a su rendimiento esperado $E(R_{p})$, es la mínima de todos los posibles portafolios que proporcionan el mismo rendimiento, y dada su varianza $\sigma_{p}^{2}$, el rendimiento esperado es el máximo de todos los posibles portafolios con la misma varianza. Este principio también se le conoce como eficiencia con respecto al criterio Media-Varianza.

En otras palabras, el portafolios $A$ es eficiente si no existe ningún otro portafolios $B$ en el conjunto de oportunidades, tal que:

$$ E(R_{B}) \geq E(R_{A}) \quad \& \quad \sigma_{B}^{2} \leq \sigma_{A}^{2} $$

Es así que la curva AB representa el {\bf conjunto de portafolios eficientes} también llamado {\bf frontera eficiente}. En el punto $B$ se encuentra el portafolios de mínimo riesgo, y en el punto $A$ el de máximo rendimiento. Los inversionistas elegirán únicamente los portafolios que se encuentran en la frontera eficiente, es en este conjunto que se encuentran los portafolios más deseables y cualquier aumento en el rendimiento implica una penalización en riesgo.

Con esto se encuentran especificados los criterios más importantes para la selección de un portafolios. El conjunto de oportunidades que se ha manejado está compuesto únicamente por activos riesgosos; el caso en que se incluye un activo de renta fija\footnote{La renta fija se refiere a las inversiones en donde se conoce con anticipación cual será el rendimiento que nos proporcionará un activo. Ejemplo de renta fija son los bonos y las cuentas de ahorro. Por otro lado, la renta variable son las inversiones en donde no se sabe cual será el rendimiento que obtendremos, como es el caso de las acciones.} no está contemplado debido a que nuestro principal interés son las acciones, además de que su incorporación en el modelo es fácil de realizar\footnote{La inclusión de un activo libre de riesgo no afecta al cálculo de la varianza del portafolio, sólo se consideran los $w_i$ que involucran activos riesgosos, y en el cálculo del rendimiento se utilizan todos los activos.}.

\chapter{Modelo Media-Varianza}

{\em
Recordando: suponemos que un inversionista racional prefiere un alto rendimiento y es averso al riesgo. Así, dado un nivel de rendimiento, el inversionista escogerá aquel portafolios que minimice el riesgo. Esto es llamado, eficiencia con respecto al criterio Media-Varianza. Más específicamente, un portafolios es eficiente si, dado un nivel de rendimiento, no existe otro portafolios con menor riesgo e igual o mayor rendimiento, o dado un nivel de riesgo, no existe un portafolios con mayor rendimiento e igual o menor riesgo. El conjunto de todos los portafolios eficientes es llamado frontera eficiente. 

Debido a que cada inversionista tendrá diferentes preferencias en cuanto el rendimiento o riesgo a tomar, no podemos hablar de preferencias exactas. Sin embargo, todas las elecciones que sean tomadas, deberán estar en la frontera eficiente.

Así, el modelo de Markowitz consiste principalmente en dividir el proceso de selección de portafolios en dos pasos, donde primero se determina el conjunto de portafolios eficientes, y después el inversionista escoge de este conjunto aquel portafolios que cumple mejor sus preferencias.

Estos dos pasos se unen en un problema de programación matemática\footnote{Debido a que no se permiten las ventas en corto ($w_i \geq 0, \quad \forall i = 1, \cdots, n$), se debe utilizar un método de programación matemática, de otra forma se puede utilizar los ``Multiplicadores de Lagrange'' para resolver el modelo de Markowitz.}, cuyo objetivo es maximizar el rendimiento y minimizar el riesgo, lo cual permitirá satisfacer, en cierta medida, las preferencias del inversionista.
}

\section{Markowitz mediante programación cuadrática}

Dos formas de plantear la solución al modelo de Markowitz son: maximizar el rendimiento dado un nivel de riesgo y minimizar el riesgo dado un nivel de rendimiento.

Para el primer método, sea $\epsilon$ el máximo nivel de riesgo que el inversionista está dispuesto a aceptar. La formulación para este problema de programación matemática es:

\begin{eqnarray}
\mbox{maximizar} \quad E({\bm w}'{\bm R}) \\
 \nonumber \\
s.a. \quad \sqrt{ {\bm w}'{\bm \Sigma}{\bm w} } \leq \epsilon \nonumber \\
{\bm w}'{\bm 1} = 1 \nonumber \\
{\bm w} \geq {\bm 0} \nonumber 
\end{eqnarray}

donde ${\bm w} \geq {\bm 0}$ se interpreta como $w_i \geq 0, \quad \forall i = 1, \cdots, n$. Además, la desigualdad $\sqrt{ {\bm w}'{\bm \Sigma}{\bm w} } \leq \epsilon$ define un conjunto convexo de variables ${\bm w}$ y las restricciones restantes son lineales en ${\bm w}$. Por lo tanto, este es un problema de optimización de una función lineal, sobre un conjunto convexo.

Para el segundo método, minimizar el riesgo dado un nivel de rendimiento, la formulación es:

\begin{eqnarray}
\mbox{minimizar} \quad {\bm w}'{\bm \Sigma}{\bm w} \label{quadp} \\
 \nonumber \\
s.a. \quad E({\bm w}'{\bm R}) \geq \beta \nonumber \\
{\bm w}'{\bm 1} = 1 \nonumber \\
{\bm w} \geq {\bm 0} \nonumber 
\end{eqnarray}

donde $\beta$ es el rendimiento mínimo que el inversionista está dispuesto a aceptar. La función objetivo ${\bm w}'{\bm \Sigma}{\bm w}$ es una función cuadrática convexa con restricciones lineales, por lo que se trata de un problema de programación cuadrática, de la forma:

\begin{eqnarray}
\mbox{minimizar} \quad -{\bm \delta}'{\bm x} + \frac{1}{2} {\bm x}'{\bm D}{\bm x} \nonumber \\
 \nonumber \\
s.a. \quad {\bm A}'{\bm x} \geq \lambda \nonumber \\
{\bm x}'{\bm 1} = 1 \nonumber \\
{\bm x} \geq {\bm 0} \nonumber 
\end{eqnarray}

donde la matriz ${\bm D}$ es semidefinida positiva y ${\bm \delta}$ es un vector que aparece en la función cuadrática para ser minimizado. La matriz ${\bm A}$ define las constantes bajo las que queremos minimizar la función cuadrática y ${\bm x}$ es un vector (solución) desconocido que minimiza la función cuadrática, bajo las restricciones proporcionadas.

Así, en el modelo de Markowitz, ${\bm \delta} = {\bm 0}$, la matriz de la función cuadrática a minimizar es ${\bm \Sigma}$ y ${\bm A} = E({\bm R})$. Por lo tanto, $\frac{1}{2} {\bm w}'{\bm \Sigma}{\bm w}$ es la función a minimizar o en su lugar ${\bm w}'{\bm \Sigma}{\bm w}$, ya que evidentemente la misma solución ${\bm w}$ minimiza ambas funciones.

El modelo de programación cuadrática en (\ref{quadp}) será el que utilizaremos en adelante para seleccionar portafolios por medio del criterio Media-Varianza.

\section{Utilidad y el criterio Media-Varianza}

Las decisiones tomadas en el Mercado de Valores involucran riesgo, por lo cual, el seleccionar el portafolios de máximo rendimiento no es la decisión más apropiada. Un inversionista que busca el máximo rendimiento esperado jamás escogerá un portafolios diversificado, sin embargo, las diferentes actitudes de los inversionistas frente al riesgo involucran diferentes elecciones de portafolios, para dar explicación a esta situación se utiliza la Teoría de Utilidad y el criterio de máxima utilidad esperada.

En el modelo de Markowitz se asume que se conoce la distribución de los rendimientos; el rendimiento de un portafolios es cuantificado por medio de su valor esperado y su riesgo por medio de la varianza, lo cual haría parecer que Markowitz simplemente presenta así su modelo, sin embargo esto no se hace sin justificación. En general, la introducción de la varianza es presentada como una aproximación cuadrática a una función de utilidad general.

Una función de utilidad es una función real que mide la ``satisfacción'' o ``utilidad'' obtenida mediante diferentes bienes y servicios. 

Es así que, de acuerdo con la Teoría de Utilidad, cada individuo posee preferencias con respecto a un conjunto de alternativas, dichas preferencias pueden expresarse por medio de una función de utilidad.

Una característica de toda función de utilidad, es ser creciente, por lo que su primer derivada cumple con la desigualdad $U'(X)>0$. 

\begin{figure}[h]
\begin{center}
	\includegraphics[width=11cm]{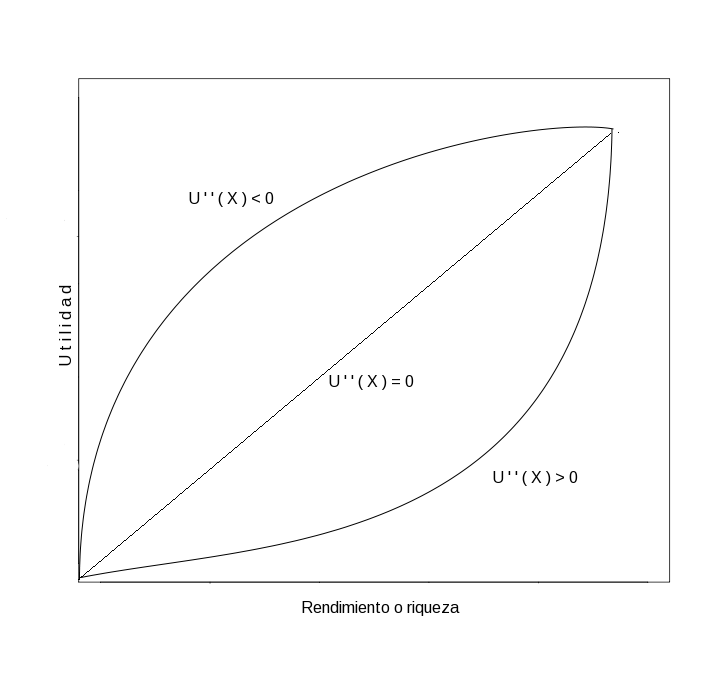} 
	\caption{Los tres tipos básicos de funciones de utilidad ($U'(X)>0$)} \label{utility}
\end{center}
\end{figure}

Con respecto al riesgo tenemos tres tipos básicos de funciones de utilidad, que se muestran en la Figura \ref{utility} y son:
\\
\begin{itemize}
\item[$\bullet$] {\em \bf Cuando $U''(X)<0$} : Significa que el incremento de la función de utilidad es cada vez menor, por lo que se trata de una función cóncava (cóncava hacia abajo). Es la función que representa al inversionista {\bf averso al riesgo}, debido que a incrementos iguales de riqueza en su función de utilidad (recordar que el rendimiento está altamente correlacionado con el riesgo), el incremento en su satisfacción (utilidad), es cada vez menor.

\item[$\bullet$] {\em \bf Cuando $U''(X)=0$} : Significa que el crecimiento de la función de utilidad es constante, por lo que su gráfica es una recta. Esta función representa al inversionista {\bf indiferente al riesgo}, ya que el nivel de satisfacción que obtiene es directamente proporcional a la riqueza que obtiene.

\item[$\bullet$] {\em \bf Cuando $U''(X)>0$} : Significa que el incremento de la función de utilidad es cada vez mayor, por lo que se trata de una función convexa (cóncava hacia arriba). Es la función que representa al inversionista {\bf adicto al riesgo}, debido a que a incrementos iguales de riqueza en su función de utilidad, el incremento en su satisfacción es cada vez mayor.
\end{itemize}

En cuanto al modelo de Markowitz, la función de utilidad es una función cuadrática\footnote{Esta función, como puede verse en la ecuación (\ref{quaduf}), corresponde a una función de utilidad cóncava, por lo que representa al inversionista averso al riesgo.}, la cual está dada por:

\begin{eqnarray}
 U = U(R_p) = R_p - AR^2_p \label{quaduf}
\end{eqnarray}

donde $A>0$ y $R_p \leq \frac{1}{2A}$, por lo que la utilidad esperada es:

\begin{eqnarray}
E(U(R_p)) &=&  E(R_p - AR^2_p) = E(R_p) - AE(R^2_p) = E(R_p) - A \left( \sigma^2_p + E(R_p)^2 \right) \nonumber \\
          &=&  E(R_p) - AE(R_p)^2 - A \sigma^2_p \label{uesp}
\end{eqnarray} 

Nótese que la utilidad esperada en (\ref{uesp}) puede interpretarse como una función del rendimiento del portafolios, que es penalizado con el riesgo del mismo\footnote{La variable $A$ representaría la proporción en que deseamos se penalice el riesgo.}, es decir, el portafolios de máxima utilidad esperada no será necesariamente el de mayor rendimiento, ya que, dependiendo de la aversión al riesgo del inversionista, si éste es proporcionalmente más grande que el de otros portafolios, se elegirá un portafolios de rendimiento medio, y poco riesgo.

Matemáticamente es posible modelar la conducta de un inversionista ``racional'' mediante una función de utilidad. A pesar de esto, en la práctica no conocemos la función de utilidad del inversionista, sin embargo, si los inversionistas tienen preferencias similares, sus funciones de utilidad serán similares. Es así que es adecuado establecer un criterio de eficiencia ({\it M-V}), para determinar un conjunto de opciones eficientes (Frontera eficiente), para que cada inversionista determine cual de ellas satisface mejor sus preferencias.

El criterio Media-Varianza, asume que el inversionista tendrá una función de utilidad cuadrática, o en su defecto, una función cóncava con una distribución elíptica\footnote{Markowitz muestra que en este caso el criterio {\it M-V} también es adecuado} de los rendimientos, como por ejemplo la distribución normal. Esto es, la función de utilidad debe ser cóncava; aunque en 1952 Markowitz publicó otro artículo en el que asume una función de utilidad no-cóncava\footnote{Markowitz, H.M. (1952b), ``The Utility of Wealth'', \textit{Journal of Political Economy, 60}, 151-156.}.

\section{Medidas alternativas de tendencia central}

La media es el nombre técnico de un simple promedio, es así que la media de una serie de datos se le llama promedio y a la media de una variable aleatoria se le llama valor esperado. Recordemos que la media de una variable aleatoria es el centro de gravedad definido por la función de densidad de la misma variable aleatoria. Ya que consideramos como una variable aleatoria al rendimiento de una empresa ($R_i$) y por consiguiente al rendimiento del portafolios ($R_p$), su valor esperado (el rendimiento esperado) es una medida de dónde los valores de los rendimientos están ``centrados''.

Sin embargo existen otras medidas de centralidad además de la media, las cuales, dependiendo de algunas características de los datos (como la distribución de probabilidad), pueden ser más adecuadas en un análisis. 
\\
\\
Las principales medidas de tendencia central son:

\begin{itemize}
\item[$\bullet$] {\em \bf Media} : La media es una medida adecuada cuando el conjunto de datos es relativamente homogéneo, de tal forma que su varianza sea pequeña; por lo que la media será una medida representativa de la población.

\item[$\bullet$] {\em \bf Mediana} : La mediana es una medida apropiada cuando en los datos existen valores extremos. En el caso de los rendimientos, puede haber alguno muy pequeño o muy grande que afecte considerablemente a una medida como la media, pero no afectar a la mediana al ser una medida en la que el 50\% de la probabilidad es asociada con números menores o iguales a éste y el otro 50\% está asociado a valores mayores o iguales.

\item[$\bullet$] {\em \bf Moda} : La moda es el valor que aparece con mayor probabilidad, o en el caso continuo aquel valor donde se encuentra el máximo de la función de densidad (si es que existe). Es una medida más adecuada cuando la mayor o una buena parte de los datos se encuentran agrupados en una zona especifica del rango posible de valores.
\end{itemize}

Una medida de tendencia central será mejor que otra si genera mejores portafolios eficientes. Nótese que para algunas distribuciones de probabilidad, como la normal, estas tres medidas se sitúan en el mismo punto, por lo que los criterios de decisión, Media-Varianza, Mediana-Varianza y Moda-Varianza, generan el mismo conjunto eficiente de portafolios.

También puede cambiarse la medida utilizada para el riesgo en el modelo de Markowitz, por ejemplo la probabilidad de pérdida o la semivarianza; esta ultima será el tema que se aborda en el siguiente capítulo.

En adelante sólo utilizaremos la media (rendimiento esperado) como medida del rendimiento de un portafolios, esto debido a que nuestro propósito no es probar cada una de las opciones presentadas, sino mostrar algunos de los distintos métodos en la construcción y optimización de portafolios, sus características, funcionamiento, variaciones, resultados, ventajas y desventajas.

\section{Aplicando el modelo de Markowitz}

Para la aplicación del modelo se consideraron 95 empresas, de las cuales la mayor parte son de gran importancia en el mercado de valores y otras se eligieron por su relación media-varianza. El periodo de análisis comprende de Enero de 2005 a Diciembre de 2007, por lo que una característica común de las empresas a estudiar es que cotizan al menos desde el año 2002\footnote{Consideramos deseable 3 años de cotización en la Bolsa para que los precios de una empresa se estabilicen y marquen una tendencia relativamente confiable.}. 

Para analizar el comportamiento del modelo, se emplea el intervalo temporal de los años 2005 a 2007, antes de la crisis económica de 2008, ya que, comparado con los años recientes, los datos de 2005--2007 presentan un entorno menos distorsionado por políticas monetarias extremas (como los tipos de interés muy bajos o la expansión cuantitativa), que han influido fuertemente en los mercados a partir de 2008. Esto permite una evaluación más clara de los mecanismos subyacentes del mercado, sin que estén tan afectados por intervenciones externas masivas. Además, no hay influencia de eventos disruptivos globales como la pandemia de COVID-19. Al usar datos anteriores a 2008, el análisis se enfoca en patrones clásicos de mercado, sin la distorsión introducida por fenómenos como la masificación de trading algorítmico, las criptomonedas o la digitalización acelerada del trading, que han cambiado la dinámica de los mercados en años recientes. 

Este enfoque es particularmente valioso cuando se busca evaluar la capacidad subyacente del modelo para predecir dinámicas tradicionales de mercado, ya que posteriormente el modelo podría mostrar mejores resultados de lo que es realista de esperar, debido al propio crecimiento acelerado que ha experimentado el mercado desde el año 2009.

\noindent Las empresas a utilizar son:

\begin{multicols}{2}
\begin{verbatim}
1	AirTran Holdings, Inc.
2	Apple Inc.
3	Advanced Battery Technologies
4	ARCA biopharma, Inc.
5	Ambac Financial Group, Inc.
6	Aluminum Corporation of China 
7	Axcelis Technologies, Inc.
8	Adobe Systems
9	American International Group
10	AAR Corp.
11	ATP Oil & Gas Corporation
12	Abraxas Petroleum Corporation 
13	Bank of America Corporation
14	Blockbuster Inc.
15	Best Buy
16	Bank of New York Mellon Corp.
17	Citigroup, Inc.
18	Cisco Systems, Inc. 
19	Cell Therapeutics, Inc. 
20	E.I. du Pont de Nemours
21	Dell
22	Walt Disney Company
23	Dendreon Corporation 
24	Dow Chemical Company
25	Ford Motor Company
26	Federal Home Loan Mortgage
27	General Electric Company 
28	Home Depot, Inc.
29	Honeywell International Inc.
30	Hewlett-Packard 
31	Intuitive Surgical, Inc.
32	JDS Uniphase Corporation
33	Johnson & Johnson
34	Juniper Networks, Inc. 
35	Kellogg Company 
36	KeyCorp 
37	Krispy Kreme Doughnuts, Inc.
38	Kinder Morgan Energy Partners
39	Coca-Cola
40	Loews Corporation 
41	Laboratory Corporation of America
42	Eli Lilly and Company
43	Macy's Inc 
44	Mattel
45	Massey Energy Company
46	New M&I Corporation 
47	3M
48	Altria Group, Inc.
49	Merck & Company, Inc.
50	Marathon Oil Corporation
51	Microsoft
52	Nike
53	Nokia
54	NVIDIA Corporation
55	New York Community Bancorp
56	Telecom Corporation
57	Realty Income Corporation
58	Oracle Corporation
59	Occidental Petroleum
60	Palm, Inc.
61	Pepsico
62	Pfizer, Inc.
63	Procter & Gamble Company
64	Potash Corporation
65	PetroChina
66	QUALCOMM Incorporated
67	PowerShares Exchange-Traded
68	Quantum Corporation 
69	Quantum Fuel Systems
70	Rite Aid Corporation
71	Regions Financial Corporation
72	Research In Motion Limited
73	U.S. Concrete, Inc.
74	Sprint Nextel Corporation
75	Sirius XM Radio Inc.
76	STEC, Inc. 
77	AT&T
78	Toyota Motor Corporation
79	Time Warner Inc. 
80	Texas Instruments Incorporated
81	VODAFONE
82	Verizon Communications Inc.
83	Walmart
84	United States Steel Corporation
85	Exxon Mobil Corporation
86	Xerox
87	Alleghany Corporation
88	Yahoo!
89	YRC Worldwide
90	Yum! Brands
91	Zions Bancorporation 
92	Zix Corporation
93	Zale Corporation
94	Zoltek Companies
95	Quiksilver 
\end{verbatim}
\end{multicols}

El objetivo detrás del análisis de estas empresas por medio del modelo de Markowitz es obtener un portafolios que proyecte nuestras expectativas de rendimiento para el periodo de Enero-Diciembre de 2007, es decir, utilizaremos los rendimientos de 2005 y 2006 para crear un portafolios que utilicemos durante 2007\footnote{Consideramos adecuado utilizar el doble de información del pasado, con respecto al futuro al que se proyecta. En este caso queremos construir un portafolios para un año, 2007, por lo que utilizamos dos años de rendimientos como información previa, 2005 y 2006; los cuales nos mostraran el comportamiento de los rendimientos que podemos esperar a futuro.}.

La base de datos consta de los precios de cierre diarios ajustados\footnote{Yahoo! reporta sus precios de cierre ajustados como aquellos que han sido corregidos por motivo de splits y dividendos.} (Adj.Close) de las 95 empresas. En caso de que alguna empresa no operara en determinada fecha, y otra sí, se utiliza el precio de cierre ajustado del día inmediatamente anterior. Es así que para cada empresa tenemos 503 observaciones que corresponden a sus precios de cierre ajustados de 2005-2006, los cuales se encuentran en el archivo que llamaremos ``prices-2005-2006.csv''.

En adelante usaremos el paquete estadístico ``R'' y el código programado en el archivo ``markowitz.R'', el cual se encuentra en el apéndice \ref{codf}\footnote{Se ha desarrollado un código propio, especifico para los modelos abordados en esta investigación, el cual se puede adaptar fácilmente  para distintas necesidades, por ejemplo permitir ventas en corto; sin embargo puede utilizarse la biblioteca ``fPortfolio'' en R, para un trabajo más completo con portafolios, al igual que para consultar su código.}.

\subsection{Dispersión}

Una gráfica importante es la dispersión entre la desviación estándar y el rendimiento esperado de cada empresa, ya que nos mostrará la relación individual entre riesgo y rendimiento de las empresas que compondrán el futuro portafolios.

\begin{verbatim}
# importamos los precios, y los guardamos en la variable 'prices'
prices<-read.table('prices-2005-2006.csv', header=T, sep=',')  

# obtenemos los rendimientos y los guardamos en 'R'
R<-assetsReturn(prices)

# graficamos 
plot( sqrt(diag(var(R))), colMeans(R), xlab="Desviación estándar", 
  ylab="Rendimiento esperado" )
\end{verbatim}

\begin{figure}[h]
\begin{center}
	\includegraphics[width=11cm]{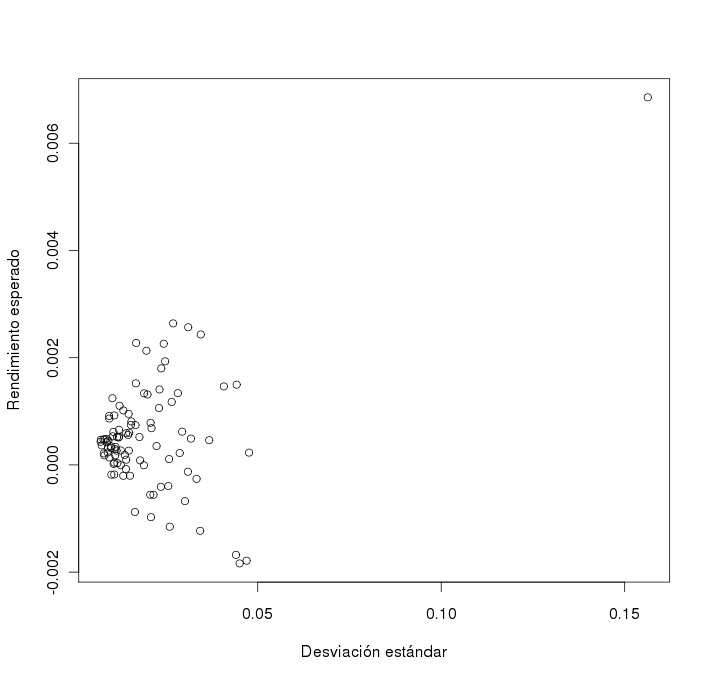} 
	\caption{Gráfica de dispersión de las empresas} \label{disp}
\end{center}
\end{figure}

El resultado puede verse en la Figura \ref{disp}, la cual muestra que existen empresas con desviación estándar similar, pero dentro de éstas algunas tienen mayor rendimiento esperado que otras y de la misma forma existen empresas con un nivel similar de rendimiento, pero con diferentes riesgos. También puede observarse la relación riesgo-rendimiento, siendo forzoso aceptar mayor riesgo para encontrar empresas con mayor rendimiento; en este caso la empresa con mayor rendimiento igualmente es la que posee mayor riesgo.

Hablando en términos matemáticos, el rendimiento esperado de una empresa podría ser tan grande como se desee y su varianza ser muy pequeña o incluso cero, sin embargo, en la vida real los rendimientos de una empresa normalmente no se distribuirán de forma que esto sea posible, ya que habrá ganancias y pérdidas de diferentes tamaños, por lo que serán necesarios incrementos de mayor tamaño en los rendimientos de una empresa para que éste sea mayor que el de otra en el mismo periodo de tiempo, lo que se traduce a mayor varianza. Es así que en la realidad encontraremos una estrecha relación entre los rendimientos y el riesgo de un activo.

\subsection{Conjunto de oportunidades y frontera eficiente}

El conjunto de oportunidades es una gráfica de gran importancia, esto debido a que podemos observar claramente como se comporta el riesgo mediante la diversificación, la frontera eficiente, y en qué nivel o rango de rendimientos nos conviene movernos. Esta gráfica la podemos obtener mediante:

\begin{verbatim}
# leemos el archivo de funciones 'markowitz.R'
source('markowitz.R') 

# importamos los precios, y los guardamos en la variable 'prices'
prices<-read.table('prices-2005-2006.csv', header=T, sep=',')  

# obtenemos los rendimientos y los guardamos en 'R'
R<-assetsReturn(prices)

# nuevo objeto 'portafolios', con medida de riesgo 'var'
pf<-newPortfolio(R)

# graficamos el conjunto de oportunidades
plotPortfolio(pf)
\end{verbatim}

\begin{figure}[h]
\begin{center}
	\includegraphics[width=11cm]{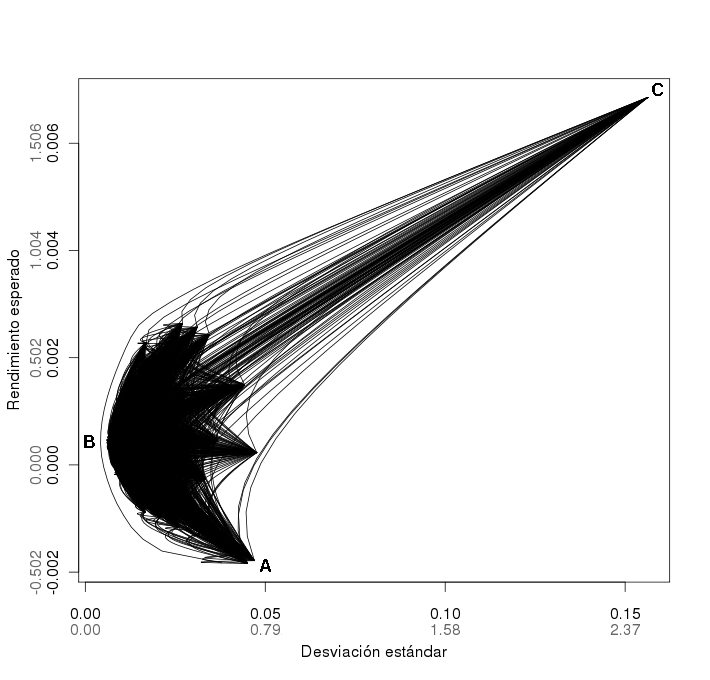} 
	\caption{Conjunto de oportunidades} \label{front3}
\end{center}
\end{figure}

El resultado se muestra en la Figura \ref{front3}. En negro se denotan los valores de los rendimientos-desviaciones estándar diarios y en gris los anuales. Las curvas que se encuentran en la gráfica representan todas las posibles combinaciones de portafolios conformados por dos empresas. Estas curvas, junto con el conjunto de varianza mínima\footnote{El conjunto de varianza mínima es aquel conjunto de portafolios que tienen la menor varianza para cada nivel de rendimiento posible.} (curva ABC), contendrán al conjunto de oportunidades, esto es, cualquier portafolios posible estará dentro de los límites conformados por estas curvas.

Podemos ver que existe cierta relación con la gráfica de dispersión, ya que si la sobreponemos al conjunto de oportunidades, los vértices que conforman las uniones de las curvas corresponden con portafolios de una sola empresa, de los cuales, su desviación estándar y rendimiento son los mismos que en la gráfica de dispersión.

La frontera eficiente se presenta en la curva BC, en la cual se observa un punto a partir del que el riesgo crece rápidamente conforme aumenta el rendimiento. En adelante seleccionaremos algunos portafolios sobre esta curva para estudiar como se comportan a futuro.

En la Figura \ref{front4} podemos ver el mismo conjunto de oportunidades, pero con $100,000$ portafolios aleatorios. La mayor parte de estos portafolios se encuentran donde se cruzan las curvas, ya que estos lugares corresponde a mayores combinaciones de portafolios (más densas), por lo que tienen mayor probabilidad de aparecer en estas zonas. La gráfica la obtenemos con la sentencia:

\begin{verbatim}
plotPortfolio(pf, points=100000)
\end{verbatim}

\begin{figure}[h]
\begin{center}
	\includegraphics[width=11cm]{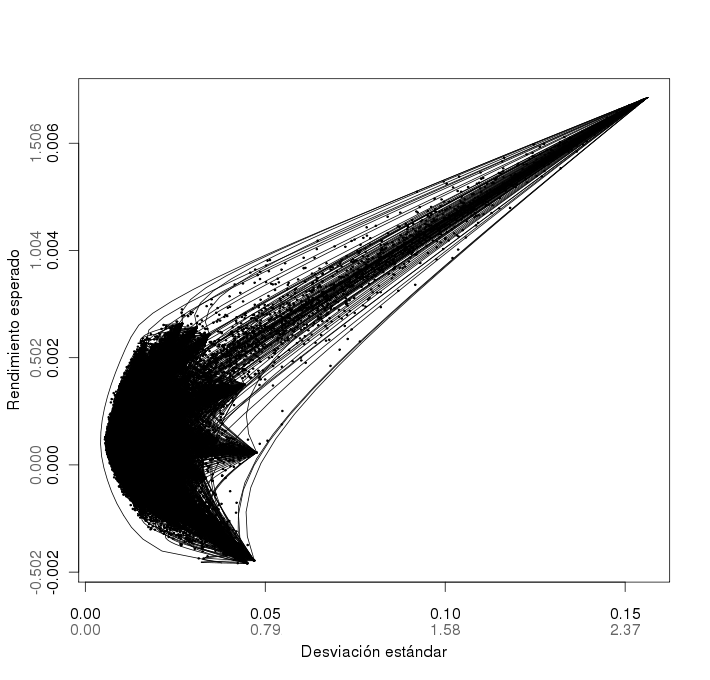} 
	\caption{10,0000 portafolios aleatorios en el conjunto de oportunidades} \label{front4}
\end{center}
\end{figure}

\subsection{Portafolios de varianza mínima}

Para obtener el portafolios de varianza mínima utilizamos el mismo modelo de programación cuadrática, ignorando el rendimiento, esto es:

\begin{eqnarray}
\mbox{minimizar} \quad {\bm w}'{\bm \Sigma}{\bm w} \\
 \nonumber \\
s.a. \quad {\bm w}'{\bm 1} = 1 \nonumber \\
{\bm w} \geq {\bm 0} \nonumber 
\end{eqnarray}

Para encontrar este portafolios, en nuestro programa utilizamos:

\begin{verbatim}
# Encontramos el portafolios de varianza mínima y almacenamos el 
# resultado en 'p'
p <- markowitzPortfolio( pf )

# imprimimos en pantalla el contenido de p
print( p )

# er=rendimiento esperado, sd=desviación estándar, weights=proporciones
# de compra para cada empresa, portfolio=identificadores de las empresas
# que compraremos acciones, y la proporción de compra para cada una.

$call
markowitzPortfolio(pObject = pf)

$er
[1] 0.0004407315

$sd
[1] 0.004236929

$weights
 [1] 0.0000 0.0000 0.0018 0.0000 0.0000 0.0000 0.0000 0.0000 0.0000 0.0000
[11] 0.0000 0.0000 0.0041 0.0000 0.0000 0.0000 0.0243 0.0000 0.0000 0.0000
[21] 0.0000 0.0000 0.0000 0.0000 0.0000 0.0000 0.0000 0.0000 0.0000 0.0000
[31] 0.0000 0.0000 0.1172 0.0000 0.1176 0.0000 0.0000 0.1558 0.0117 0.0000
[41] 0.0735 0.0183 0.0000 0.0000 0.0000 0.0000 0.0000 0.0265 0.0000 0.0000
[51] 0.0162 0.0370 0.0000 0.0000 0.0196 0.0606 0.0000 0.0000 0.0000 0.0000
[61] 0.1333 0.0000 0.0000 0.0000 0.0013 0.0000 0.0000 0.0000 0.0000 0.0000
[71] 0.0000 0.0000 0.0000 0.0000 0.0004 0.0000 0.0310 0.0000 0.0157 0.0000
[81] 0.0000 0.0000 0.0138 0.0000 0.0000 0.0000 0.1202 0.0000 0.0000 0.0000
[91] 0.0000 0.0000 0.0000 0.0000 0.0000

$portfolio
ABAT_Adj.Close  BAC_Adj.Close    C_Adj.Close  JNJ_Adj.Close    K_Adj.Close 
        0.0018         0.0041         0.0243         0.1172         0.1176 
 KMP_Adj.Close   KO_Adj.Close   LH_Adj.Close  LLY_Adj.Close   MO_Adj.Close 
        0.1558         0.0117         0.0735         0.0183         0.0265 
MSFT_Adj.Close  NKE_Adj.Close  NYB_Adj.Close  NZT_Adj.Close  PEP_Adj.Close 
        0.0162         0.0370         0.0196         0.0606         0.1333 
 PTR_Adj.Close SIRI_Adj.Close    T_Adj.Close  TWX_Adj.Close  WMT_Adj.Close 
        0.0013         0.0004         0.0310         0.0157         0.0138 
   Y_Adj.Close 
        0.1202 
\end{verbatim}

De esta forma, nuestro portafolios de varianza mínima tiene un rendimiento esperado diario de 0.000440 \mbox{({\it p\$er=0.0004407315})}. Tenemos 502 rendimientos en el periodo 2006-2007, por lo que consideramos 251 rendimientos al año, es así que calculamos un rendimiento esperado anual de 11.0\% \mbox{({\it p\$er*251=0.1106236})} para el portafolios de varianza mínima.

Para conocer el rendimiento que obtendremos en 2007 con este portafolios, utilizamos el archivo de precios de cierre ajustados para las 95 empresas en este periodo (``prices-2008.csv''), el cual consta de 251 observaciones, esto es, 250 rendimientos.

\begin{verbatim}
# importamos los precios de 2007, y los guardamos en la variable 'prices2'
prices2 <- read.table('prices-2007.csv', header=T, sep=',')  

# obtenemos los rendimientos y los guardamos en 'R2'
R2 <- assetsReturn(prices2)

# producto cruzado entre las proporciones del portafolios 2005-2006 y los 
# rendimientos esperados de 2007
r <- p$weights %*% colMeans(R2)

print( r )
             [,1]
[1,] 0.0004532749

print( r * 250 )
          [,1]
[1,] 0.1133187
\end{verbatim}

Es así que, con el portafolios de varianza mínima obtenemos un rendimiento de 11.3\% en 2007 \mbox{({\it r*250=0.1133187})}, muy cercano del 11.0\% que proyectamos conseguir conforme el resultado del modelo de Markowitz.

Este portafolios lo podemos graficar en el conjunto de oportunidades mediante la comando:

\begin{verbatim}
# graficando un portafolios con rendimiento 'er' 
# y desviación estándar 'sd'
plotPortfolio(pf, er = p$er, sd = p$sd)
\end{verbatim}

\begin{figure}[h]
\begin{center}
	\includegraphics[width=11cm]{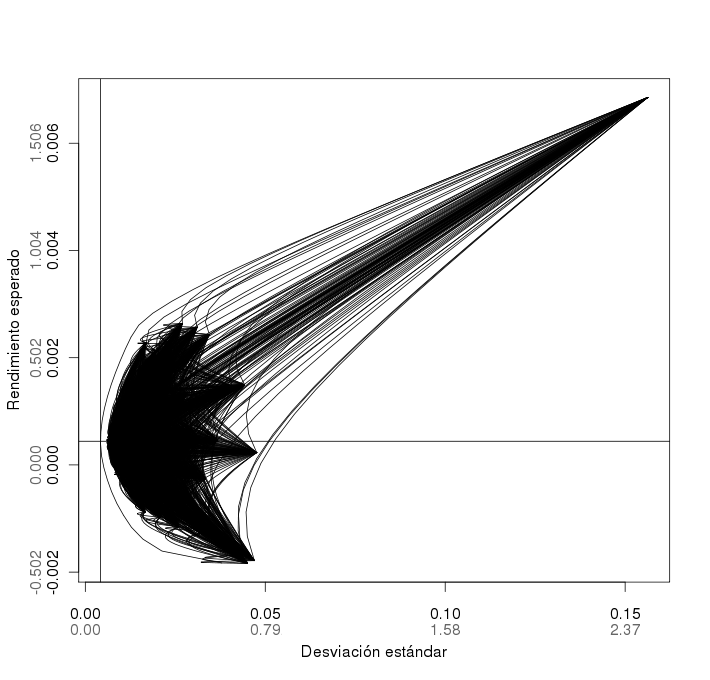} 
	\caption{Portafolios de varianza mínima} \label{front5}
\end{center}
\end{figure}

El resultado se muestra en la Figura \ref{front5}. El portafolios de varianza mínima se encuentra donde se intersectan la recta vertical y horizontal. Es este punto el de menor riesgo dentro de la frontera eficiente, es decir, la frontera eficiente comprende la curva entre los puntos donde se encuentra el portafolios de varianza mínima y el portafolios de máximo rendimiento.

Las empresas que se les asigna mayor proporción de la inversión en el portafolios de varianza mínima son: Kinder Morgan Energy Partners (KMP) con 15.5\%, Pepsico (PEP) con 13.3\%, Alleghany Corporation con 12\%, Johnson \& Johnson (JNJ) y Kellogg Company (K) con 11.7\%.

Aunque el riesgo de este portafolios (en este caso) es muy pequeño, el rendimiento también lo es, por lo que un inversionista podría preferir inversiones en renta fija que proporcionen un rendimiento cercano, sin riesgo\footnote{Por ejemplo, para 2007 los CETES ofrecían un rendimiento anual por arriba del 7\%, por lo que esta pudo ser una mejor opción para un inversionista con gran aversión al riesgo.}.

\subsection{Especificando el rendimiento}

Anteriormente hemos visto el portafolios de varianza mínima, donde no conocemos de antemano el rendimiento esperado que tendrá. Sin embargo este rendimiento será muy pequeño la mayoría de las ocasiones, por lo que es deseable resolver el problema de optimización especificando el rendimiento que deseamos, el cual debe estar entre los límites de la ecuación (\ref{rendminmax}) para que el modelo tenga solución.

Seleccionaremos tres niveles de rendimiento con los que trabajaremos, (1) de riesgo bajo, (2) de riesgo medio, y (3) de riesgo alto. Estos niveles son escogidos conforme al modelo de optimización (\ref{lquadp}), el cual se vera más adelante y en el que $\lambda$ es un factor de aversión al riesgo, por lo que al variar $\lambda$ obtenemos portafolios de mayor riesgo asociados a un nivel de rendimiento.

\subsubsection{Portafolios de riesgo bajo}

Para este portafolios utilizaremos 0.002 como el rendimiento esperado diario que deseamos. Puede verse que en este nivel de rendimiento la curva de la frontera eficiente aun no crece rápidamente con respecto a la desviación estándar.

\begin{verbatim}
# Encontramos el portafolios de varianza mínima con el rendimiento
# deseado y almacenamos el resultado en 'p'
p <- markowitzPortfolio( pf, 0.002 )

# imprimimos en pantalla el contenido de p
print( p )

# er=rendimiento esperado, sd=desviación estándar, weights=proporciones
# de compra para cada empresa, portfolio=identificadores de las empresas
# que compraremos acciones, y la proporción de compra para cada una.
$call
markowitzPortfolio(pObject = pf, eRet = 0.002)

$er
[1] 0.002

$sd
[1] 0.01006362

$weights
 [1] 0.0000 0.0780 0.0208 0.0000 0.0000 0.0000 0.0000 0.0000 0.0000 0.0196
[11] 0.0000 0.0000 0.0000 0.0000 0.0000 0.0000 0.0000 0.0000 0.0000 0.0000
[21] 0.0000 0.0000 0.0000 0.0000 0.0000 0.0000 0.0000 0.0000 0.0000 0.1085
[31] 0.0394 0.0000 0.0000 0.0000 0.0000 0.0000 0.0000 0.0000 0.0000 0.1092
[41] 0.0377 0.0000 0.0000 0.0000 0.0000 0.0000 0.0000 0.0147 0.0515 0.0489
[51] 0.0000 0.0000 0.0000 0.0467 0.0000 0.0000 0.0000 0.0000 0.0000 0.0000
[61] 0.0000 0.0000 0.0000 0.0000 0.2882 0.0000 0.0000 0.0000 0.0000 0.0000
[71] 0.0000 0.0078 0.0000 0.0000 0.0000 0.0865 0.0426 0.0000 0.0000 0.0000
[81] 0.0000 0.0000 0.0000 0.0000 0.0000 0.0000 0.0000 0.0000 0.0000 0.0000
[91] 0.0000 0.0000 0.0000 0.0000 0.0000

$portfolio
AAPL_Adj.Close ABAT_Adj.Close  AIR_Adj.Close  HPQ_Adj.Close ISRG_Adj.Close 
        0.0780         0.0208         0.0196         0.1085         0.0394 
   L_Adj.Close   LH_Adj.Close   MO_Adj.Close  MRK_Adj.Close  MRO_Adj.Close 
        0.1092         0.0377         0.0147         0.0515         0.0489 
NVDA_Adj.Close  PTR_Adj.Close RIMM_Adj.Close STEC_Adj.Close    T_Adj.Close 
        0.0467         0.2882         0.0078         0.0865         0.0426 
\end{verbatim}

De esta manera nuestro portafolios tiene un rendimiento esperado diario de 0.002 \mbox{({\it p\$er=0.002})}, como se ha fijado, con lo que obtenemos un rendimiento esperado anual de 50.2\% \mbox{({\it p\$er*251=0.502})} en el periodo 2005-2006.
\\
\\
Calculamos el rendimiento para 2007:

\begin{verbatim}
# producto cruzado entre las proporciones del portafolios 2005-2006 y los 
# rendimientos esperados de 2007
r <- p$weights %*% colMeans(R2)

print( r )
            [,1]
[1,] 0.001646671

print( r * 250 )
          [,1]
[1,] 0.4116678
\end{verbatim}

Por medio de este portafolios obtenemos un rendimiento anual de 41.1\% \mbox{({\it r*250=0.4116678})} de la inversión para 2007 (un rendimiento diario de \mbox{{\it r=0.001646}}). A pesar de que nuestra expectativa era de 50.2\%, el rendimiento que obtuvimos representa el 82\% de lo que originalmente planeamos conseguir, siendo que no sabemos como se comportarán a futuro los rendimientos de una empresa, éste es un ``buen'' resultado.

\begin{figure}[h]
\begin{center}
	\includegraphics[width=11cm]{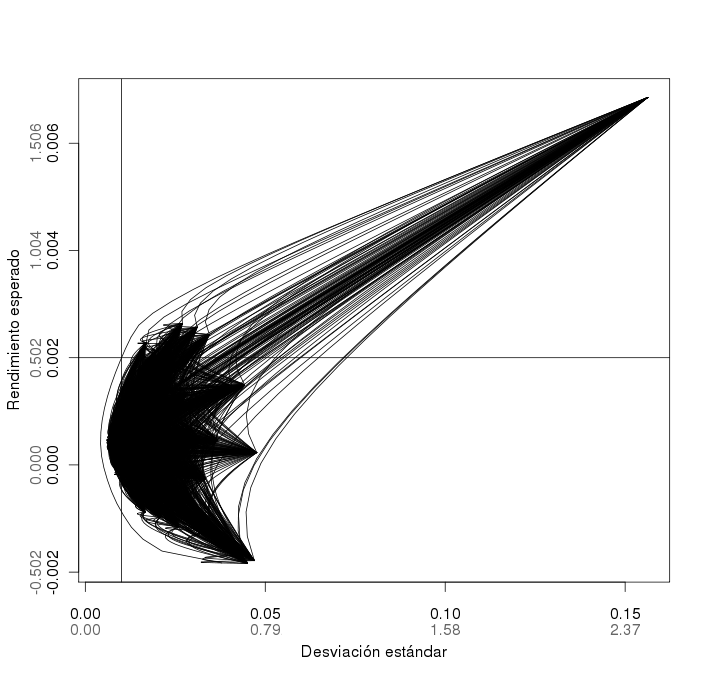} 
	\caption{Portafolios de riesgo bajo} \label{front6}
\end{center}
\end{figure}

En la Figura \ref{front6} se muestra la gráfica de este portafolios en el conjunto de oportunidades:

\begin{verbatim}
plotPortfolio(pf, er = p$er, sd = p$sd)
\end{verbatim}

En cuanto a las empresas que conforman nuestro portafolios, las más relevantes son: PetroChina (PTR) con un 28.8\% de inversión de nuestro capital, Loews Corporation (L) con 10.9\%, y Hewlett-Packard (HPQ) con 10.8\%.

En este caso obtuvimos un portafolio con poco riesgo (en comparación con otros), que genero un buen rendimiento que puede ser suficiente para muchos inversionistas que únicamente busquen incrementar el rendimiento de su ahorro.

\subsubsection{Portafolios de riesgo medio}

En este caso utilizaremos un rendimiento esperado diario de 0.004, éste se encuentra en la frontera eficiente en una zona de equilibrio entre rendimiento-desviación estándar, es decir, aunque el riesgo a este nivel de rendimiento empieza a crecer rápidamente, este punto tiene un rendimiento mayor a la media del máximo que podemos obtener y sin embargo su desviación estándar es menor al punto medio de ésta.
\\
\\
Para encontrar este portafolios utilizamos:

\begin{verbatim}
# Encontramos el portafolios de varianza mínima con el rendimiento
# deseado y almacenamos el resultado en 'p'
p <- markowitzPortfolio( pf, 0.004 )

# imprimimos en pantalla el contenido de p
print( p )

# er=rendimiento esperado, sd=desviación estándar, weights=proporciones
# de compra para cada empresa, portfolio=identificadores de las empresas
# que compraremos acciones, y la proporción de compra para cada una.
$call
markowitzPortfolio(pObject = pf, eRet = 0.004)

$er
[1] 0.004

$sd
[1] 0.052987

$weights
 [1] 0.0000 0.0000 0.3265 0.0000 0.0000 0.0000 0.0000 0.0000 0.0000 0.0000
[11] 0.0000 0.0000 0.0000 0.0000 0.0000 0.0000 0.0000 0.0000 0.0000 0.0000
[21] 0.0000 0.0000 0.0000 0.0000 0.0000 0.0000 0.0000 0.0000 0.0000 0.0000
[31] 0.0000 0.0000 0.0000 0.0000 0.0000 0.0000 0.0000 0.0000 0.0000 0.0000
[41] 0.0000 0.0000 0.0000 0.0000 0.0000 0.0000 0.0000 0.0000 0.0000 0.0000
[51] 0.0000 0.0000 0.0000 0.4550 0.0000 0.0000 0.0000 0.0000 0.0000 0.0000
[61] 0.0000 0.0000 0.0000 0.0000 0.0000 0.0000 0.0000 0.0000 0.0000 0.0000
[71] 0.0000 0.0000 0.0000 0.0000 0.0000 0.2185 0.0000 0.0000 0.0000 0.0000
[81] 0.0000 0.0000 0.0000 0.0000 0.0000 0.0000 0.0000 0.0000 0.0000 0.0000
[91] 0.0000 0.0000 0.0000 0.0000 0.0000

$portfolio
ABAT_Adj.Close NVDA_Adj.Close STEC_Adj.Close 
        0.3265         0.4550         0.2185 
\end{verbatim}

Ya que elegimos un rendimiento esperado de 0.004 el portafolios generado tiene un rendimiento esperado anual de 100.4\% \mbox{({\it p\$er*251=1.004})} para el periodo 2005-2006.

\begin{verbatim}
# producto cruzado entre las proporciones del portafolios 2005-2006 y los 
# rendimientos esperados de 2007
r <- p$weights %*% colMeans(R2)

print( r )
            [,1]
[1,] 0.003903566

print( r * 250 )
          [,1]
[1,] 0.9758916
\end{verbatim}

\begin{figure}[h]
\begin{center}
	\includegraphics[width=11cm]{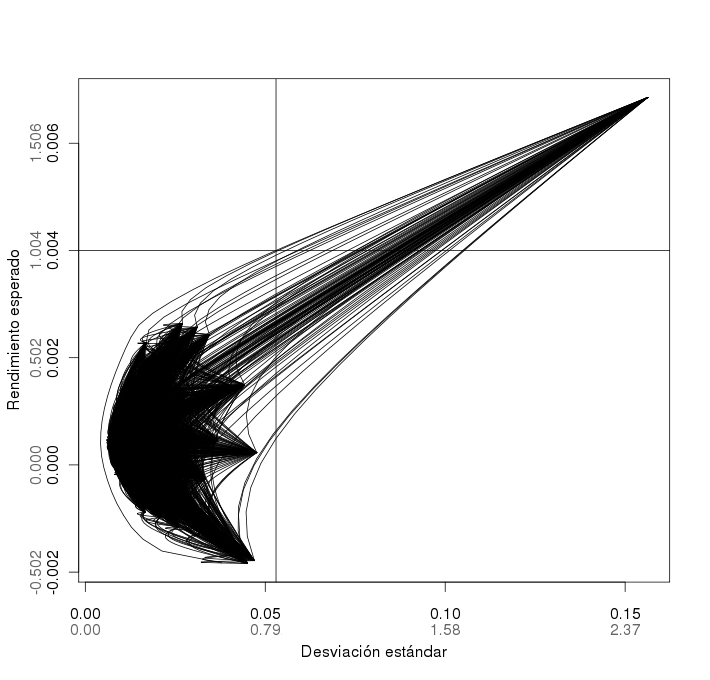} 
	\caption{Portafolios de riesgo medio} \label{front7}
\end{center}
\end{figure}

El rendimiento anual para 2007 que conseguimos con este portafolios es de 97.5\% de la inversión (un rendimiento diario de \mbox{{\it r=0.003903}}), el cual es un resultado muy cercano al deseado. Este portafolios se muestra en la Figura \ref{front7}.

\begin{verbatim}
plotPortfolio(pf, er = p$er, sd = p$sd)
\end{verbatim}

El portafolios de riesgo medio está conformado por sólo tres empresas, esto debido a que se requiere de empresas con mayores rendimientos y por supuesto estas son menores en cantidad que aquellas de poco rendimiento. 

NVIDIA (NVDA) representa el 45.5\% de inversión, Advanced Battery Technologies (ABAT) el 32.65\% y STEC (STEC) el 21.85\%. Aquí se puede observar parte del riesgo del portafolios, ya que consta de tres empresas y todas ellas tienen una proporción de inversión importante, si al menos una de ellas tuviera problemas en el futuro nuestra inversión podría verse afectada de forma importante.

\subsubsection{Portafolios de riesgo alto}

Ahora utilizaremos un rendimiento esperado diario de 0.006, ésto es un rendimiento anual de 150.6\%, ya que un rendimiento mayor nos daría un portafolios conformado prácticamente por una sólo empresa y un análisis de ese tipo no tendría sentido. Puede verse en la gráfica del conjunto de oportunidades que en este nivel de rendimiento el riesgo es muy alto, encontrándose en el último tercio de su rango.

\begin{verbatim}
# Encontramos el portafolios de varianza mínima con el rendimiento
# deseado y almacenamos el resultado en 'p'
p <- markowitzPortfolio( pf, 0.006 )

# imprimimos en pantalla el contenido de p
print( p )

# er=rendimiento esperado, sd=desviación estándar, weights=proporciones
# de compra para cada empresa, portfolio=identificadores de las empresas
# que compraremos acciones, y la proporción de compra para cada una.
$call
markowitzPortfolio(pObject = pf, eRet = 0.006)

$er
[1] 0.006

$sd
[1] 0.1247029

$weights
 [1] 0.0000 0.0000 0.7971 0.0000 0.0000 0.0000 0.0000 0.0000 0.0000 0.0000
[11] 0.0000 0.0000 0.0000 0.0000 0.0000 0.0000 0.0000 0.0000 0.0000 0.0000
[21] 0.0000 0.0000 0.0000 0.0000 0.0000 0.0000 0.0000 0.0000 0.0000 0.0000
[31] 0.0000 0.0000 0.0000 0.0000 0.0000 0.0000 0.0000 0.0000 0.0000 0.0000
[41] 0.0000 0.0000 0.0000 0.0000 0.0000 0.0000 0.0000 0.0000 0.0000 0.0000
[51] 0.0000 0.0000 0.0000 0.2029 0.0000 0.0000 0.0000 0.0000 0.0000 0.0000
[61] 0.0000 0.0000 0.0000 0.0000 0.0000 0.0000 0.0000 0.0000 0.0000 0.0000
[71] 0.0000 0.0000 0.0000 0.0000 0.0000 0.0000 0.0000 0.0000 0.0000 0.0000
[81] 0.0000 0.0000 0.0000 0.0000 0.0000 0.0000 0.0000 0.0000 0.0000 0.0000
[91] 0.0000 0.0000 0.0000 0.0000 0.0000

$portfolio
ABAT_Adj.Close NVDA_Adj.Close 
        0.7971         0.2029 
\end{verbatim}

En este caso tenemos un portafolios conformado principalmente por Advanced Battery Technologies (ABAT), con el 79.71\% y 20.29\% para NVIDIA (NVDA). La mayor parte de la inversión se realiza en ABAT ya que es la empresa con mayor rendimiento esperado (0.006855) y el portafolios que consta sólo de esta empresa está en el extremo de la frontera eficiente, que es hacia donde nos acercamos.

\begin{verbatim}
# producto cruzado entre las proporciones del portafolios 2005-2006 y los 
# rendimientos esperados de 2007
r <- p$weights %*% colMeans(R2)

print( r )
            [,1]
[1,] 0.008331887

print( r * 250 )
          [,1]
[1,] 2.082972
\end{verbatim}

Con el portafolios se obtiene un rendimiento anual para 2007 de 208.2\% de la inversión; a pesar de ser bastante más de lo que esperábamos, es un resultado engañoso, ya que el rendimiento del portafolios es afectado en gran medida sólo por ABAT, por lo que su futuro define también el del portafolios. 

La gran variabilidad en los rendimientos de esta empresa en este caso ha aumentado el rendimiento del portafolios considerablemente, pero de esta misma manera pudo haberlo disminuido en igual proporción, éste es precisamente el riesgo de un portafolios de alto rendimiento. 

En general cuando empresas como ABAT (con rendimientos y varianzas altas) representan una proporción significativa del portafolio, podremos observar este comportamiento en el que se presentan grandes rendimientos o grandes pérdidas, bajo esta incertidumbre la decisión dependerá de nuestra aversión al riesgo.

\begin{figure}[h]
\begin{center}
	\includegraphics[width=11cm]{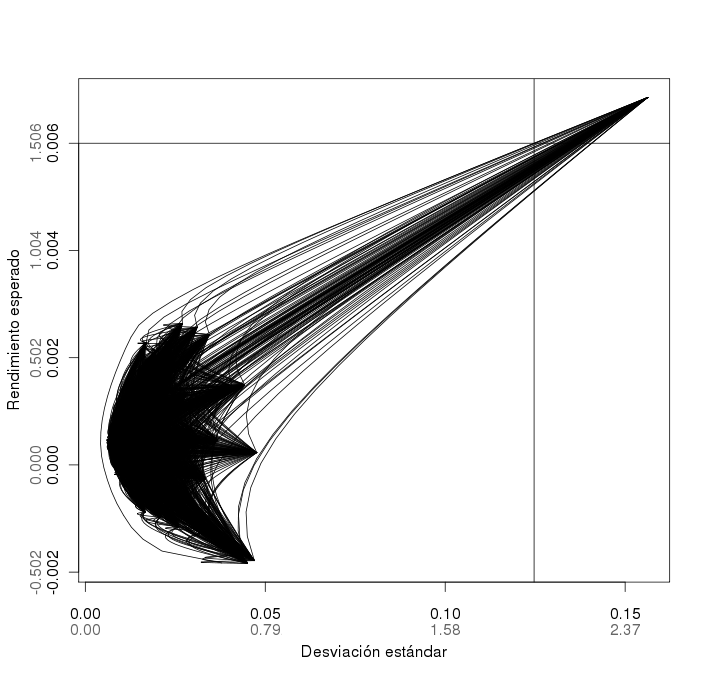} 
	\caption{Portafolios de riesgo alto} \label{front8}
\end{center}
\end{figure}

Por último, podemos observar la gráfica de este portafolios, la cual aparece en la Figura \ref{front8}.

\begin{verbatim}
plotPortfolio(pf, er = p$er, sd = p$sd)
\end{verbatim}

\subsection{Precios}

\begin{figure}[h]
\begin{center}
	\includegraphics[width=11cm]{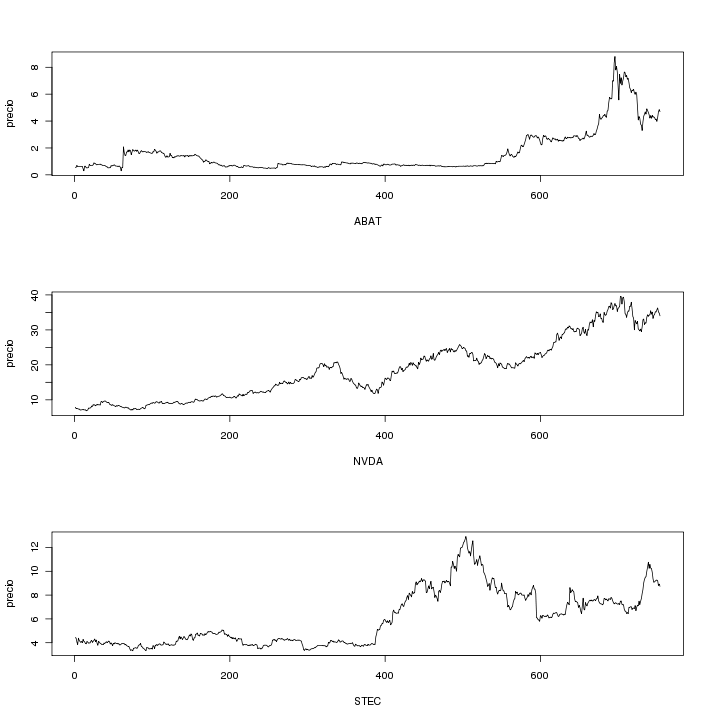} 
	\caption{Precios para ABAT, NVDA y STEC de 2005-2007} \label{prices}
\end{center}
\end{figure}

Para comprender mejor los últimos dos portafolios podemos apreciar en la Figura \ref{prices} la gráfica de los precios de cierre diarios ajustados\footnote{Yahoo!: enero 1, 2005 - diciembre 31, 2007} de las empresas que los componen (ABAT, NVDA y STEC), para el periodo 2005-2007.

En el portafolios de riesgo medio, NVDA representa la mayor proporción con el 45.5\% de inversión, en su gráfica de precios podemos notar que posee una tendencia más acentuada y cambios no muy bruscos, así, esta empresa es la de menor riesgo entre las tres y la de rendimiento medio, lo que también describe de mejor manera a este portafolios.

Para el portafolios de riesgo alto, STEC ya no aparece entre las empresas que constituyen al portafolios, esto debido a que entre las tres es la de menor rendimiento. ABAT, al que se le asigna 79.71\% de la inversión, es la que representa el portafolios, posee alto rendimiento y riesgo alto; puede verse en su gráfica de precios como en el último año creció rápidamente en comparación con los dos primeros años, razón por la que obtuvimos un rendimiento mayor del que esperábamos para este portafolios.

\subsection{Calidad del análisis}

Sabemos que entre más cercano es el resultado que nos proporciona nuestro modelo con respecto al real, mejor es el modelo. Es así que una buena forma de conocer que tan bueno es nuestro análisis es comparar los rendimientos esperados de los portafolios en la frontera eficiente, con el rendimiento que realmente nos proporcionan, similar a lo que hicimos anteriormente. Para comparar todos los portafolios de la frontera eficiente utilizamos el comando:

\begin{verbatim}
# comparativo de los rendimientos esperados contra los que obtuvimos

efficientFrontierFit(pf, R2)
# diario               error medio      error medio de sub-estimación 
                      0.0009241558                       0.0003265892

# anual                error medio      error medio de sub-estimación 
                        0.22997690                         0.08396657 
\end{verbatim}

\begin{figure}[h]
\begin{center}
	\includegraphics[width=11cm]{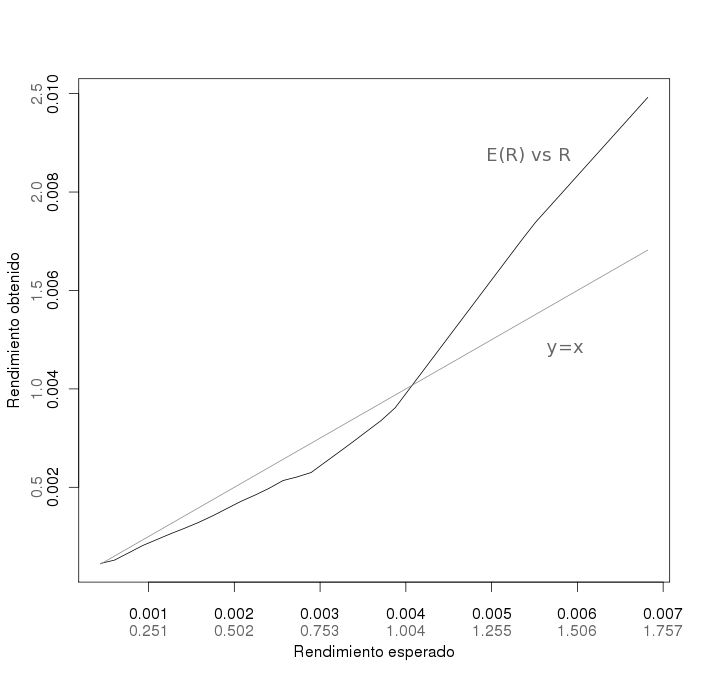} 
	\caption{Ajuste de la frontera eficiente} \label{effit}
\end{center}
\end{figure}

El resultado se muestra en la Figura \ref{effit}. En el eje de las $x$, se encuentra el rendimiento esperado diario que nos proporciona el modelo Media-Varianza para el periodo de 2005-2006 (en gris el anual). En el eje de las $y$ el rendimiento obtenido en 2007 con el portafolios construido anteriormente.

Evidentemente, mientras mejor se ajuste la gráfica de los pares de rendimientos a la recta $y=x$, mejor se comporta el modelo con respecto al conjunto de datos actual.

En este caso la gráfica se encuentra cercana a la recta en la primer parte, aproximadamente hasta $x=0.004$, después se aleja cada vez más, esto debido a que aumenta la presencia de ABAT, que es muy variable; es decir, nuestros portafolios son menos confiables conforme aumenta el riesgo.

La función {\it efficientFrontierFit} devuelve dos valores, el error medio, que mide la diferencia promedio entre el rendimiento esperado contra el obtenido y el error medio de sub-estimación, que proporciona la diferencia promedio anterior para el caso en donde el rendimiento obtenido es menor al esperado. Estos valores nos serán especialmente útiles en adelante para la comparación de modelos de selección de portafolios.

En este estudio se obtuvo un error medio de estimación en el rendimiento anual de 23\% , es decir, en promedio existe una diferencia de $\pm 0.23$ entre el rendimiento esperado y el obtenido, siendo este error medio de $\pm 0.08$ en el caso en que se consiguió un rendimiento menor al esperado.

Podemos notar que el modelo de Markowitz es especialmente eficiente cuando las empresas que conforman el portafolios elegido no varían mucho en el futuro (tienen una tendencia mejor definida), y para que esto suceda es mejor seleccionar portafolios de riesgo medio o bajo.

Aunque esta gráfica no nos es posible obtenerla sino hasta terminar el periodo de estudio, nos permite conocer la calidad del análisis y es una buena referencia al conformar portafolios con el mismo grupo de empresas (o empresas con características similares) en el futuro.

\section{Características del modelo}

El modelo de Markowitz es la base para la Teoría Moderna de Portafolios (TMP) cuya principal característica es su simplicidad y la facilidad de implementación del mismo.

A pesar de esto existen otros modelos que su objetivo es simplificar el de Markowitz, por ejemplo para reducir el poder de cómputo necesario. Este es el caso del modelo de índice simple de Sharpe (1963) el cual sólo requiere el cálculo de $3N+2$ parámetros, esto es $3(95)+2=287$, en lugar del modelo de Markowitz que requiere del cálculo de la matriz de covarianzas, que presenta $\frac{N(N+1)}{2}$ parámetros, en nuestro caso  $\frac{95(95+1)}{2}=4560$.

Sin embargo, dado el poder de cómputo actual y los métodos para la solución de problemas de programación cuadrática, el modelo de Markowitz tiene una resultado casi inmediato, por ejemplo la optimización de un portafolios al que lo componen 500 empresas puede solucionarse en un par de segundos con tan sólo un equipo común. 

Otras características a destacar del modelo de Markowitz son:

\begin{itemize}
\item[$\bullet$] En el modelo de Markowitz se asume un mercado perfecto donde no hay costos por transacción y las proporciones de inversión en las empresas son infinitamente divisibles, es decir, se pueden realizar compras y ventas sin pérdidas de nuestro capital y se puede comprar un número cualquiera de acciones de una empresa (proporción de inversión), lo cual no sucede en el mercado, pues deben adquirirse lotes\footnote{Se denomina lote a la cantidad mínima de títulos que comúnmente se intercambian en una transacción. En la BMV, un lote se integra de 100 títulos. Para títulos con precio mayor a 200 pesos un lote se compone de 5 acciones.}.

\item[$\bullet$] El criterio para determinar el conjunto de portafolios eficientes concuerda con la Teoría de Utilidad, sin embargo la función de utilidad del inversionista debe ser cuadrática o en su defecto una función de utilizad cóncava con rendimientos que se distribuyen normalmente.

\item[$\bullet$] El modelo de Markowitz está realizado para una inversión de un sólo periodo, es decir, es estático, sin embargo se puede modificar para hacerlo dinámico y así realizar operaciones de compra-venta que maximicen nuestra función de utilizad en un periodo de inversión.
\end{itemize}

Algunos de los problemas mencionados se abordan en los siguientes capítulos para darles solución.

\chapter{Modelo Media-Semivarianza}

{\em Debido a que las técnicas de análisis de portafolios están altamente relacionadas con el modelo Media-Varianza, otros modelos con características convenientes frecuentemente son ignorados a causa de sus diferencias con el modelo de Markowitz o por su complejidad.

El modelo Media-Semivarianza emplea una medida alternativa de riesgo, sin embargo este modelo comúnmente no es utilizado debido a la complicación del cálculo de la semivarianza y de su algoritmo de optimización. En este capítulo abordamos el modelo con un enfoque heurístico que nos permitirá darle solución de la misma forma que con el modelo Media-Varianza.
}

\section{Planteamiento del modelo}

Markowitz, en su publicación de 1959\cite{mark2} favorece a otra medida de riesgo con respecto a la varianza: la semivarianza de los rendimientos.

La semivarianza es definida como el valor esperado del cuadrado de las desviaciones negativas (o positivas) con respecto a un valor crítico seleccionado por el inversionista (generalmente la media o cero). Es decir, si $R$ es una variable aleatoria de los rendimientos de una empresa, su semivarianza está definida como:

$$ S_B = E(\mbox{min}\{ R-B,0 \}^2) $$

donde $B$ es un punto crítico seleccionado por el inversionista. Así, la semivarianza de un conjunto de datos puede calcularse por medio de $S_B = \frac{1}{T} \sum_{t} (r_t-B)^2$, para aquellos $r_t \leq B$, donde $r_t$ son los rendimientos observados en un periodo de tiempo, con $r = 1,\cdots, T$.

Nótese que la semivarianza es una medida similar a la varianza, con la diferencia de que sólo incluye en su suma de cuadrados a aquellos rendimiento que tienen un desempeño menor al deseado.

Markowitz argumenta que el criterio Media-Semivarianza ($\mbox{{\it M-S}}_B$) tiende a producir mejores portafolios que los basados en el criterio Media-Varianza ({\it M-V}). Esto debido a que la varianza considera a valores extremos, altos y bajos, igual de indeseables, un análisis basado en la varianza trata de eliminar ambos extremos. Sin embargo una análisis basado en la semivarianza $\mbox{{\it S}}_B$ se centra en reducir pérdidas (pérdidas por debajo de $B$). Es decir, en la realidad un inversionista se preocupa más por el sub-desempeño que por el sobre-desempeño de su inversión, por lo que la semivarianza es una medida más apropiada para representar el riesgo de un portafolios.

En cuanto a la distribución de los rendimientos, si ésta es simétrica, entonces la semivarianza para el caso $B=E(R)$, es equivalente a la varianza, esto es $V=2 \cdot S_{E(R)} $, ya que las desviaciones que se encuentran sobre la media, también se encuentran bajo la misma. En esta situación, el criterio {\it M-V} y $\mbox{{\it M-S}}_{E(R)}$ producen el mismo conjunto de portafolios eficientes.

En cuanto al cálculo de la semivarianza de un portafolios Markowitz sugiere utilizar la siguiente expresión:

\begin{equation}
\Sigma_{pB}^2 = \sum_{i=1}^n \sum_{j=1}^n w_i w_j S_{ijB} \label{semimark}
\end{equation}
donde
$$ S_{ijB} = \frac{1}{T} \sum_{t=1}^k (R_{it} - B) (R_{jt} - B) $$

Los periodos $1$ a $k$ son aquellos en que el portafolios tiene un rendimiento menor a $B$, con $T$ número total de periodos. 

Esta definición de semivarianza de un portafolios nos provee de la ventaja de una estimación exacta, sin embargo, a pesar de su ventaja como medida de riesgo, existen también desventajas para su uso. 

Principalmente la semivarianza requiere mayor poder de cómputo para su cálculo, ya que en un análisis basado en la varianza sólo es necesario calcular las medias y la matriz de covarianzas (variables exógenas), pero un análisis basado en la semivarianza requiere de la distribución conjunta de los rendimientos, es decir, la semivarianza es una medida que depende de las proporciones de inversión de los rendimientos (variable endógena), por lo que ésta medida varía dependiendo de las proporciones que conformen nuestro portafolios. Todo esto convierte al criterio Media-Semivarianza en un modelo de optimización que requiere de algoritmos complejos para su manejo y solución.

Estrada\cite{semiv} sugiere una muy buena aproximación a la semivarianza con respecto al valor crítico $B$, que se define como:

\begin{equation}
\Sigma_{pB}^2 \approx \sum_{i=1}^n \sum_{j=1}^n w_i w_j \Sigma_{ijB} \label{semiest}
\end{equation}
donde
\begin{eqnarray*}
\Sigma_{ijB} &=& E[\mbox{min}(R_i-B,0) \cdot \mbox{min}(R_j-B,0)] \\
&=& \frac{1}{T} \sum_{t=1}^T \mbox{min}(R_{it}-B,0) \cdot \mbox{min}(R_{jt}-B,0)
\end{eqnarray*}

Esta aproximación nos proporciona una matriz fija simétrica de semicovarianzas (exógena), la cual nos permite solucionar el modelo mediante programación cuadrática, de la misma manera que con el modelo Media-Varianza. Esto es:

\begin{eqnarray}
\mbox{minimizar} \quad {\bm w}' \cdot {\bm \Sigma}_{pB-h} \cdot {\bm w} \\
 \nonumber \\
s.a. \quad E({\bm w}'{\bm R}) \geq \beta \nonumber \\
{\bm w}'{\bm 1} = 1 \nonumber \\
{\bm w} \geq {\bm 0} \nonumber 
\end{eqnarray}

Así podemos determinar nuestra frontera eficiente de la misma manera que lo hacemos con el ya bien conocido modelo Media-Varianza.

\section{Aplicando el modelo Media-Semivarianza}

La implementación de este modelo se realiza en el código proporcionado por el archivo ``markowitz.R'' del apéndice \ref{codf}. Compararemos los resultados de este modelo con los obtenidos en el anterior capítulo.

\subsection{Dispersión}

En este caso la dispersión es entre las semidesviaciones (raíz cuadrada de las semivarianzas) y los rendimientos. Por conveniencia en adelante se utilizará el valor crítico cero ($B=0$).

\begin{verbatim}
# leemos el archivo de funciones 'markowitz.R'
source('markowitz.R') 

# importamos los precios, y los guardamos en la variable 'prices'
prices<-read.table('prices-2005-2006.csv', header=T, sep=',')  

# obtenemos los rendimientos y los guardamos en 'R'
R<-assetsReturn(prices)

# graficamos, scov es función de markowitz.R
plot( sqrt(diag(scov(R))), colMeans(R), xlab="Semi-desviación", 
  ylab="Rendimiento esperado" )
\end{verbatim}

\begin{figure}[h]
\begin{center}
	\includegraphics[width=11cm]{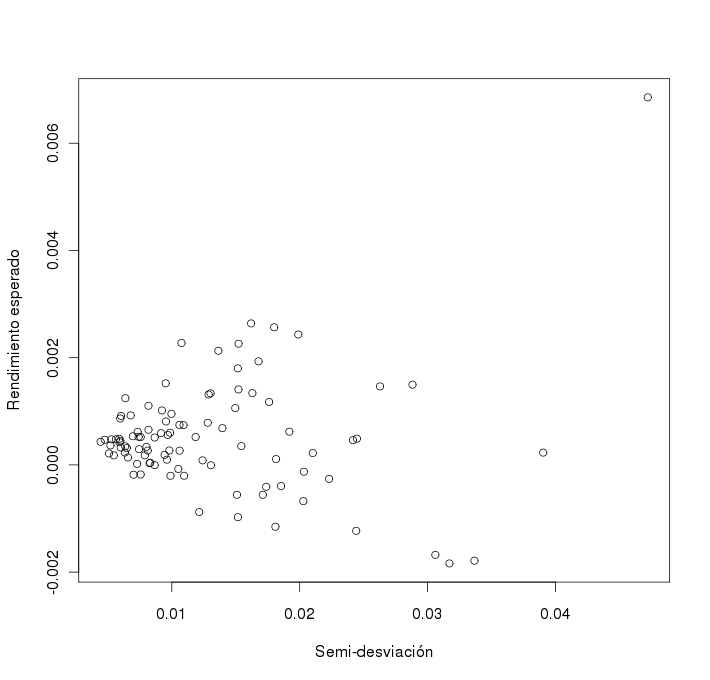} 
	\caption{Gráfica de dispersión de las empresas (semidesviación)} \label{disps}
\end{center}
\end{figure}

El resultado puede verse en la Figura \ref{disps}. La diferencia más notable con el modelo de Media-Varianza ({\it M-V}) es que los datos se encuentran mejor dispersos en el caso del modelo Media-Semivarianza ({\it M-S}), esto es, en ambos gráficas se presenta un valor extremo de mayor riesgo y rendimiento, pero en el modelo {\it M-V} los datos parecen concentrarse en un pequeño rango de riesgo mientras que en el modelo {\it M-S} los datos se distribuyen a lo largo de sus posibles valores.

Esta diferencia se presenta debido a que la semivarianza es una medida de riesgo más discriminaste. Como ya vimos la varianza considera valores sobre y bajo la media de igual forma, mientras que la semivarianza trata de diferente manera a valores bajo $B$ que sobre éste.

\subsection{Conjunto de oportunidades y frontera eficiente}

Como sabemos, el conjunto de oportunidades es una gráfica de gran importancia, ya que podemos observar más claramente como se comporta el riesgo mediante la diversificación, la frontera eficiente, y en que nivel o rango de rendimientos nos conviene movernos. Esta gráfica la podemos obtener mediante:

\begin{verbatim}
# leemos el archivo de funciones 'markowitz.R'
source('markowitz.R') 

# importamos los precios, y los guardamos en la variable 'prices'
prices<-read.table('prices-2005-2006.csv', header=T, sep=',')  

# obtenemos los rendimientos y los guardamos en 'R'
R<-assetsReturn(prices)

# nuevo objeto 'portafolios', con medida de riesgo 'svar'
pf<-newPortfolio(R, risk="svar")

# graficamos el conjunto de oportunidades
plotPortfolio(pf)
\end{verbatim}

\begin{figure}[h]
\begin{center}
	\includegraphics[width=11cm]{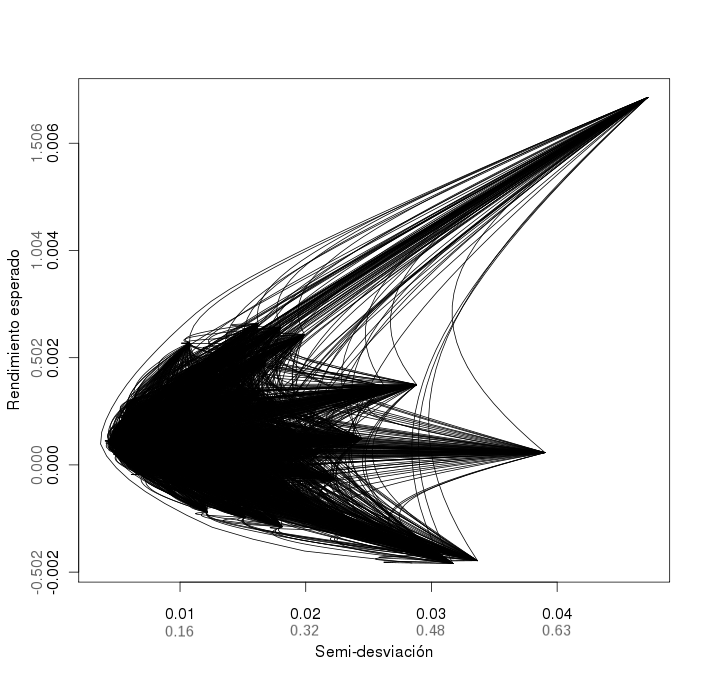} 
	\caption{Conjunto de oportunidades (semidesviación)} \label{front3s}
\end{center}
\end{figure}

El resultado\footnote{Debe observarse que el uso de la semivarianza como medida de riesgo se realiza a través de la orden `{\it pf$<$-newPortfolio(R, risk=``svar'')}', cuando se define al nuevo objeto `portafolios'.} se muestra en la Figura \ref{front3s}. En negro se denotan los valores de los rendimientos-semidesviaciones estándar diarios y en gris los anuales. Nuevamente, si sobreponemos la gráfica de dispersión al conjunto de oportunidades observamos una relación entre estas; los vértices que conforman las uniones de las curvas corresponden con portafolios de una sola empresa, presentes en la gráfica de dispersión.

Puede notarse una diferencia importante con la forma del conjunto de oportunidades del modelo {\it M-V}. Al existir una dispersión mayor en el modelo {\it M-S}, se produce un conjunto de oportunidades más amplio.

\subsection{Portafolios de semivarianza mínima}

El portafolios de semivarianza mínima es aquel portafolios de mínimo riesgo en la frontera eficiente; podemos obtenerlo con los comandos:

\begin{verbatim}
# Encontramos el portafolios de semivarianza mínima y almacenamos el 
# resultado en 'p'
p <- markowitzPortfolio( pf )

# imprimimos en pantalla el contenido de p
print( p )

# er=rendimiento esperado, sd=semidesviación, weights=proporciones
# de compra para cada empresa, portfolio=identificadores de las empresas
# que compraremos acciones, y la proporción de compra para cada una.

$call
markowitzPortfolio(pObject = pf)

$er
[1] 0.0004402566

$sd
[1] 0.00365887

$weights
 [1] 0.0000 0.0000 0.0000 0.0000 0.0000 0.0000 0.0000 0.0000 0.0000 0.0000
[11] 0.0000 0.0000 0.0045 0.0000 0.0000 0.0000 0.0260 0.0000 0.0000 0.0000
[21] 0.0000 0.0000 0.0000 0.0000 0.0000 0.0000 0.0000 0.0000 0.0000 0.0000
[31] 0.0000 0.0000 0.1409 0.0000 0.1527 0.0000 0.0000 0.1592 0.2432 0.0000
[41] 0.0297 0.0000 0.0000 0.0000 0.0000 0.0000 0.0000 0.0000 0.0000 0.0000
[51] 0.0000 0.0000 0.0000 0.0000 0.0000 0.0000 0.0000 0.0000 0.0000 0.0000
[61] 0.1652 0.0000 0.0000 0.0000 0.0000 0.0000 0.0000 0.0000 0.0000 0.0000
[71] 0.0000 0.0000 0.0000 0.0000 0.0000 0.0000 0.0204 0.0000 0.0000 0.0000
[81] 0.0000 0.0000 0.0000 0.0000 0.0000 0.0000 0.0581 0.0000 0.0000 0.0000
[91] 0.0000 0.0000 0.0000 0.0000 0.0000

$portfolio
BAC_Adj.Close   C_Adj.Close JNJ_Adj.Close   K_Adj.Close KMP_Adj.Close 
       0.0045        0.0260        0.1409        0.1527        0.1592 
 KO_Adj.Close  LH_Adj.Close PEP_Adj.Close   T_Adj.Close   Y_Adj.Close 
       0.2432        0.0297        0.1652        0.0204        0.0581 
\end{verbatim}

Dado el rendimiento esperado diario que obtenemos, el portafolios de semivarianza mínima generado tiene un rendimiento esperado anual de al menos\footnote{En este modelo castigamos el sub-desempeño, por lo que decimos que el rendimiento esperado es de al menos el que esperamos obtener.} 11.05\%.

Puede verse que el rendimiento esperado anual es muy cercano al del modelo {\it M-V} que es de 11.06\%, pero el portafolios generado por {\it M-S} es más simple. Para conocer el rendimiento que obtendremos en 2007 se utiliza:

\begin{verbatim}
# importamos los precios de 2007, y los guardamos en la variable 'prices2'
prices2 <- read.table('prices-2007.csv', header=T, sep=',')  

# obtenemos los rendimientos y los guardamos en 'R2'
R2 <- assetsReturn(prices2)

# producto cruzado entre las proporciones del portafolios 2005-2006 y los 
# rendimientos esperados de 2007
r <- p$weights %*% colMeans(R2)

print( r )
             [,1]
[1,] 0.0005892042

print( r * 250 )
          [,1]
[1,] 0.1473011
\end{verbatim}

Es así que con el portafolios de semivarianza mínima se obtiene un rendimiento de 14.7\% en 2007, que es ligeramente mayor al que esperábamos y al que obtuvimos con el modelo {\it M-V}. Sin embargo, ya que con el modelo de {\it M-S} nuestra intención es castigar el sub-desempeño, sobrepasar el rendimiento esperado no está necesariamente relacionado con la posibilidad de que la ganancia con respecto al rendimiento esperado se transforme en pérdida, por lo que es un buen resultado.

En este portafolio a las empresas que se les asigna mayor proporción de inversión son: Coca-Cola (KO) con 24.32\%, Pepsico (PEP) con 16.52\%, Kinder Morgan Energy Partners (KMP) con 15.92\%, Kellogg Company (K) con 15.27\% y Johnson \& Johnson (JNJ) con 14.09\%.

\subsection{Especificando el rendimiento}

Así como en el modelo {\it M-V} nuevamente utilizamos tres niveles de rendimiento en relación al riesgo, (1) riesgo bajo, (2) riesgo medio, y (3) riesgo alto, para así poder comparar los portafolios obtenidos anteriormente.

\subsubsection{Portafolios de riesgo bajo}

Usamos 0.002 como el rendimiento esperado diario que al menos deseamos obtener, esto es un rendimiento esperado anual de al menos 50.2\%. Igualmente (como en {\it M-V}) este nivel de rendimiento se presenta en la frontera eficiente en una zona donde la semidesviación aun no crece rápidamente con relación al rendimiento.

\begin{verbatim}
# Encontramos el portafolios de semivarianza mínima con el rendimiento
# deseado y almacenamos el resultado en 'p'
p <- markowitzPortfolio( pf, 0.002 )

# imprimimos en pantalla el contenido de p
print( p )

# er=rendimiento esperado, sd=semidesviación, weights=proporciones
# de compra para cada empresa, portfolio=identificadores de las empresas
# que compraremos acciones, y la proporción de compra para cada una.
$call
markowitzPortfolio(pObject = pf, eRet = 0.002)

$er
[1] 0.002

$sd
[1] 0.007577653

$weights
 [1] 0.0000 0.0216 0.0541 0.0000 0.0000 0.0000 0.0000 0.0000 0.0000 0.0000
[11] 0.0000 0.0000 0.0000 0.0000 0.0000 0.0000 0.0000 0.0000 0.0000 0.0000
[21] 0.0000 0.0000 0.0000 0.0000 0.0000 0.0000 0.0000 0.0000 0.0000 0.0942
[31] 0.0000 0.0000 0.0000 0.0000 0.0000 0.0000 0.0000 0.0000 0.0000 0.2675
[41] 0.0595 0.0000 0.0000 0.0000 0.0000 0.0000 0.0000 0.0000 0.0000 0.0000
[51] 0.0000 0.0000 0.0000 0.0278 0.0000 0.0000 0.0000 0.0000 0.0000 0.0000
[61] 0.0000 0.0000 0.0000 0.0000 0.3359 0.0000 0.0000 0.0000 0.0000 0.0000
[71] 0.0000 0.0000 0.0000 0.0000 0.0000 0.0533 0.0861 0.0000 0.0000 0.0000
[81] 0.0000 0.0000 0.0000 0.0000 0.0000 0.0000 0.0000 0.0000 0.0000 0.0000
[91] 0.0000 0.0000 0.0000 0.0000 0.0000

$portfolio
AAPL_Adj.Close ABAT_Adj.Close  HPQ_Adj.Close    L_Adj.Close   LH_Adj.Close 
        0.0216         0.0541         0.0942         0.2675         0.0595 
NVDA_Adj.Close  PTR_Adj.Close STEC_Adj.Close    T_Adj.Close 
        0.0278         0.3359         0.0533         0.0861


# producto cruzado entre las proporciones del portafolios 2005-2006 y los 
# rendimientos esperados de 2007
r <- p$weights %*% colMeans(R2)

print( r )
            [,1]
[1,] 0.001585825

print( r * 250 )
          [,1]
[1,] 0.3964562
\end{verbatim}

El rendimiento anual obtenido en 2007 es de 39.6\% de la inversión. Nos encontramos por debajo de la expectativa de 50.2\%, obteniendo el 78.8\% de lo que originalmente planeamos conseguir, aunque este resultado es muy cercano al obtenido por medio del modelo {\it M-V} de 41.1\% anual.

En cuanto a las empresas que lo conforman, este portafolios es más simple que el de {\it M-V}, las más relevantes son: PetroChina (PTR) con un 33.59\% , Loews Corporation (L) con 26.75\%, y Hewlett-Packard (HPQ) con 9.42\%, que igualmente son las empresas más representativas del portafolios para el modelo {\it M-V}, de ahí la similitud del resultado.

\subsubsection{Portafolios de riesgo medio}

En este caso el rendimiento esperado diario a utilizar es de 0.004, siendo un rendimiento anual de al menos 100.4\%. Similarmente que en la frontera eficiente del modelo {\it M-V}, este se encuentra en una zona de equilibrio entre rendimiento-semidesviación.

\begin{verbatim}
# Encontramos el portafolios de semivarianza mínima con el rendimiento
# deseado y almacenamos el resultado en 'p'
p <- markowitzPortfolio( R, 0.004 )

# imprimimos en pantalla el contenido de p
print( p )

# er=rendimiento esperado, sd=desviación estándar, weights=proporciones
# de compra para cada empresa, portfolio=identificadores de las empresas
# que compraremos acciones, y la proporción de compra para cada una.
$call
markowitzPortfolio(pObject = pf, eRet = 0.004)

$er
[1] 0.004

$sd
[1] 0.01970659

$weights
 [1] 0.0000 0.0000 0.3536 0.0000 0.0000 0.0000 0.0000 0.0000 0.0000 0.0000
[11] 0.0000 0.0000 0.0000 0.0000 0.0000 0.0000 0.0000 0.0000 0.0000 0.0000
[21] 0.0000 0.0000 0.0000 0.0000 0.0000 0.0000 0.0000 0.0000 0.0000 0.0000
[31] 0.0000 0.0000 0.0000 0.0000 0.0000 0.0000 0.0000 0.0000 0.0000 0.0000
[41] 0.0000 0.0000 0.0000 0.0000 0.0000 0.0000 0.0000 0.0000 0.0000 0.0000
[51] 0.0000 0.0000 0.0000 0.1756 0.0000 0.0000 0.0000 0.0000 0.0000 0.0000
[61] 0.0000 0.0000 0.0000 0.0000 0.3292 0.0000 0.0000 0.0000 0.0000 0.0000
[71] 0.0000 0.0000 0.0000 0.0000 0.0000 0.1416 0.0000 0.0000 0.0000 0.0000
[81] 0.0000 0.0000 0.0000 0.0000 0.0000 0.0000 0.0000 0.0000 0.0000 0.0000
[91] 0.0000 0.0000 0.0000 0.0000 0.0000

$portfolio
ABAT_Adj.Close NVDA_Adj.Close  PTR_Adj.Close STEC_Adj.Close 
        0.3536         0.1756         0.3292         0.1416 


# producto cruzado entre las proporciones del portafolios 2005-2006 y los 
# rendimientos esperados de 2007
r <- p$weights %*% colMeans(R2)

print( r )
            [,1]
[1,] 0.004252065

print( r * 250 )
          [,1]
[1,] 1.063016
\end{verbatim}

Puede verse que obtenemos para 2007 un rendimiento mayor al esperado, un rendimiento anual de 106.3\%, por lo que es un buen resultado, que es muy cercano al esperado. 

La diferencia más importante con el portafolios de {\it M-V} es la inclusión de PetroChina (PTR), la cual también tiene una participación importante en el portafolios de riesgo bajo.

\subsubsection{Portafolios de riesgo alto}

Para el portafolios de riesgo alto utilizamos un rendimiento esperado diario de 0.006, esto es un rendimiento anual de al menos 150.6\%  para el periodo 2005-2006. Puede verse en la frontera eficiente que en este nivel de rendimiento el riesgo es muy alto en comparación con otros portafolios.

\begin{verbatim}
# Encontramos el portafolios de varianza mínima con el rendimiento
# deseado y almacenamos el resultado en 'p'
p <- markowitzPortfolio( pf, 0.006 )

# imprimimos en pantalla el contenido de p
print( p )

# er=rendimiento esperado, sd=semidesviación, weights=proporciones
# de compra para cada empresa, portfolio=identificadores de las empresas
# que compraremos acciones, y la proporción de compra para cada una.
$call
markowitzPortfolio(pObject = pf, eRet = 0.006)

$er
[1] 0.006

$sd
[1] 0.03857378

$weights
 [1] 0.0000 0.0000 0.7983 0.0000 0.0000 0.0000 0.0000 0.0000 0.0000 0.0000
[11] 0.0000 0.0000 0.0000 0.0000 0.0000 0.0000 0.0000 0.0000 0.0000 0.0000
[21] 0.0000 0.0000 0.0000 0.0000 0.0000 0.0000 0.0000 0.0000 0.0000 0.0000
[31] 0.0000 0.0000 0.0000 0.0000 0.0000 0.0000 0.0000 0.0000 0.0000 0.0000
[41] 0.0000 0.0000 0.0000 0.0000 0.0000 0.0000 0.0000 0.0000 0.0000 0.0000
[51] 0.0000 0.0000 0.0000 0.1272 0.0000 0.0000 0.0000 0.0000 0.0000 0.0000
[61] 0.0000 0.0000 0.0000 0.0000 0.0000 0.0000 0.0000 0.0000 0.0000 0.0000
[71] 0.0000 0.0000 0.0000 0.0000 0.0000 0.0745 0.0000 0.0000 0.0000 0.0000
[81] 0.0000 0.0000 0.0000 0.0000 0.0000 0.0000 0.0000 0.0000 0.0000 0.0000
[91] 0.0000 0.0000 0.0000 0.0000 0.0000

$portfolio
ABAT_Adj.Close NVDA_Adj.Close STEC_Adj.Close 
        0.7983         0.1272         0.0745 


# producto cruzado entre las proporciones del portafolios 2005-2006 y los 
# rendimientos esperados de 2007
r <- p$weights %*% colMeans(R2)

print( r )
            [,1]
[1,] 0.008144225

print( r * 250 )
          [,1]
[1,] 2.036056
\end{verbatim}

Nuevamente el portafolios está conformado principalmente por Advanced Battery Technologies (ABAT), con el 79.83\% y 12.72\% para NVIDIA (NVDA). En comparación con el resultado del modelo {\it M-S} se incluye también a STEC (STEC) con 7.45\%. La participación en gran proporción de ABAT en el portafolios es evidente ya que esta empresa es la de mayor rendimiento, además de ser la única que se encuentra sobre 0.006 de rendimiento esperado diario, por lo que es un resultado necesario.

El rendimiento anual que obtenemos para 2007 es de 203.6\%. Igualmente que en el modelo {\it M-V} sobrepasamos el esperado por bastante, sin embargo el resultado obtenido a través del modelo {\it M-S} es ligeramente menor que el de {\it M-V}.

Como se menciono, el sobrepasar el rendimiento esperado puede ser resultado de la propia característica de la semivarianza como medida de riesgo, aunque no necesariamente se debe a ésto el rendimiento adicional obtenido. En este caso asumimos un mayor riesgo para obtener mayor rendimiento, ABAT que representa la mayor proporción del inversión posee un gran riesgo, por lo que el riesgo de una gran pérdida también esta presente. En general este portafolio es mejor que el obtenido con el modelo {\it M-V} ya que nos alejamos menos del rendimiento esperado.

\begin{figure}[h]
\begin{center}
	\includegraphics[width=11cm]{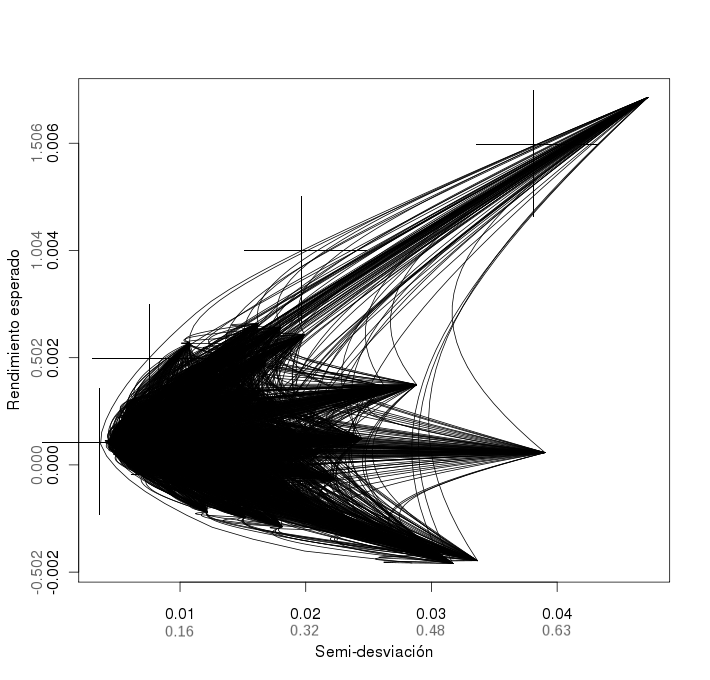} 
	\caption{Portafolios de riesgo: mínimo, bajo, medio y alto} \label{front4s}
\end{center}
\end{figure}

Por último puede observarse en la Figura \ref{front4s} la gráfica en el conjunto de oportunidades de los cuatro portafolios anteriores, nótese que todos ellos se encuentran en la frontera eficiente.

\subsection{Calidad del análisis}

Debido a la característica de la semivarianza, en la comparación de la gráfica de los rendimientos esperados contra los que realmente obtuvimos un buen análisis está representado por una recta superpuesta sobre $y=x$ ó cualquier función que su contradominio se encuentre en $y \geq x$. Sin embargo podemos comparar esta gráfica y su error medio con la obtenida en el modelo {\it M-V} para ver si existe una mejora en cuanto al ajuste a $y=x$.

\begin{verbatim}
# comparativo de los rendimientos esperados contra los que obtuvimos
efficientFrontierFit(pf, R2)

# diario               error medio      error medio de sub-estimación 
                      0.0008587857                       0.0003173200
                      
# anual                error medio      error medio de sub-estimación 
                        0.21296826                         0.08160426 
\end{verbatim}

\begin{figure}[h]
\begin{center}
	\includegraphics[width=11cm]{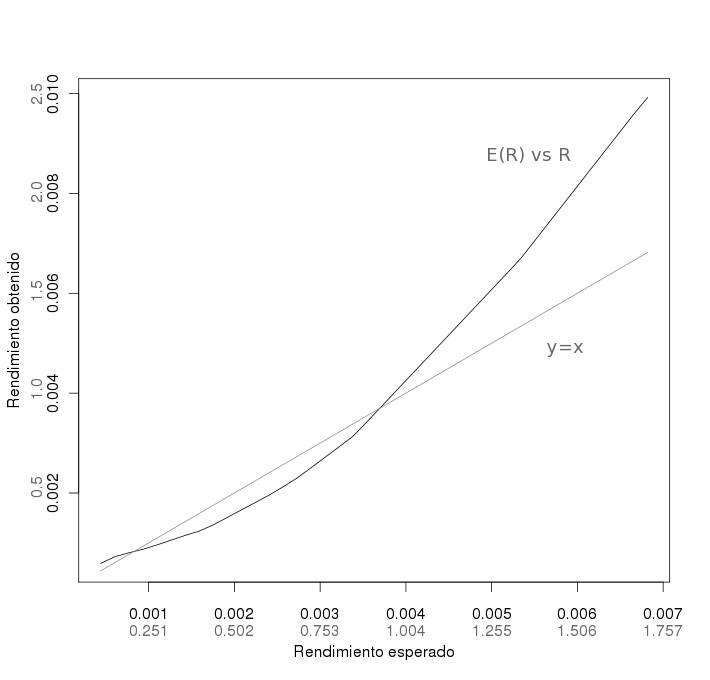} 
	\caption{Ajuste de la frontera eficiente ({\it M-S})} \label{effits}
\end{center}
\end{figure}

La gráfica obtenida es muy similar a la del modelo {\it M-V}; la diferencia más importante es que se aleja más lentamente de la recta $y=x$ que la del modelo {\it M-V}. Esto puede apreciarse mejor al comparar el error medio de ambas, es decir, el promedio de las diferencias entre el rendimiento esperado y el obtenido. 

En el modelo {\it M-V} el error medio es de $0.23$ y en el modelo {\it M-S} es $0.213$, por lo que nuestro error medio se redujo en 7\% al utilizar el modelo {\it M-S} , es decir, el modelo {\it M-S} se encuentra más cercano a la recta $y=x$.

En cuando al error de sub-estimación, que nos proporciona el promedio de diferencias entre el rendimiento esperado $E(R_p)$ y el obtenido $R_p$, donde \mbox{$E(R_p)>R_p$}, el modelo {\it M-V} posee un valor de $0.0839$ y {\it M-S} de $0.0816$, por lo que el error de sub-estimación se redujo en 2.8\%. Es así que el modelo Media-Semivarianza ha mostrado producir (en promedio) mejores portafolios que el modelo Media-Varianza.

\chapter{Algoritmos genéticos}

{\em La naturaleza frecuentemente se enfrenta a diversos y complicados problemas para los que ésta encuentra soluciones eficientes. En la actualizad existen ramas de la ciencia como la Biomimética\footnote{La Biomimética (imitar la vida) plantea que la naturaleza lleva millones de años solucionando problemas de manera eficiente mediante ensayo y error. De esta forma, en lugar de resolver un problema desde cero, se estudia como lo ha resuelto la naturaleza. Es así que seres vivos como las termitas o las aves inspiran tecnologías eficientes como sistemas de ventilación de edificios y alas de avión o impulsores de hélice basados en moluscos, etc.} que se encarga de buscar soluciones a problemas o necesidades humanas basándose en modelos de la naturaleza.

Los algoritmos genéticos (AG) son métodos de búsqueda adaptativa inspirados por la naturaleza, más específicamente se basan en el proceso genético y evolutivo de los seres vivos. Estos algoritmos pueden ser usados para encontrar soluciones a problemas de optimización donde no hay métodos de búsqueda adecuados o en problemas muy complejos que requieren de una ``buena'' solución.

Darwin (1859) postuló que a lo largo de las generaciones las poblaciones evolucionan conforme el principio de selección natural. Por imitación de este proceso los algoritmos genéticos son capaces de encontrar soluciones a problemas del mundo real, seleccionando las mejores por medio de una función de adaptación. Aunque las simulaciones por computadora comenzaron mucho antes, fue hasta 1975 (Holland) donde se describen los principios de los algoritmos genéticos como tales.

}

\section{Principios de los algoritmos genéticos}

Los algoritmos genéticos son un poderosa técnica de optimización que puede tratar con éxito una gran variedad de problemas de diferentes áreas, incluso aquellos en que otros métodos tienen dificultades. Aunque no se garantiza que el algoritmo encuentre el óptimo global, genera soluciones muy cercanas a éste, en un tiempo competitivo con otros algoritmos combinatorios, convergiendo al óptimo global conforme aumenta el número de generaciones.

El proceso básico de los algoritmos genéticos consta de generar una población inicial de posibles soluciones, de las cuales se seccionarán las mejores mediante su función de adaptación. Posteriormente se define un proceso de evolución para producir nuevos candidatos a solución, la idea es combinar buenas soluciones (aquellas con una adaptación alta), para obtener mejores soluciones, en las que los individuos con un mayor valor de adaptación tendrán más probabilidad de reproducirse (combinación) para producir una nueva población de posibles soluciones. También debe incluirse un proceso de mutación (modificar al azar parte de una solución candidata) que nos permitirá explorar diferentes espacios de soluciones para así acercarnos al óptimo global. Es así que este proceso se repite hasta que el algoritmo converge a una solución después de cierto número de generaciones.

De esta forma los algoritmos genéticos han ganado popularidad debido a que permiten una gran eficiencia en relación al tiempo de ejecución con respecto otros algoritmos tradicionales de búsqueda, permitiendo resolver problemas altamente complejos.

Sin embargo, si existe una técnica determinada para resolver un problema de optimización, es muy probable que supere a un algoritmo genético en tiempo y eficiencia. El campo de uso de los algoritmos genéticos está relacionado con aquellos problemas que no tienen una técnica especializada de solución.

\subsection{Un algoritmo genético simple}

Todo algoritmo genético posee un proceso de selección, cruce y mutación, siendo estos operadores de gran importancia para el desempeño del algoritmo. Para tener una mejor idea de este proceso en la Figura \ref{pcodag} se presenta el pseudocódigo de un AG simple.

\begin{figure}[h]
\begin{center}
\begin{verbatim}	
BEGIN 
   - Generar una población aleatoria inicial de posibles soluciones.
   - Calcular el valor de la función de adaptación (función objetivo) 
   para cada individuo de la población inicial.
   WHILE NOT convergencia DO
   BEGIN /* Producir una nueva generación a partir de la anterior */
        FOR Tamaño de la población/2 DO
        BEGIN /* Reproducción */
               - Seleccionar dos individuos de la anterior generación,
               con probabilidad proporcional a su valor de adaptación.
               - Mutar a uno de los individuos antes de cruzar si se
               cumple cierta condición de probabilidad.
               - Cruzar los dos individuos.
               Insertar los dos descendientes en la nueva generación.
        END
        IF     Mejor individuo no cambia a lo largo de las generaciones.
        THEN   Reportar individuo con mayor adaptación.
        AND    EXIT
   END
END
\end{verbatim}
	\caption{Pseudocódigo de un algoritmo genético simple} \label{pcodag}
\end{center}
\end{figure}

La explicación del pseudocódigo es:

\begin{enumerate}
\item Se genera una población inicial aleatoria de soluciones posibles de tamaño $N$.
\item Se calcula el valor de adaptación de cada individuo de la población.
\item Se toman aleatoriamente dos individuos del la población con probabilidad de selección proporcional a su valor de adaptación (calculado en el punto 2).
\item Se repite el anterior punto hasta generar una nueva población de tamaño $N$.
\item Antes de cada cruce se muta a un individuo del par seleccionado si se cumple una condición específica de probabilidad. 
\item En adelante se utiliza la nueva población o se crea otra población con los $N$ mejores individuos de las dos ultimas poblaciones.
\item Nuevamente se calcula el valor de adaptación de cada individuo de la nueva población.
\item Se selecciona al mejor individuo de la nueva población, si es el mismo después de cierto número de generaciones o se ha alcanzado un número máximo de generaciones, se detiene el proceso, de lo contrario regresamos al punto 3.
\end{enumerate}

Es de esperar por el tipo de operaciones que realiza un algoritmo genético (cruces y mutaciones) que la representación de un individuo viene dada por una codificación del mismo, así, esta codificación debe ser capaz de modificarse para representar las diferentes soluciones posibles.

Otra característica importante del AG es la función de adaptación, que determina el proceso de evolución y las características de la solución que será seleccionada como óptima. Esta función debe ser seleccionada correctamente o de lo contrario estaremos resolviendo un problema distinto al que en realidad queremos resolver.

En esta sección se presentó un AG simple, pero hay muchas formas de elaborar un AG según los distintos elementos que lo componen. En adelante veremos con más detalle cada una de las características que conforman un AG y sus variaciones.

\subsection{Codificación}

Una solución posible (individuo, el cual simboliza un cromosoma o cromosomas) está representada por una codificación especial, en la que cada elemento es un gen. La codificación de un individuo debe ser capaz de representar sus características para posteriormente poder manipularlas. Para esto existen tres tipos básicos de codificaciones que podemos utilizar:

\begin{itemize}
\item[$\bullet$] {\bf Codificación binaria}. En esta codificación un gen (característica) está representado por 0 o 1, por lo que un individuo está compuesto por una cadena binaria, por ejemplo, 11010001. Cada gen puede representar la presencia o ausencia de alguna característica del individuo.
\item[$\bullet$] {\bf Codificación entera}. La codificación entera se aplica cuando las características de un individuo requiere representarse de esta forma, por ejemplo, si los primeros 4 genes tienen un rango de 1-3, y los últimos 4 de 1-6, representando características o valores numéricos, un individuo posible sería: 13211563.
\item[$\bullet$] {\bf Codificación real}. Aquí cada gen está representado por un número real, esta codificación puede ser necesaria si las características de un individuo requieren un rango decimal de valores, por ejemplo una proporción, así un individuo posible es: $\frac{1}{2} \frac{1}{3} \frac{2}{3} \frac{5}{6} \frac{9}{10} \frac{1}{9}$.
\end{itemize}

Ejemplo de un problema de codificación: supongamos que tenemos 10 empleados y queremos repartirlos en 4 áreas distintas para aumentar la productividad, en este caso cada gen puede representar el número de empleados en cada área, por lo que una solución posible es: 3322. Sin embargo, hay que notar que existen combinaciones como 2343, que no es una solución ya que su suma es mayor a 10.

Cada una de estas representaciones será usada según el tipo y número de características que requieren ser representadas, en general es posible utilizar cualquier codificación, por ejemplo letras, esto siempre que la función de adaptación sea capaz de evaluar el código seleccionado.

\subsection{Función de adaptación}

Cuando utilizamos los algoritmos genéticos en problemas de optimización, la función de adaptación de los AG no es otra que la función objetivo, es decir, si nuestro problema de optimización es {\it maximizar} $f(x)$, su nivel de adaptación será $f(x)$. En caso de ser un problema de minimización, {\it minimizar} $g(y)$, podemos utilizar {\it maximizar} $-g(y)$, que es un problema equivalente; esto para asociar altos valores de la función objetivo con mejor adaptación.

En caso que de no conocer la función objetivo, la función de adaptación debe construirse de tal forma que los individuos que posean características más deseables a las que buscamos, tengan un mayor valor de adaptación. Por ejemplo, si hay restricciones en el modelo y existe la posibilidad de generar individuos que no cumplan con estas\footnote{En general es mejor generar individuos que cumplen con las restricciones del modelo, ya que la solución es más confiable y la convergencia del algoritmos mucho más rápida.}, la función objetivo debe asignar un menor valor a aquellos individuos que no cumplen con las restricciones para que estos sean desechados a lo largo de las generaciones.

\subsection{Operación de selección}

La operación de selección consta de escoger a los individuos que se reproducirán en las próximas generaciones. De esta forma, la operación de selección debe escoger para la reproducción a los individuos favoreciendo a aquellos mejor adaptados, incluyendo, aunque con menor probabilidad, a los individuos no tan bien adaptados. De esta forma preservamos a lo largo de las generaciones las mejores características y los individuos menos adaptados nos permitirán mantener cierto grado de diversidad en cada población, donde estas características adicionales podrán producir mejores soluciones.

Una forma de hacer ésto es asignar a cada individuo $x_i$, una probabilidad de la forma $p(x_i) = \frac{f(x_i)}{\sum_i f(x_i)}$, donde $f(x_i)$ es el valor de adaptación del individuo $x_i$. Posteriormente definimos $g(x_i)$ tal que $g(x_i) = \sum_{j=0}^i p(x_i)$ donde $p(x_0) = 0$. Por último generamos un número aleatorio $r$, uniforme entre 0 y 1, si $r$ es tal que $g(x_{i-1}) < r \leq g(x_i)$, seleccionaremos $x_i$ para la operación de cruce.

Otra forma de selección consiste en ponderar la probabilidad de cruce de cada individuo, aumentando o disminuyendo la probabilidad de selección de los individuos mejor adaptados, para ser más o menos elitistas (ser demasiado elitistas puede llevarnos a óptimos locales).

Dependiendo de la proporción de la población que queramos substituir con la nueva generación, seleccionaremos $N^*/2$ pares de individuos para la operación de cruzamiento, siendo $N^*$ el número de individuos que obtendremos después del cruce.

\subsection{Operación de cruzamiento}

De la operación de selección recibiremos dos soluciones, las cuales cruzaremos para generar una nueva solución. El punto de cruce, que es el elemento en la codificación a partir del cual se intercambian los genes de las soluciones padre, puede elegirse al azar o utilizar el punto medio, así dos padres combinan sus cromosomas para generar un nuevo individuo. 

En el ejemplo anterior de los empleados repartidos en 4 diferentes áreas podemos tener dos individuos a cruzar, 1243 y 2134. Si utilizamos para cruzar el punto medio (mitad de los genes de cada padre), los nuevos individuos generados son, 1234 y 2143. De esta forma tenemos dos nuevos individuos con sus características combinadas.

Nótese que durante la operación de cruzamiento las soluciones generadas pueden no satisfacer las restricciones del modelo. Si tenemos las soluciones, 2242 y 3232, su cruzamiento genera a 2232 y 3242, las cuales no son soluciones factibles ya que la suma de empleados debe ser 10. Por lo que en estos casos el AG debe contar con una operación de corrección de los individuos, en este ejemplo la solución 3232 puede ser corregida tratando de mantener las proporciones lo más posiblemente parecidas, así la solución corregida, 3222, será la que se incorpore a la nueva población.

\subsection{Operación de mutación}

La operación de mutación consta de la modificación aleatoria de uno o varios genes de la solución. Por supuesto esta operación debe hacerse a uno de los individuos a cruzar o después del cruzamiento y antes de la corrección de la solución.

Esta se realiza con cierta probabilidad, es decir, en cada operación de cruzamiento se genera un número aleatorio uniforme $rm$, entre 0 y 1. Si $rm<p$, donde $p$ es la probabilidad de mutación, se efectúa la operación de mutación. Así, una posible mutación de la solución 1234 es 2234 (se cambia el primer gen por un 2), después de la corrección tendremos la solución 2233.

A primera vista el operador de mutación puede no parecer de vital importancia, pero esta operación asegura que ningún punto en el espacio de búsqueda tenga probabilidad cero de ser examinado.

De esta forma, la operación de mutación nos permite explorar diferentes espacios de solución, ya que sin ésta nuestro AG podría converger a un óptimo local.

Una buena característica a agregar en un AG consiste en aumentar la probabilidad de mutación conforme crece el número de generaciones, ya que la población será más homogénea con cada generación nueva y la mutación nos permitirá ``saltar'' a mejores soluciones, si estas existen.

\subsection{Restricciones}

Los problemas de optimización frecuentemente contienen restricciones, por lo que estas deben ser incluidas en el algoritmo genético, para esto existen dos métodos principales.

El primero consiste en asignar una penalización en la función de adaptación a aquellos individuos que no cumplen con las restricciones establecidas. De esta forma las soluciones que violen las restricciones serán eliminadas a lo largo de las generaciones. Sin embargo, la generación de soluciones que cumplan con las restricciones puede ser lenta, por lo que el algoritmo puede requerir mucho tiempo en ejecución para su convergencia.

El segundo método consiste en generar un mecanismo de corrección, el cual se menciono en la operación de cruzamiento. Este mismo mecanismo nos permitirá generar soluciones aleatorias para la población inicial. Es así que siempre que sea posible éste es el método más adecuado para cumplir con las restricciones del AG, debido a que tendrá un comportamiento más eficiente.

\subsection{Tamaño de la población}

El tamaño de la población con que trabajemos es especialmente importante, si esta es muy pequeña existe la posibilidad de no cubrir adecuadamente el espacio de búsqueda, por lo que la población convergerá rápidamente en dirección a un óptimo local. Contrariamente, si la población es muy grande el procesamiento de la población puede significar un costo grande de poder de cómputo.

Empíricamente observamos que un buen tamaño de población puede ser $N=n$, donde $n$ es el número de genes en un cromosoma, es decir, el número de caracteres que requerimos para codificar una solución a nuestro problema. Con este tamaño de población podemos explorar en cada individuo valores de cada gen. Si $n<30$ entonces utilizaremos $N=30$, que consideramos el tamaño de población mínimo efectivo.

\subsection{Población inicial}

La población inicial generalmente es aleatoria, sin embargo, en ocasiones un población inicial seleccionada explícitamente puede permitir una convergencia más rápida del algoritmo, por ejemplo utilizar soluciones que cumplan con alguna característica deseada. Pero si la población inicial es muy similar ésto puede también llevarnos a una convergencia prematura del algoritmo, que nos daría como resultado un óptimo local. Para evitar ésto se incrementa la probabilidad de mutación, que como ya mencionamos, nos permitirá alcanzar el óptimo global.

\subsection{Reemplazo de la población}

El reemplazo de la población puede seguir varios métodos, estos provocaran diferentes velocidades de convergencia en el AG, por lo que en cada caso se usara un menor o mayor grado de mutación para evitar que la convergencia sea a un óptimo local. Algunas formas de reemplazo de la población son:

\begin{itemize}
\item[$\bullet$] Cada vez que se genera una nueva solución mediante el cruce y mutación, ésta se coloca independientemente de su valor de adaptación en la anterior población, eliminado al resultado con menor valor de adaptación (en cada generación se puede crear un número nuevo de soluciones menor al de la población).
\item[$\bullet$] Otro método de reemplazo es utilizar la nueva generación reemplazando completamente a la anterior, este es uno de los métodos más utilizados. Aunque abra una mayor diversidad a lo largo de las generaciones, la convergencia puede ser lenta.
\item[$\bullet$] Si queremos ser más elitistas en la selección, uniremos la nueva y la anterior población, eliminando los resultados con el valor de adaptación menor, por lo que el individuo mejor adaptado permanece a lo largo de las generaciones. Este método requiere de un aumento en la mutación ya que la diversidad de resultados decae rápidamente.
\end{itemize}

En general el reemplazo de elementos de la población puede hacerse como mejor se crea conveniente para el problema, ya sea utilizando estos métodos u otros, siempre permitiendo algo de diversidad en las poblaciones.

\subsection{Criterio de parada}

Un algoritmo genético puede crear generaciones nuevas indefinidamente, por lo que tenemos que determinar un criterio que nos permita detener este proceso. Algunos de estos son:

\begin{itemize}
\item[$\bullet$] Cuando el valor promedio de adaptación entre la población de cada generación se diferencie en menos del 1\% con el mejor valor obtenido, podemos considerar que la población ha convergido y detendremos el AG.
\item[$\bullet$] Si después de cierto número de generaciones no hemos tenido una mejora en el miembro de la población con mejor adaptación podemos detener el proceso de búsqueda.
\item[$\bullet$] Podemos definir un número máximo de generaciones $m$, en el que conozcamos de antemano que la población converge a lo mucho en $k<m$.
\end{itemize}

En general una combinación de estas tres técnicas sería una buena condición de parada, es decir, si la población converge a un óptimo o no hemos tenido mejora en el resultado, detener el algoritmo; en caso contrario, pararlo después de cumplir un mínimo número de generaciones y reportar el mejor resultado.

La convergencia de la población puede ser temporal, por lo que el número de generaciones que revisemos antes de observar si existe una mejora importante debe ser suficientemente grande, es decir, al menos 30 generaciones.

\subsection{Ventajas y desventajas}

La ventaja de los algoritmos genéticos es básicamente su método de búsqueda, que le da gran poder frente a otros algoritmos de búsqueda heurística. Estos funcionan bien principalmente en grandes espacios de solución, con muchas dimensiones y óptimos locales. Donde otros algoritmos fallan, los AG escapan de estos óptimos locales, incluso con funciones objetivo complejas, por ejemplo cuando existen condiciones y valores de diferente tipo.

Es así que, con operaciones de cruce y mutación, podemos determinar que parte de una solución es la que le da un buen valor de adaptación y así combinarla con otras buenas características. De otra forma cada solución iría por su cuenta, buscando otra mejor.

Una ventaja muy importante de los algoritmos genéticos es que no es necesario especificar todos los detalles del problema en cuestión. Las soluciones potenciales son evaluadas por la función de adaptación, que representa el problema que deseamos resolver, seleccionando aquellas soluciones a lo largo de las generaciones con mejor valor de adaptación. Es así que los AG han producido muy buenos resultados en la ingeniería, ya que los AG pueden encontrar una solución que a priori hubiese sido desechada por una persona y sin embargo un AG se encargaría de explorar.

Otra ventaja de los AG es la capacidad de manipular varios parámetros simultáneamente, lo cual es necesario para definir muchos problemas de la vida real.

La principal desventaja de los AG viene dada en la representación del problema, tanto la codificación de las soluciones como la función de adaptación deben estar bien definidas. Si la codificación es incorrecta, la cruza podría darnos resultados aleatorios o sin sentido. Una mala selección de la función de adaptación nos llevara a resolver un problema diferente. Sin embargo esta desventaja tiene solución por medio de una examinación adecuada y cuidadosa del problema.

Por último, existe una desventaja en la necesidad de definir adecuadamente los parámetros del AG, como lo son el tamaño de la población, la probabilidad de mutación y el reemplazo. Pero podemos determinar los valores de estos parámetros al variarlos y observar como se comporta el algoritmo. En la mayoría de los casos buenos valores de los parámetros se encontrarán rápidamente.

\section{Un algoritmo genético para el modelo Media-Varianza}

En esta sección se presenta un algoritmo genético para obtener portafolios eficientes con respecto al criterio Media-Varianza, esto con el fin de mostrar la efectividad de los algoritmos genéticos comparando el resultado obtenido mediante programación cuadrática y el resultado de un AG.

En el modelo (\ref{quadp}), visto en el capítulo 3, definimos el rendimiento deseado y al variarlo obteníamos la frontera eficiente. 
\\
\\
Una alternativa al modelo (\ref{quadp}) es:

\begin{eqnarray}
\mbox{minimizar} \quad (1-\lambda ) {\bm w}'{\bm \Sigma}{\bm w} - \lambda E({\bm w}'{\bm R}) \label{lquadp} \\
 \nonumber \\
s.a. \quad {\bm w}'{\bm 1} = 1 \nonumber \\
{\bm w} \geq {\bm 0} \nonumber \\
0 \leq \lambda \leq 1 \nonumber
\end{eqnarray}

En el modelo (\ref{lquadp}) incluimos el rendimiento del portafolios en la función objetivo, de esta forma minimizamos el riesgo mientras maximizamos el rendimiento. El parámetro $\lambda$ puede ser considerado como una medida de aversión al riesgo, donde $\lambda = 0$ (portafolios de varianza mínima) representa la mayor aversión al riesgo y $\lambda = 1$ (portafolios de máximo rendimiento) representa la menor aversión o adicción a éste. Es así que al utilizar diferentes valores de $\lambda$ podemos construir la frontera eficiente anteriormente obtenida con (\ref{quadp}).

Por facilidad de programación utilizaremos el modelo (\ref{lquadp}) para el algoritmo genético que se desarrolla en esta sección.

Debido a que el problema que se quiere resolver mediante AG es de optimización, la definición de los elementos de éste es inmediata.

\begin{itemize}
\item[$\bullet$] {\bf Codificación:} El valor que deseamos determinar con el AG es ${\bm w}$, el cual es el vector de proporciones del capital invertidas en cada empresa. Es así que la codificación de las posibles soluciones es este mismo vector, en el cual podremos hacer operaciones de cruce y mutación fácilmente mediante la corrección $w_i^* = w_i / \sum_i w_i$, para cumplir con la restricción ${\bm w}'{\bm 1} = 1$.

\item[$\bullet$] {\bf Función de adaptación:} Es la función objetivo del problema de optimización, sin embargo, ya que nos interesa relacionar valores altos de la función con una mayor adaptación, utilizaremos: $f({\bm w}) = \lambda E({\bm w}'{\bm R}) - (1-\lambda ) {\bm w}'{\bm \Sigma}{\bm w}$, el cual es un problema de maximización.

\item[$\bullet$] {\bf Tamaño de población:} El tamaño de población que utilizaremos es 94 (número de empresas = 95), ya que este debe ser par, debido a que las operaciones de cruce generan soluciones pares. Cada nueva población estará conformada por las mejores soluciones de cada generación.
\end{itemize}

Estas son las características más importantes a definir del AG que se implementará. En cuanto al cruzamiento, se seleccionará un punto aleatorio de cruce. La mutación consistirá en cambiar aleatoriamente la proporción de inversión $w_i$ para algún $i$ también aleatorio. La población inicial será generada al azar, y para cumplir con la restricción ${\bm w} \geq {\bm 0}$, toda proporción generada $w_i$ será una variable aleatoria distribuida uniformemente $U(0,1)$.

El código R correspondiente a este algoritmo genético se encuentra en al apéndice \ref{GA1}, representado por la función {\bf GAlambdaPortfolio}. Equivalentemente la función {\bf lambdaPortfolio} del apéndice \ref{codf} da solución al modelo (\ref{lquadp}) mediante programación cuadrática.

\begin{figure}[h]
\begin{center}
	\includegraphics[width=11cm]{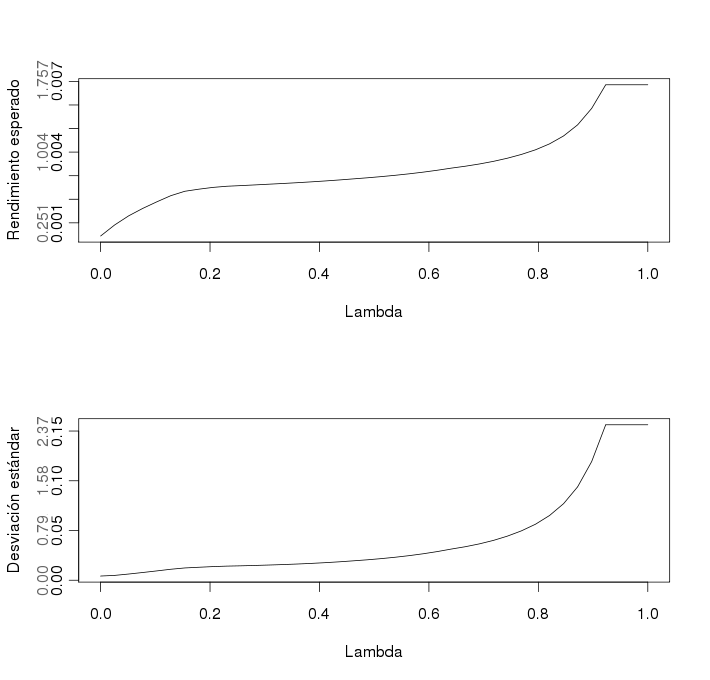} 
	\caption{Rendimiento esperado y desviación estándar del portafolios al variar $\lambda$} \label{lambda3}
\end{center}
\end{figure}

Haremos una comparación entre estos dos métodos de solución con $\lambda = 0.5$, como un punto medio entre la relación riesgo-rendimiento. Ya que esta relación no es lineal, el valor de $\lambda$ seleccionado en realidad no se encuentra en el medio del rango del rendimiento o del riesgo, sino que su localización dependerá de la forma de la frontera eficiente, sin embargo, en este caso $\lambda = 0.5$ nos proporciona un rendimiento adecuado con escaso riesgo. Este comportamiento puede apreciarse mejor en la Figura \ref{lambda3}.

El número de generaciones que se utilizarán en el AG es de $m=300$. 

\begin{verbatim}
# Solución por programación cuadrática, pf=objeto portafolios
lambdaPortfolio(pf, lambda=0.5)

$er
[1] 0.002934608

$sd
[1] 0.02111648

$portfolio
ABAT_Adj.Close ISRG_Adj.Close NVDA_Adj.Close  PTR_Adj.Close STEC_Adj.Close 
        0.1017         0.1031         0.3060         0.2623         0.2269 
\end{verbatim}
\medskip 
Para el algoritmo genético:
\\
\begin{verbatim}
# Solución por algoritmos genéticos, pf=objeto portafolios
# m=número de generaciones
GAlambdaPortfolio(pf, lambda=0.5, m=300)

$er
            [,1]
[1,] 0.002915660

$sd
           [,1]
[1,] 0.02094783

$portfolio
AAPL_Adj.Close ABAT_Adj.Close ISRG_Adj.Close NVDA_Adj.Close  PTR_Adj.Close 
        0.0213         0.1013         0.0928         0.3008         0.2518 
STEC_Adj.Close 
        0.2211 
\end{verbatim}

Podemos observar que la solución obtenida mediante algoritmos genéticos es muy cercana al óptimo global. La programación cuadrática (PC) nos proporciona un portafolios con un rendimiento esperado diario de 0.002934 y el AG de 0.002915, en cuando a la desviación estándar del portafolios, PC reporta 0.021116 y AG 0.020947. 

También la composición de los portafolios en ambas soluciones es similar, con la diferencia de que el AG incluye a Apple (AAPL) en el portafolios\footnote{En el portafolios generado mediante AG sólo se presentan aquellas empresas que tienen una proporción de inversión $w_i>0.005$, ya que hay muchas empresas con inversiones insignificantes de $0.0001$.}. 

Es así que el algoritmo genético nos ha proporcionado una muy buena solución, la cual es bastante cercana al óptimo global con tan sólo 300 generaciones.

\begin{figure}[h]
\begin{center}
	\includegraphics[width=11cm]{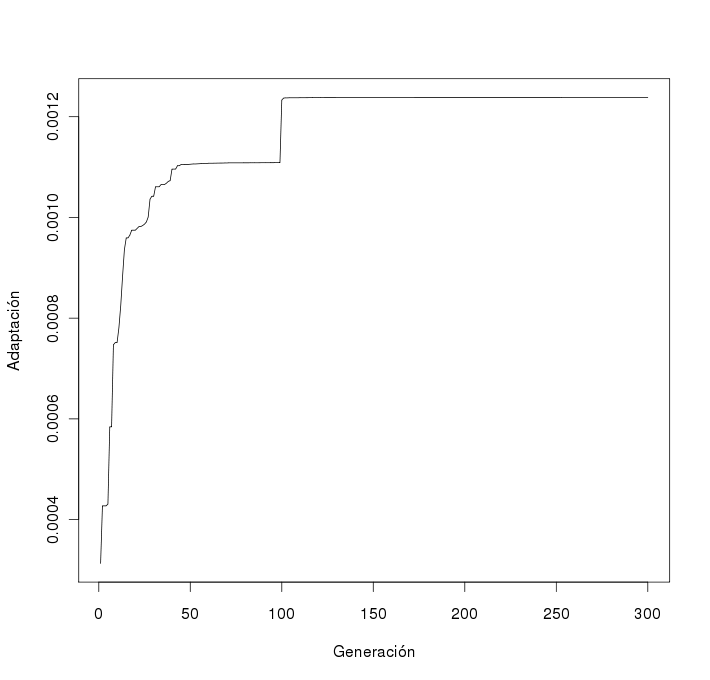} 
	\caption{Convergencia del algoritmo genético} \label{gaInter}
\end{center}
\end{figure}

La función {\bf GAlambdaPortfolio} también nos devuelve una gráfica de la convergencia de la población, la cual se presenta en la Figura \ref{gaInter}.

En esta gráfica podemos ver que la convergencia al mejor adaptado sucede desde la generación 100, sin existir mejora en las iteraciones posteriores. 

En el primer sexto del proceso hay un rápido crecimiento en la adaptación de las soluciones, después aparece una convergencia temporal y por último un salto al óptimo que alcanzo el algoritmo. Estos grandes saltos posteriores a una convergencia generalmente son causados por la mutación, de aquí su importancia; si sólo utilizamos el cruzamiento el óptimo se vería altamente afectado por la población inicial.

Aunque el tiempo de ejecución del algoritmo genético utilizado fue corto, se apreció una gran diferencia con el método de programación cuadrática, que fue inmediato.

\begin{figure}[h]
\begin{center}
	\includegraphics[width=11cm]{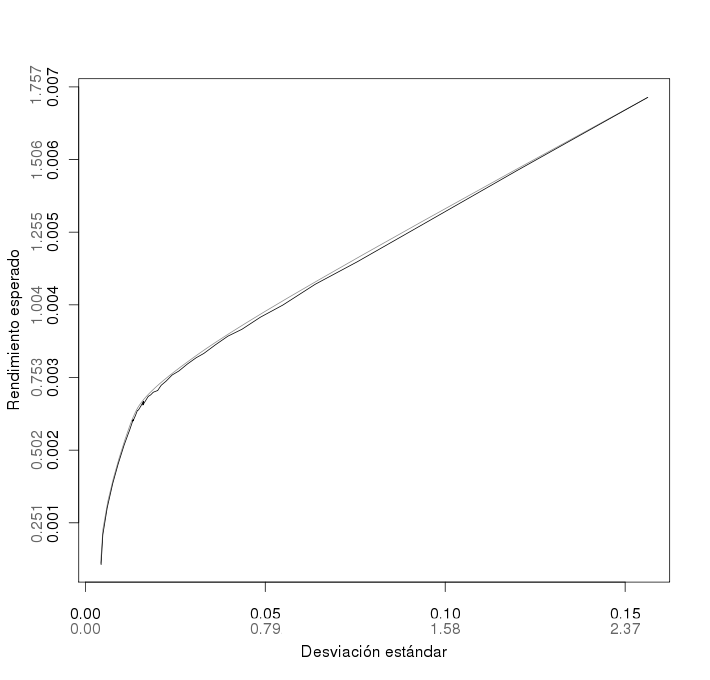} 
	\caption{Frontera eficiente - algoritmos genéticos} \label{gafront}
\end{center}
\end{figure}

Por último, para realizar una mejor comparación de los resultados de algoritmos genéticos y programación cuadrática, graficamos la frontera eficiente obtenida mediante PC (gris) y la de AG (negro) con 500 generaciones cada punto. El resultado se muestra en la Figura \ref{gafront}.

\begin{verbatim}
# graficamos la frontera eficiente mediante programación cuadrática
plot(lambdaFrontier(pf), xlab="Desviación estándar", 
  ylab="Rendimiento esperado", type="l", col=c("gray"))
  
# mediante algoritmos genéticos
lines(GAlambdaFrontier(pf, 500), type="l")
\end{verbatim}

Puede apreciarse como la diferencia entre ambas fronteras es prácticamente imperceptible, especialmente en los extremos aparentan ser una misma línea. Es así que el algoritmo genético programado nos proporciona soluciones muy cercanas al óptimo global.

En conclusión, los algoritmos genéticos son una buena técnica de aproximación a los óptimos globales de aquellos problemas que son difíciles de solucionar con métodos deterministas (lo cual nos será muy útil en el próximo capítulo). Sin embargo, si existe un método de optimización para abordar el problema a resolver, es mejor utilizar éste, ya que en general un algoritmo genético será más lento y la solución no será tan buena como la proporcionada por el método de optimización.

\chapter{Costos por transacción y restricciones de enteros}

{\em Al final del capítulo 2 mencionamos dos de las principales debilidades del modelo de Markowitz, una es que no incluye costos por transacción y la otra que las proporciones de inversión que devuelve son infinitamente divisibles.

El costo por transacción es una comisión que debe cubrirse para efectuar un intercambio económico, en este caso, por la compra o venta de activos de una empresa.

Las restricciones de enteros hacen referencia al número de acciones que podemos comprar de una empresa con el capital disponible, en lugar de las proporciones.

Puede verse que cuando se desea realizar una inversión, tendremos que afrontar estos dos problemas. Es así que la inclusión de los costos por transacción y las restricciones de enteros son elementos esenciales en cualquier modelo practico de optimización de portafolios.

Sin embargo, cuando tomamos en cuenta los costos por transacción y/o las restricciones de enteros, el problema de optimización se torna mucho más complicado. En este capítulo incluiremos estas características en el modelo de Markowitz y veremos el impacto que provocan en éste.

Para la solución de este problema algunos métodos heurísticos que se han desarrollado muestran ser efectivos. Uno de estos métodos son los algoritmos genéticos.
}

\section{El modelo de optimización}

Comúnmente los costos por transacción son proporcionales al volumen comerciado, estos costos varían linealmente con la cantidad de acciones compradas y vendidas. Es así que el modelo de optimización que utilicemos debe considerar que los costos por transacción que serán pagados, son proporcionales, permitiendo diferentes costos para cada empresa y diferentes costos para comprar y vender. Así, $c_{bi}$ y $c_{si}$ representan las proporciones del costo por transacción de comprar y vender una acción de la empresa $i$, respectivamente, donde $i \in \{1,\cdots,N\}$.

Sea $n_i$ el número de activos que adquirimos, tal que $n_i \in \mathbb{N}_0^+ $ y $p_i$ es el precio actual del activo $i$. Entonces, el costo total por comprar $n_i$ acciones de la empresa $i$, es $n_i \cdot p_i + n_i \cdot p_i \cdot c_{bi}$, que en adelante expresaremos como $n_i \cdot p_i + C_{bi}$, tal que $C_{bi} = n_i \cdot p_i \cdot c_{bi}$.

Si el inversionista posee un capital inicial $K$; debido a la indivisibilidad de las proporciones por las restricciones enteras $n_i$, el capital gastado en la compra de activos no es necesariamente $K$, por lo que existe un residuo $\epsilon = K - \sum_{i=1}^N (n_i \cdot p_i + C_{bi})$, el cual puede ser utilizado en una inversión a una tasa libre de riesgo $R_f$\footnote{Aunque decimos que el residuo se invierte en un activo libre de riesgo, en realidad este activo tendrá mayor o menor participación si la solución así lo amerita, es decir, el capital que se invierta en $R_f$ no será el residuo después de invertir lo más posible en renta variable, sino que dependerá de su conveniencia.}, por ejemplo los intereses que paga un banco. Es así que $\epsilon \geq 0$, por lo que el apalancamiento\footnote{El apalancamiento se refiere a la obtención de un préstamo para financiar una compra mayor de activos y así aumentar el rendimiento de una inversión.} no está permitido.

Para el costo por transacción de compra, $C_{bi}$ ya está siendo cobrado en el proceso de adquisición de activos, es decir, si tenemos un dólar para invertir, sólo podremos utilizar parte de ese dólar en la compra de activos, ya que el resto será para pagar los costos de compra. Es así que los costos de compra se pagan al principio del periodo de inversión.

En cuanto al costo por venta, este será cobrado al final del periodo, ya que en algún momento tenemos que vender todos los activos para transformarlos en rendimiento. Es así que definimos el costo por transacción de venta como $C_{si} =  n_i \cdot ( p_i + T \cdot R_i \cdot p_i ) \cdot c_{si}$, donde $T$ es el número total de periodos $t$ y $R_i$ el rendimiento de la empresa $i$ por cada periodo $t$. Así, $p_i + T \cdot R_i \cdot p_i$ es el precio al final de los $T$ periodos (precio esperado) por cada acción de la empresa $i$. De esta forma, el costo de venta dependerá del valor de $T$, en nuestro caso $t=$ un día de Bolsa (día bursátil), por lo que $T=251$, es decir, un periodo de inversión de un año.
\\
\\
Así, el rendimiento del portafolios por cada periodo $t$, está dado por:

\begin{equation}
R_p = \frac{\sum_{i=1}^N R_i \cdot p_i \cdot n_i}{K} - \frac{C_{si}}{K \cdot T} + \frac{\epsilon \cdot R_f}{K}
\end{equation}
\\
Esto es, el rendimiento del portafolios es el rendimiento que nos proporcionan los activos adquiridos, menos la proporción del rendimiento a pagar por costos de venta, más el rendimiento obtenido por el activo libre de riesgo.
\\
\\
Además, si definimos la proporción de inversión:

$$ w_i = \frac{n_i \cdot p_i}{K} \Longleftrightarrow \sum_{i=1}^N w_i = 1 - \frac{\epsilon + \sum_{i=1}^N C_{bi}}{K} $$
\\
\\
podemos definir el riesgo del portafolios como:

\begin{equation}
\sigma_p^2 = \sum_{i=1}^N \sum_{j=1}^N w_i  w_j \sigma_{ij}
\end{equation}
\\
\\
donde $\sigma_{ij}$ es la covarianza entre los rendimientos esperados $i$ y $j$. Es así que el modelo de optimización que utilizaremos es:

\begin{eqnarray}
\mbox{maximizar} \quad \lambda E(R_p) - (1-\lambda ) {\bm w}'{\bm \Sigma}{\bm w}  \label{lquadp2} \\
 \nonumber \\
s.a. \quad w_i = \frac{n_i \cdot p_i}{K} \nonumber \\
n_i \in \mathbb{N}_0^+ \nonumber \\
\epsilon = K - \sum_{i=1}^N (n_i \cdot p_i + C_{bi}) \geq 0 \nonumber \\
0 \leq \lambda \leq 1 \nonumber
\end{eqnarray}

Puede verse que este modelo requiere de más información de la inversión, como el capital disponible $K$, el precio $p_i$ para determinar el número máximo de activos que podemos adquirir para cada empresa $i$, los costos proporcionales por transacción $c_{bi}$ y $c_{si}$, el rendimiento para el activo libre de riesgo $R_f$, así como el periodo total de inversión $T$, para calcular los costos de venta.

Es evidente que la inclusión de costos por transacción y restricciones enteras complica significativamente el problema de optimización de portafolios, por lo que requeriremos de un método de búsqueda heurística para dar solución a este modelo.

\section{Algoritmo genético}

El modelo de selección de portafolios presentado anteriormente será resuelto por medio de un algoritmo genético, que como vimos en el capítulo anterior, es un método de búsqueda con gran éxito en este tipo de problemas. Las características principales del algoritmo son:

\begin{itemize}
\item[$\bullet$] {\bf Codificación:} En este caso el valor que deseamos determinar con el AG es ${\bm n}$, el cual es el vector del número de acciones compradas de cada empresa, es decir ${\bm n} = [ n_1, n_2, \cdots, n_N ]$ tal que $n_i \in \mathbb{N}_0^+ \;  \forall \; i \in \{1,\cdots,N\}$. Es así que la codificación de las posibles soluciones es este mismo vector.

\item[$\bullet$] {\bf Función de adaptación:} La función de adaptación es la función objetivo del problema de optimización ya que se trata de un problema de maximización.

\item[$\bullet$] {\bf Tamaño de población:} El tamaño de población que utilizaremos es 94 (número de empresas = 95). Cada nueva población estará conformada por las mejores soluciones de cada generación.
\end{itemize}

En cuanto al cruzamiento, se seleccionará un punto aleatorio de cruce. La mutación consistirá en cambiar aleatoriamente el número de activos adquiridos $n_i$ para algún $i$ igualmente aleatorio. La población inicial será generada al azar.

Para cumplir con la restricción $\epsilon \geq 0$, todo número de activos generado $n_i$, debe estar entre el 0 y el máximo número posible de compra con el capital actual. Si requerimos corregir la solución generada debido a que se realizó una operación de cruzamiento o mutación, las proporciones de inversión deben conservarse lo mejor posible.

El código R correspondiente a este algoritmo genético se encuentra en al apéndice \ref{GA2}, representado por la función {\bf GAlambdaNPortfolio}.

Haremos una comparación entre la solución de PC y AG seleccionando $\lambda = 0.5$, como un punto medio entre la relación riesgo-rendimiento. El número de generaciones que utilizará el AG es de $m=300$. 

El capital disponible que utilizaremos es de $K=10,000$. En cuanto a los costos por transacción de compra y venta usaremos $c_{bi}=0.01$ y $c_{si}=0.01, \; \forall i \in \{1,\cdots,N\}$. El periodo de inversión que tenemos contemplado es de un año, por lo que $T=251$. Por último, el rendimiento para el activo libre de riesgo será $R_f = 0.07/251$, aproximadamente lo que proporciona un activo libre de riesgo en México.

Debido a que es necesario definir un capital y precios, es evidente que todos estos valores deben estar en la misma escala (moneda), en este caso son dólares.
\\
\\
Una vez definidos estos parámetros obtenemos el portafolios con cada método. Primero el resultado por programación cuadrática:

\begin{verbatim}
# Solución por programación cuadrática, pf=objeto portafolios
lambdaPortfolio(pf, lambda=0.5)

$er
[1] 0.002934608

$sd
[1] 0.02111648

$F
[1] 0.001244351

$portfolio
ABAT_Adj.Close ISRG_Adj.Close NVDA_Adj.Close  PTR_Adj.Close STEC_Adj.Close 
        0.1017         0.1031         0.3060         0.2623         0.2269 
\end{verbatim}
\medskip 
Para el algoritmo genético:
\\
\begin{verbatim}
# Solución por algoritmos genéticos, pf=objeto portafolios

# precios del periodo 2007, el primer precio es el actual
prices2<-read.table('prices-2007.csv', header=T, sep=',')

# guardamos el precio actual en pr
pr<-as.numeric( prices2[1, 2:96] )

# vector de costos por transacción proporcionales
cb<-rep(0.01, 95)
cs<-rep(0.01, 95)

# m=número de generaciones
# prices=vector de precios actuales
# cb=vector de costos proporcionales de compra
# cs=vector de costos proporcionales de venta
# cap=capital inicial
# rf=rendimiento libre de riesgo
# t=número total de periodos

p<-GAlambdaNPortfolio(pf, lambda=0.5, m=300, prices=pr, cb=cb, cs=cs, 
   cap=10000, rf=0.07/251, t=251)

print(p)

$er
            [,1]
[1,] 0.002842697

$sd
           [,1]
[1,] 0.02109197

$e
[1] 3.5149

$F
[1] 0.001198913

$n
 [1]    0    0 1569    0    0    0    0    0    0    0    0    0    0    0    0
[16]    0    0    0    0    0    0    0    0    0    0    0    0    0    0    0
[31]   14    0    0    0    0    0    0    0    0    0    0    0    0    0    0
[46]    0    0    0    0    0    0    0    0  117    0    0    0    0    0    0
[61]    0    0    0    0   21    0    0    0    0    0    0    0    0    0    0
[76]  164    0    0    0    0    0    0    0    0    0    0    0    0    0    0
[91]    0    0    0    0    0

$portfolio
ABAT_Adj.Close ISRG_Adj.Close NVDA_Adj.Close  PTR_Adj.Close STEC_Adj.Close 
        0.1036         0.1323         0.2814         0.2607         0.2119 

$nportfolio
ABAT_Adj.Close ISRG_Adj.Close NVDA_Adj.Close  PTR_Adj.Close STEC_Adj.Close 
          1569             14            117             21            164 
\end{verbatim}

Debe observarse que ambas soluciones (PC y AG) no son del todo comparables, ya que el nuevo modelo agrega restricciones, siendo $\sum_{i=1}^N w_i \leq 1$, por lo que las medidas del riesgo y el rendimiento son ligeramente diferentes. 

Es así que el óptimo que proporciona la función {\bf lambdaPortfolio}, no es en realidad el óptimo del modelo que abordamos en este capítulo. Sin embargo, la comparación entre estos resultados nos podrá dar una idea de que tan buena es la solución proporcionada por el AG, ya que los costos por transacción asignados son del 1\%, por lo que las soluciones deben ser similares.

El resultado devuelto por el AG nos proporciona un portafolios cuyo rendimiento esperado anual es de 71.3\% (esto es un rendimiento esperado diario de {\it p\$er = 0.002842697}).

Después de la compra de acciones y el pago de los costos por transacción, 3.515 ({\it p\$e = 3.5149}) dólares no se pudieron invertir (residuo), por lo que estos se invierten en renta fija, por supuesto el rendimiento diario que aportan es insignificante, por lo que no tiene efecto en el rendimiento del portafolios.

El portafolios resultante está conformado por NVIDIA (NVDA) con el 28.14\% de inversión del capital, esto es 117 acciones, seguido de PetroChina (PTR) con 26.07\% (21 acciones), STEC (STEC) con 21.19\% (164 acciones), Intuitive Surgical (ISRG) con 13.23\% (14 acciones) y Advanced Battery Technologies (ABAT) con 10.36\% (1569 acciones).

Podemos ver que las empresas que conforman al portafolios generado por AG y PC son las mismas y sus proporciones muy similares, pero en el resultado ahora se incluye la cantidad de acciones que se compran de cada empresa con el capital disponible ({\it p\$nportfolio}). Por ejemplo, ABAT representa la menor proporción de inversión, pero se compran 1569 acciones de esta empresa ya que su precio es de \$0.66 dólares.

Por el rendimiento esperado y desviación estándar de ambos portafolios, puede suponerse que la solución que nos ha devuelto el algoritmo genético es muy cercana al óptimo global.

\begin{figure}[h]
\begin{center}
	\includegraphics[width=11cm]{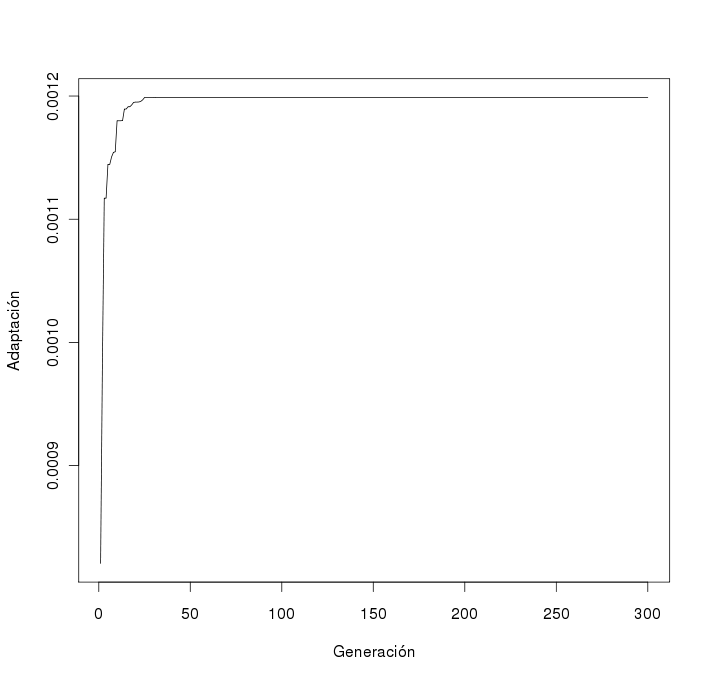} 
	\caption{Convergencia del algoritmo genético} \label{ga2Inter}
\end{center}
\end{figure}

La función {\bf GAlambdaNPortfolio} también genera la gráfica de la convergencia de la población, la cual se presenta en la Figura \ref{ga2Inter}.

Ya que el espacio de soluciones es finito, en general, con este algoritmo tendremos convergencias más rápidas, así, en el espacio de soluciones nos moveremos por valores enteros que tienen límites bien definidos, con la posibilidad de alcanzar los óptimos con un bajo número de generaciones, inclusive el óptimo global, es decir, la solución devuelta tiene gran probabilidad de ser el óptimo global.

Por último, para apreciar mejor el efecto de los costos por transacción en la optimización de portafolios, graficamos la frontera eficiente obtenida mediante PC (sin restricciones de enteros o costos por transacción) y la de AG, con diferentes valores para los costos por transacción. Los costos por transacción proporcionales que utilizaremos son: (1) $c_{bi} = c_{si} = 0.01$, (2) $c_{bi} = c_{si} = 0.05$ y (3) $c_{bi} = c_{si} = 0.1 \;\; \forall \; i \in \{1,\cdots,N\}$.

Aunque en general el residuo de la inversión será insignificante y en este caso el rendimiento del activo libre de riesgo no genera una diferencia relevante en el rendimiento del portafolios, para valores de $\lambda$ que generan portafolios de escaso rendimiento, la participación de este activo será mayor, por lo que para una comparación más adecuada el rendimiento del activo libre de riesgo será redefinido como $R_f=0$.

Para realizar estas gráficas utilizamos el siguiente código:

\begin{verbatim}
# graficamos la frontera eficiente mediante programación cuadrática
plot(lambdaFrontier(pf), xlab="Desviación estándar", 
    ylab="Rendimiento esperado", type="l", col=c("gray"))
  
# costos = 0.01
cb<-rep(0.01, 95)
cs<-rep(0.01, 95)
lines(GAlambdaNFrontier(pf, m=500, prices=pr, cb=cb, cs=cs, cap=10000, 
    rf=0, t=251), type="l")

# costos = 0.05
cb<-rep(0.05, 95)
cs<-rep(0.05, 95)
lines(GAlambdaNFrontier(pf, m=500, prices=pr, cb=cb, cs=cs, cap=10000, 
    rf=0, t=251), type="l")

# costos = 0.1
cb<-rep(0.1, 95)
cs<-rep(0.1, 95)
lines(GAlambdaNFrontier(pf, m=500, prices=pr, cb=cb, cs=cs, cap=10000, 
    rf=0, t=251), type="l")
\end{verbatim}

\begin{figure}[h]
\begin{center}
	\includegraphics[width=11cm]{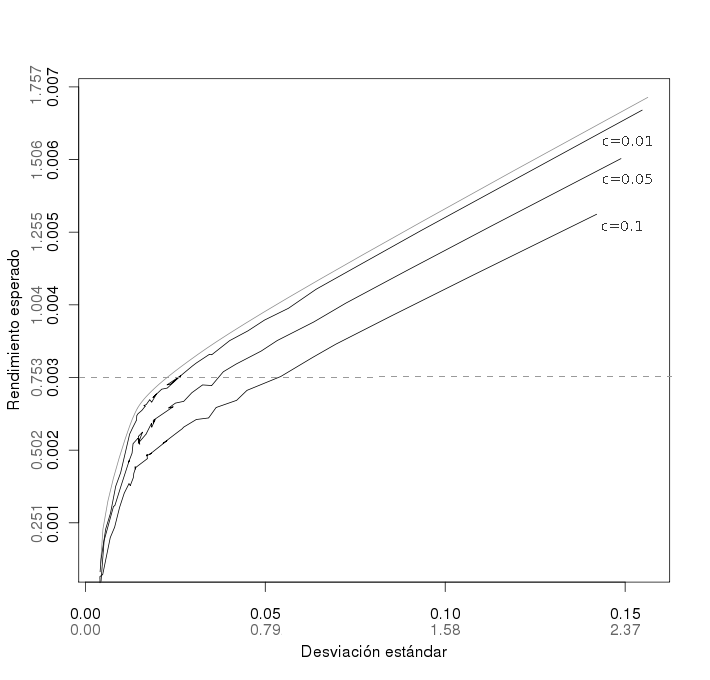} 
	\caption{Frontera eficiente por algoritmos genéticos y diferentes costos por transacción} \label{ga2front}
\end{center}
\end{figure}

El resultado se muestra en la Figura \ref{ga2front}. Puede verse como afectan los costos por transacción al rendimiento y al riesgo. Entre mayor es este costo, el rendimiento esperado máximo del portafolios disminuye y el riesgo para cada nivel de rendimiento es mayor.

Es así que la forma de la frontera eficiente cambiará dependiendo de los costos por transacción que debemos pagar, al igual que los portafolios que la conforman. Es evidente la importancia de la inclusión de estos costos en el problema de optimización, ya que de no ser así, en la realidad no alcanzaremos las expectativas de rendimiento que nos proporciona el modelo Media-Varianza, invirtiendo en un portafolios diferente del que debemos utilizar.

Un ejemplo es el portafolio de mínima varianza del capítulo 2. Si un inversionista selecciona este portafolios, al final del periodo de inversión tendrá un rendimiento menor al esperado debido a los costos por transacción (y también a que no puede adquirir el portafolios con las proporciones exactas), siendo este el caso, el inversionista pudo preferir invertir en renta fija.

\section{Lotes de transacción mínimos}

Dentro de las restricciones que maneja un problema de selección de portafolios puede existir un mínimo de acciones por transacción, es decir, lotes.

El manejo de lotes en el modelo (\ref{lquadp2}) es fácil de incluir, ya que se trata de una restricción de enteros. Sólo es necesario utilizar el precio por lote en lugar del precio por acción, por ejemplo, si el número mínimo de acciones por transacción para la empresa $i$, es de 100, entonces utilizaremos $p_i^* = p_i \cdot 100$, de esta forma, el resultado que nos proporcionará el algoritmo será el número de lotes a comprar de la empresa.

Si para el problema de selección anterior definimos lotes de al menos 100 acciones para todas las empresas, sólo es necesario multiplicar por 100 los precios correspondientes. Esto es:

\begin{verbatim}
# utilizamos prices=pr*100
p<-GAlambdaNPortfolio(pf, lambda=0.5, m=300, prices=pr*100, cb=cb, cs=cs, 
   cap=10000, rf=0.07/251, t=251)

print(p)

$er
           [,1]
[1,] 0.003009554

$sd
           [,1]
[1,] 0.02627683

$e
[1] 27.26

$F
[1] 0.001159541

$n
 [1]  0  0 18  0  0  0  0  0  0  0  0  0  0  0  0  0  0  0  0  0  0  0  0  0  0
[26]  0  0  0  0  0  0  0  0  0  0  0  0  0  0  0  0  0  0  0  0  0  0  0  0  0
[51]  0  0  0  2  0  0  0  0  0  0  0  0  0  0  0  0  0  0  0  0  0  0  0  0  0
[76]  3  0  0  0  0  0  0  0  0  0  0  0  0  0  0  0  0  0  0  0

$portfolio
ABAT_Adj.Close NVDA_Adj.Close STEC_Adj.Close 
        0.1188         0.4810         0.3876 

$nportfolio
ABAT_Adj.Close NVDA_Adj.Close STEC_Adj.Close 
            18              2              3 
\end{verbatim}

\begin{figure}[h]
\begin{center}
	\includegraphics[width=11cm]{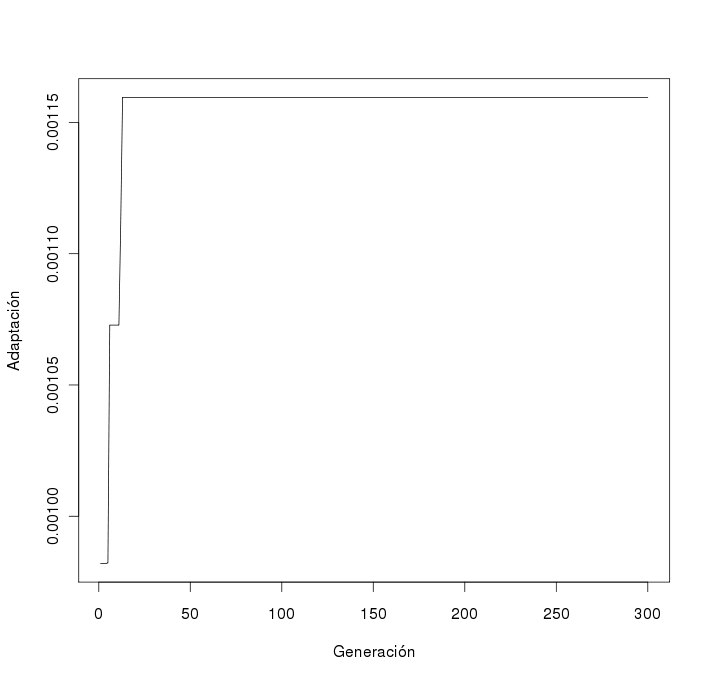} 
	\caption{Convergencia del algoritmo genético (lotes)} \label{ga3}
\end{center}
\end{figure}

El portafolios resultante posee un rendimiento esperado anual de 75.5\%, el cual está conformado por: 18 lotes para Advanced Battery Technologies (ABAT), 2 lotes para NVIDIA (NVDA) y 3 lotes de STEC (STEC). 

En cuanto al riesgo, éste es un portafolios de mayor riesgo que el anterior (aunque el rendimiento también es mayor, proporcionalmente el riesgo creció más), evidentemente, al tener una restricción adicional (lotes), el portafolios resultante es a lo más tan buen como el portafolios sin esta restricción.

El residuo en este caso es de 27.26 dólares ({\it p\$e = 27.26}). Es de esperar que el residuo aumente, ya que el dinero necesario para comprar un lote es mucho mayor que para acciones.

El modelo Media-Varianza, visto en el capítulo 2, supone que podemos aspirar a cualquier expectativa de rendimiento dentro de los límites de la frontera eficiente, sin importar el precio de las acciones. El efecto de los lotes está relacionado con el rendimiento del portafolios que en realidad nos es posible alcanzar. Por ejemplo, si tenemos poco capital y requerimos comprar en mayor proporción acciones de determinada empresa, pero no es posible debido al precio de los lotes, entonces la expectativa de rendimiento que podemos alcanzar es menor. 

Así, la presencia de lotes tiene menor efecto en la selección de portafolios conforme el capital disponible es mayor, ya que de esta forma podemos aproximarnos a las proporciones de inversión óptimas.

\begin{figure}[h]
\begin{center}
	\includegraphics[width=11cm]{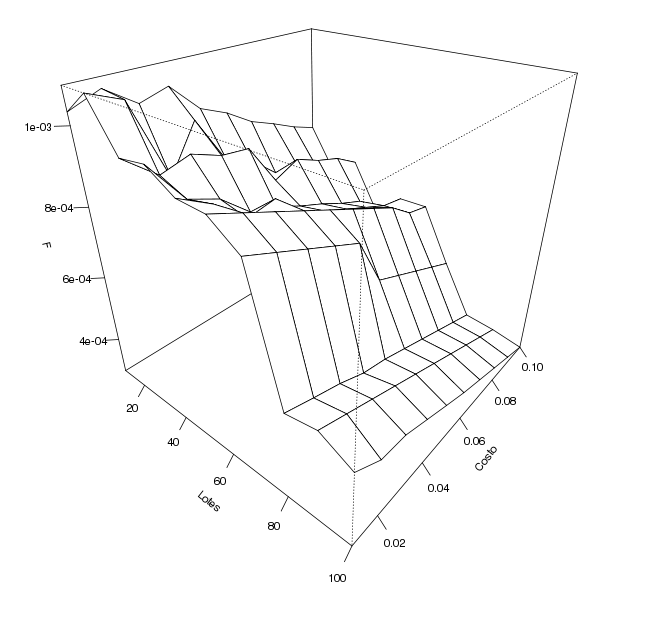} 
	\caption{Efecto al aumentar el tamaño de lotes y costos por transacción} \label{lot3d}
\end{center}
\end{figure}

Para entender mejor el efecto de los lotes cuando el capital es escaso, en la Figura \ref{lot3d} se muestra el comportamiento conjunto sobre la función objetivo al aumentar el tamaño de los lotes y de los costos por transacción\footnote{Para esto se utilizó $\lambda=0.5$ y $K=1,000$, con lotes que van desde 10 hasta 100 acciones y costos por transacción de 1\% - 10\%.}; recordando que en este caso el máximo teórico para la función objetivo es de 0.001244, podemos observar que entre mayor es el número de acciones que conforman un lote o el costo por transacción, menor es el valor que esta función puede alcanzar y, por lo tanto, más deficiente la relación riesgo-rendimiento.

\section{Observaciones}

En este capítulo construimos un modelo para la selección de portafolios con costos por transacción y restricciones de enteros. Observamos que la inclusión de los costos por transacción puede tener efectos importantes en la composición de los portafolios eficientes, como en el riesgo y rendimiento de los mismos.

Así, las soluciones óptimas bajo los costos por transacción pueden diferir significativamente del caso de un mercado perfecto, donde estos costos no existen, por lo que todo modelo realista debe incluirlos.

No se ha mostrado el desempeño a futuro de los portafolios formados en este capítulo debido a que el comportamiento será similar al del capítulo 2, ya que se ha utilizado el mismo criterio de selección. Los portafolios donde ABAT represente una proporción importante de la inversión, tendrán mayor rendimiento del esperado en 2007, y aquellos donde NVDA, STEC, PTR representen la mayor proporción, tendrán un rendimiento futuro muy similar al esperado.

En cuando a la implementación del modelo, los algoritmos genéticos han sido un buen método de optimización, que nos ha proporcionado muy buenas soluciones. El modelo aquí presentado y el algoritmo genético para su solución pueden ser fácilmente adaptados para resolver el problema de rebalanceamiento, donde se genera un nuevo portafolios $\bm w^*$, a partir de uno ya existente $\bm w$, considerando los costos por transacción.

La inclusión de la restricción de enteros nos permite determinar el número exacto de acciones a comprar de cada empresa, dado el capital del que disponemos, al igual que los lotes, los cuales también tienen efecto en el rendimiento del portafolios que podemos obtener, cambiando considerablemente la composición del mismo.

Los portafolios formados en este capítulo pueden ser seleccionados con el criterio Media-Semivarianza, simplemente haciéndolo explicito en la definición del objeto {\it portfolio} con {\it pf$<$-newPortfolio(R, risk="svar")}.

Puede pensarse que la solución al modelo visto en este capítulo no varía significativamente del modelo de Markowitz, pero ésto no es así. Se sabe que en optimización el agregar restricciones a un modelo cambia la solución optima, y en este caso, entre menos capital disponible tengamos, mayor sea el costo por transacción o mayor el número de acciones que conforman un lote, el portafolios óptimo diferirá cada vez más del generado por el modelo de Markowitz, así como nuestras posibles expectativas de riesgo y rendimiento.

Como vimos, el espacio de soluciones en este modelo es finito, por lo que es alta la probabilidad de que la solución generada por el AG sea el óptimo global.

\chapter*{Conclusión} \addcontentsline{toc}{chapter}{Conclusiones}

Los modelos de selección y optimización de portafolios tienen como objetivo proporcionar al inversionista un instrumento estadístico que le permita elegir de manera razonable un portafolio de inversión futuro, con las características que prefiera. Sin embargo, todos los modelos de optimización tienen ventajas y desventajas; en esta investigación se muestran los resultados y características de los modelos abordados.

Después de analizar los rendimientos de 95 empresas mediante el modelo Media-Varianza de Markowitz, así como el modelo Media-Semivarianza, los costos por transacción y lotes, llegamos a las siguientes conclusiones, algunas ya mencionadas en sus respectivos capítulos.

\begin{itemize}
    \item[$\bullet$] En los capítulos 2 y 3 mostramos la forma de aplicar la teoría de portafolios, la importancia de la diversificación, así como el uso del criterio Media-Varianza. Aunque existen diferencias entre las preferencias de un inversionista con respecto a su aversión al riesgo, todas las decisiones que éste tome deben estar ubicadas en la frontera eficiente.
    
    \item[$\bullet$] La solución óptima de un modelo de selección es sensible a la información utilizada. En los modelos aquí abordados, la medida de rendimiento es la media, que puede verse afectada fácilmente por los datos de la muestra, causando selecciones de portafolios diferentes\footnote{V. K. Chopra y W. T. Ziemba. \emph{The effect of errors in means, variances, and covariances on optimal portfolio choice}. En: Worldwide asset and liability modeling, páginas 5361. Cambridge University Press, 1998}. Por lo tanto, la medida de rendimiento debe ser calculada adecuadamente y con suficiente información.

    \item[$\bullet$] El modelo Media-Varianza y el modelo Media-Semivarianza en general presentan un buen comportamiento, pero las elecciones de portafolios son menos confiables conforme se acepta más riesgo para obtener un mayor rendimiento.
    
    \item[$\bullet$] Es más adecuado utilizar la semivarianza como medida de riesgo, ya que mide el subdesempeño de los rendimientos que conforman el portafolio, lo cual es más cercano a lo que un inversionista considera riesgo. Así, la semivarianza nos permitirá realizar una elección de portafolios más conveniente; en esta investigación, el modelo Media-Semivarianza ha proporcionado mejores portafolios.

    \item[$\bullet$] Para realizar una inversión en bolsa también es necesario incluir algunas estrategias de inversión, como por ejemplo saber decidir el momento de compra-venta una vez que se sabe cual será nuestro portafolio, ya que una mala selección en este punto puede disminuir considerablemente el rendimiento.

    \item[$\bullet$] En la realidad siempre existen costos asociados con la compra-venta de acciones, ya sea costos por transacción o impuestos. En el capítulo 5 pudimos apreciar el impacto de estos costos y cómo integrarlos en el modelo Media-Varianza. Existen también costos por transacción fijos y costos mínimos por transacción; estos o su combinación pueden ser agregados fácilmente al modelo presentado.

    \item[$\bullet$] Hemos construido un modelo, presentado en el capítulo 5, que nos proporciona portafolios óptimos con el número exacto de lotes (de una o más acciones) a comprar, el rendimiento esperado que obtendremos después de pagar los costos por transacción y la proporción de inversión en renta fija, dada nuestra capacidad de inversión. Así, mediante este modelo podemos realizar una inversión real conociendo su posible comportamiento a futuro y el riesgo que asumimos al buscar mayor rendimiento.

    \item[$\bullet$] Las restricciones de costos por transacción y las de enteros pueden afectar considerablemente el portafolio óptimo dependiendo de varios factores, por lo que la inclusión de estas restricciones es muy importante para obtener un modelo de selección de portafolios práctico.

    \item[$\bullet$] En nuestro caso, el modelo del capítulo 5 es muy complejo para ser abordado con un método clásico de optimización; por lo tanto, usamos los algoritmos genéticos, que muestran dar muy buenos resultados con la capacidad de alcanzar el óptimo global en un corto número de iteraciones.

    \item[$\bullet$] Los modelos abordados en esta investigación son para un solo período de inversión, pero estos (en especial el del capítulo 5) pueden ser modificados para hacerlos multi-periodo y, por ejemplo, agregar el caso del rebalanceamiento.
\end{itemize}

\appendix

\chapter{Código fuente: markowitz.R} \label{codf}

El programa R aquí presentado contiene varias funciones utilizadas durante la optimización de portafolios a lo largo de esta investigación. Las proporciones de inversión que devuelve la función {\it markowitzPortfolio} pueden tener un error por redondeo de $\pm 0.0001$.

\begin{verbatim}
##-----------------------------------------------------------------------##
## Este programa es software libre; puede redistribuirse y/o modificarse ##
## bajo los términos de la Licencia Publica General (GNU) versión 2.     ##
##-----------------------------------------------------------------------##


# biblioteca de programación cuadrática, función solve.QP
library(quadprog)


# crea un objeto portafolios, que contiene las características del mismo
# como media y medida de riesgo a utilizar
newPortfolio <- function ( returns , return="mean", risk="var", B=0 ) {

    call <- match.call()

    if(return=="mean") r <- colMeans ( returns )

    if(risk=="var") { 
        s <- cov ( returns ) 
    } else if(risk=="svar") {
        s <- scov ( returns, B )
    }


    nObs <- nrow ( returns )
    nAssets <- ncol ( returns )

    p <- list("call" = call, "r" = r, "s" = s, "nAssets" = nAssets, 
       "nObs" =  nObs, "risk"=risk ) 

    return( p )
}



# genera una matriz de semicovarianzas aproximadas (Estrada 2008)
scov <- function( returns, B=0 ) {

    nAssets<-ncol(returns)
    n<-nrow(returns)

    S<-matriz(-1, nAssets, nAssets)

    for ( i in 1:nAssets ) {
        for ( j in i:nAssets ) {
           x<-ifelse( (returns[,i]-B)>0, 0, returns[,i]-B)
           y<-ifelse( (returns[,j]-B)>0, 0, returns[,j]-B)
           
           S[i,j]<- ( (x%*%y) / n )
           S[j,i]<-S[i,j]
        }
    }

    return( S )

}



# devuelve el portafolios de varianza mínima asociado a un nivel de 
# rendimiento o devuelve el portafolios de mínimo riesgo
markowitzPortfolio <- function ( pObject , eRet=-1, minvar=FALSE ) {

    call <- match.call()

    nAssets <- pObject$nAssets

    E_R <- pObject$r
    S <- pObject$s

    c <- matriz( 0, nAssets, 1 )
    A <- rbind ( matriz( 1, 1, nAssets ), diag(nAssets) )
    b <- matriz( c(1, rep(0, nAssets)), nAssets + 1, 1 )

    if( eRet != -1 ) {
       A <- rbind( A[1,], E_R, A[ 2:(nAssets+1) ,] )
       b <- rbind( b[1], eRet, as.matrix(b[ 2:(nAssets+1) ]) )
    }

    if( det( S ) > 0 ) (lambda <- 0) 
    else lambda <- diag(nAssets)*0.00000000001

    if(minvar) (meq<-2) else meq<-1
   
    solQP <- solve.QP ( S + lambda, c, t(A), b, meq = meq )

    w <- as.matrix( solQP$solution )
    er <- as.vector( t(w)%*%E_R )
    sd <- as.vector( sqrt( t(w)%*%S%*%w ) )

    weights <-round( solQP$solution, 4 )

    portfolio <- weights[ weights > 0 ]
    names(portfolio) <- names( pObject$r[ weights > 0 ] )


    results <- list( "call" = call, "er" = er, "sd" = sd, 
        "weights" = weights, "portfolio" = portfolio)

    return( results )
}



# determina el portafolios eficiente para el modelo: 
# minimizar (1-lambda)*var(R) - lambda*E(R)
lambdaPortfolio <- function (pObject , eRet=-1, minvar=FALSE, lambda=0.5) {

    call <- match.call()

    nAssets <- pObject$nAssets

    E_R <- pObject$r
    S <- pObject$s

    c <- matriz( 0, nAssets, 1 )
    A <- rbind ( matriz( 1, 1, nAssets ), diag(nAssets) )
    b <- matriz( c(1, rep(0, nAssets)), nAssets + 1, 1 )

    if( eRet != -1 ) {
       A <- rbind( A[1,], E_R, A[ 2:(nAssets+1) ,] )
       b <- rbind( b[1], eRet, as.matrix(b[ 2:(nAssets+1) ]) )
    }

    if( det( S ) > 0 ) (lbd <- 0) 
    else lbd <- diag(nAssets)*0.00000000001

    if(minvar) (meq<-2) else meq<-1
   
    solQP <- solve.QP ( S*((1-lambda)*2) + lbd, E_R*lambda, t(A), b, 
              meq = meq )

    w <- as.matrix( solQP$solution )
    er <- as.vector( t(w)%*%E_R )
    sd <- as.vector( sqrt( t(w)%*%S%*%w ) )

    weights <-round( solQP$solution, 4 )

    portfolio <- weights[ weights > 0 ]
    names(portfolio) <- names( pObject$r[ weights > 0 ] )


    results <- list( "call" = call, "er" = er, "sd" = sd, 
                  "weights" = weights, "portfolio" = portfolio)

    return( results )
}



########### Funciones gráficas ###########


# genera nRand portafolios aleatoriamente y los gráfica
plotPortfolioPoints <- function( pObject, nRand = 10000 ) {


    nAssets <- pObject$nAssets

    E_R <- pObject$r
    S <- pObject$s

    pReturn <- rep( 0, nRand )
    pRisk <- rep( 0, nRand )

    for ( i in 1:nRand ) {

       w <- matriz(0, nAssets, 1)

       for ( j in 1:nAssets ) {

           if( j == 1) w[j] <- runif ( 1, 0, 1 )
           else if( j == nAssets) w[j] <- 1 - sum(w)
           else w[j] <- runif ( 1, 0, 1 - sum(w) )

       }

    w <- w[ sample(1:nAssets,nAssets) ]

    pReturn[i] <- t(w) %*% E_R
    pRisk[i] <- sqrt( t(w) %*% S %*% w )

    }

    lines(pRisk, pReturn, type="p", pch=19, cex = 0.25, col="gray")


}



# genera los pares desviación estándar-rendimiento, para graficar el
# conjunto de varianza mínima
efficientFrontier <- function( pObject ) {

    n <- 40

    retM <- rep(0, n)
    sdM <- rep(0, n)


    E_R <- pObject$r

    min <- min(E_R) + abs(min(E_R)) * .005
    max <- max(E_R) - max(E_R) * .005
    range <- seq( min, max, ( max - min ) / ( n-1 ) )

    range <- round( range, 6 )

    for( i in 1:n) {

       p <- markowitzPortfolio( pObject, range[i], minvar=TRUE )
       retM[i] <- p$er
       sdM[i] <- p$sd

    }

    return( cbind(sdM, retM ) )

}



# genera los pares desviación estándar-rendimiento, para graficar
# la frontera eficiente del modelo:
# minimizar (1-lambda)*var(R) - lambda*E(R)
lambdaFrontier <- function( pObject ) {

    n <- 40

    retM <- rep(0, n)
    sdM <- rep(0, n)


    E_R <- pObject$r

    range <- seq( 0, 1, ( 1 ) / ( n-1 ) )

    range <- round( range, 6 )

    for( i in 1:n) {

       p <- lambdaPortfolio( pObject, lambda=range[i] )
       retM[i] <- p$er
       sdM[i] <- p$sd

    }

    return( cbind(sdM, retM ) )

}



# gráfica el conjunto de oportunidades de todos los pares posibles
# de empresas
opportunitySetLines <- function( pObject ) {


    nAssets <- pObject$nAssets

    E_R <- pObject$r
    S <- pObject$s


    for ( i in 1:(nAssets - 1) ) {
        for ( j in (i + 1):nAssets ) {
            index = c(i, j)

            lines( twoAssetsEF( E_R[index], S[index,index] ), type="l" )
        }
    }


}



# gráfica el conjunto de oportunidades de un par de empresas
twoAssetsEF <- function( E_R, S ) {

    n <- 30

    retM <- rep(0, n)
    sdM <- rep(0, n)

    w<-seq( 0, 1, 1 / ( n-1 ) )


    for( i in 1:n) {
       W <- as.matrix(c(w[i], 1-w[i]))
       retM[i] <- t(W)%*%E_R
       sdM[i] <- sqrt( t(W)%*%S%*%W ) 

    }

    return( cbind(sdM, retM ) )

}



# gráfica los rendimientos esperados vs los rendimientos obtenidos
efficientFrontierFit <- function ( pObject, returns2 ) {


    n <- 40

    retM <- rep(0, n)


    E_R <- pObject$r
    E_R2 <- colMeans( returns2 )

    min <- markowitzPortfolio( pObject )$er
    max <- max(E_R) - max(E_R) * .005

    range <- seq( min, max, ( max - min ) / ( n-1 ) )
    range <- round( range, 6 )

    for( i in 1:n) {

       p <- markowitzPortfolio( pObject, range[i] )
       retM[i] <- p$weights%*%E_R2

    }



    plot( range, retM, type="l", xlab="Rendimiento esperado", 
        ylab="Rendimiento obtenido" )
    lines( range, range, type="l", col=c("#8c8c8c") )


    x<-c( mean(abs(range-retM)), mean( (range-retM)[(range-retM)>0] ) )
    y<-c( "error medio", "     error medio de sub-estimación" )  
    names(x)<-y

    print(x)

}



# gráfica el conjunto de oportunidades para un conjunto de empresas
plotPortfolio <- function ( pObject, er=-1, sd=-1, points=-1 ) {

    call <- match.call()

    efficientFrontier <- efficientFrontier( pObject )

    E_R <- pObject$r

    Xlim <- range( efficientFrontier[,1], max(sqrt(diag(pObject$s))) )
    Ylim <- c( min( E_R ), max( E_R ) )


    if(pObject$risk=="var") Xlab<-"Desviación estándar"
    else if(pObject$risk=="svar") Xlab<-"Semi-desviación"
    plot( efficientFrontier, xlim=Xlim, ylim=Ylim, xlab=Xlab, 
       ylab="Rendimiento esperado", type="l" )
    opportunitySetLines( pObject )

    if( points != -1 ) plotPortfolioPoints( pObject , points )

    if( er!= -1 && sd != -1 ) {
       abline( v = sd )
       abline( h = er )
    }

}



# obtiene los rendimientos a partir de los precios
# prices debe ser de una matriz, en la que los precios
# se encuentran a partir de la segunda columna

assetsReturn <- function ( prices ) {

    call <- match.call()

    nrows <- nrow( prices ) 
    ncols <- ncol( prices ) 

    R <- prices[2:nrows, 2:ncols] / prices[1:(nrows-1), 2:ncols] - 1

    return( R )
}
\end{verbatim}

\chapter{Código fuente: Algoritmo genético - Modelo M-V} \label{GA1}

Aquí se presenta el código R para solucionar el modelo de optimización: maximizar $\lambda E({\bm w}'{\bm R}) - (1-\lambda ) {\bm w}'{\bm \Sigma}{\bm w}$, presentado en el capítulo 4, por medio de algoritmos genéticos. Debe recordarse que al utilizar precios, estos deben estar en la misma escala (moneda).

\begin{verbatim}
##-----------------------------------------------------------------------##
## Este programa es software libre; puede redistribuirse y/o modificarse ##
## bajo los términos de la Licencia Publica General (GNU) versión 2.     ##
##-----------------------------------------------------------------------##

# optimización de portafolios por medio de algoritmos genéticos
GAlambdaPortfolio<-function(pObject, lambda=0.5, m=500, plot=TRUE) {

    # m=número de generaciones

    # tasa de mutación
    mr<-0.2

   
    # número de empresas
    nAssets <- pObject$nAssets

    # Tamaño de la población, número par
    N<-trunc(nAssets/2)*2

    # rendimiento y varianza
    E_R <- pObject$r
    S <- pObject$s

    Fg<-rep(0,m)
    W<-matriz( 0 , N, nAssets)
    Wtmp<-matrix( 0 , N, nAssets)


    # generamos la población inicial
    for( i in 1:N) {
       w <- matrix(0, nAssets, 1)
       w<-runif(nAssets)
       w<-w/sum(w)
       W[i,]<-w
    }

    
    # vector con los rendimientos de la población generada
    Rp<-W%*%(as.matrix(E_R))
    
    # vector con las varianzas de la población generada
    Sp<-as.matrix(diag(W%*%S%*%t(as.matrix(W))))

    # vector con los valores de adaptación de la población generada
    F<-lambda*Rp-(1-lambda)*Sp
    
    x<-order(F)

    # ordenamos F y W, de peor a mejor adaptado
    F<-as.matrix((F)[x])
    W<-W[(1:N)[x],]

    # guardamos el elemento mejor adaptado de esta población
    Fg[1]<-F[N,]


    # generamos las m-1 generaciones faltantes
    for(j in 2:m) {

       # calculamos la probabilidad de cruzamiento, según adaptación
       pb<-(F/sum(F))
       z<-rep(0, N)
       for(i in 1:N) z[i]<-sum(pb[1:i,])

       # obtenemos elementos a cruzar
       crossIdx<-rep(0, N)
       for(i in 1:N) {
          crossIdx[i]<-((1:N)[runif(1)<z])[1]
       }


       # proceso de cruzamiento
       for(i in seq(1,N,2)) {
       
          # soluciones a cruzar
          w1<-W[crossIdx[i],]
          w2<-W[crossIdx[i+1],]

          # mutamos, la tasa de mutación aumenta con las generaciones 
          if(runif(1)<(mr+j/m*.5)) w1[sample(1:nAssets, 1)]<-runif(1, 0, 2)

          # punto de cruce aleatorio
          wx<-sample(1:(nAssets-1), 1)
 
          # obtenemos los nuevos elementos y los guardamos en una población 
          # temporal
          Wtmp[i,]<-c(w1[1:wx], w2[(wx+1):nAssets])
          Wtmp[i,]<-Wtmp[i,]/sum(Wtmp[i,])

          Wtmp[i+1,]<-c(w2[1:wx], w1[(wx+1):nAssets])
          Wtmp[i+1,]<-Wtmp[i+1,]/sum(Wtmp[i+1,])

       }


       # rendimiento y varianza de la nueva población
       Rptmp<-Wtmp%*%(as.matrix(E_R))
       Sptmp<-as.matrix(diag(Wtmp%*%S%*%t(as.matrix(Wtmp))))

       # adaptación de la nueva población
       Ftmp<-lambda*Rptmp-(1-lambda)*Sptmp
    
       x<-order(Ftmp)

       Ftmp<-as.matrix((Ftmp)[x])
       Wtmp<-Wtmp[(1:N)[x],]

       #ordenamos del peor al mejor adaptado
       F2<-rbind(F, Ftmp)
       x<-order(F2)
 
       # mantenemos en la población sólo a los N mejor adaptados
       # de la población anterior y la nueva
       F<-as.matrix(((F2)[x])[(N+1):(2*N)])
       W<-(rbind(W,Wtmp)[ (1:(2*N))[x] ,])[(N+1):(2*N), ]

       # guardamos al mejor adaptado de esta generación
       Fg[j]<-F[N,]

    }


    # graficamos a los elementos mejor adaptados de cada generación
    if(plot) plot(Fg, type="l", xlab="Generación", ylab="Adaptación")

    # obtenemos a al mejor adaptado de todas las generaciones
    # y calculamos su rendimiento y desviación estándar
    w<-W[N,]
    er<-w%*%(as.matrix(E_R))
    sd<-sqrt(w%*%S%*%w)

    weights<-round(w, 4)

    portfolio <- weights[ weights > 0 ]
    names(portfolio) <- names( pObject$r[ weights > 0 ] )

    # presentamos el resultado
    results <- list("er" = er, "sd" = sd, "weights" = round(w, 4), 
        "portfolio" = portfolio)
        
    return( results )
}




# obtenemos la frontera eficiente variando lambda
GAlambdaFrontier <- function( pObject, ng=500) {

    # número de valores a graficar
    n <- 40

    retM <- rep(0, n)
    sdM <- rep(0, n)

    E_R <- pObject$r

    # valores de lambda
    range <- seq( 0, 1, ( 1 ) / ( n-1 ) )

    range <- round( range, 6 )

    # obtenemos el rendimiento y desviación estándar para cada lambda
    for( i in 1:n) {

       p <- GAlambdaPortfolio( pObject, lambda=range[i], plot=FALSE, m=ng)
       retM[i] <- p$er
       sdM[i] <- p$sd

    }

    # devolvemos los pares desviación estándar-rendimiento para graficar
    return( cbind(sdM, retM ) )

}
\end{verbatim}

\chapter[Código fuente: AG - Costos y enteros]{Código fuente: Algoritmo genético - Costos por transacción y restricciones de enteros} \label{GA2}

Aquí se presenta el código R para solucionar el modelo de optimización: $\mbox{maximizar} \quad \lambda E(R_p) - (1-\lambda ) {\bm w}'{\bm \Sigma}{\bm w}$, con costos por transacción y restricciones de enteros, presentado en el capítulo 5, por medio de algoritmos genéticos. Debe recordarse que al utilizar precios, estos deben estar en la misma escala (moneda).

\begin{verbatim}
##-----------------------------------------------------------------------##
## Este programa es software libre; puede redistribuirse y/o modificarse ##
## bajo los términos de la Licencia Publica General (GNU) versión 2.     ##
##-----------------------------------------------------------------------##


# optimización de portafolios por medio de algoritmos genéticos
GAlambdaNPortfolio <- function(pObject, lambda=0.5, m=500, plot=TRUE, 
                                      prices, cb=0, cs=0, cap, rf=0, t=1) {

    # m=número de generaciones

    # tasa de mutación
    mr<-0.3

    # número de empresas
    nAssets <- pObject$nAssets

    # vector de precios por cada acción, de cada empresa
    p<-prices

    # si no se mencionan los costos por transacción, hacerlos ceros
    if(length(cb)==0 && cb==0) c<-rep(0, nAssets)
    else c<-cb
    if(length(cs)==0 && cs==0) c2<-rep(0, nAssets)
    else c2<-cs

    # variables, capital, rendimiento libre de riesgo, total de periodos
    n0<-rep(0, nAssets)
    k<-cap
    rf<-rf
    t<-t

    # Tamaño de la población, número par
    N<-trunc(nAssets/2)*2

    # rendimiento y varianza
    E_R <- pObject$r
    S <- pObject$s

    # máximo número de acciones que se pueden comprar
    limn<-trunc(k/min(p))

    Fg<-rep(0,m)
    n<-matrix(0, N, nAssets)
    ntmp<-matrix( 0 , N, nAssets)
    
    
    # generamos la población inicial    
    for(i in 1:N) {
       ke<-k
       for(j in sample(1:nAssets)) {

            n[i, j]<-sample(0:trunc( ke/(p[j] * (1+c[j])) ), 1)
            ke<-( k - ( n[i, ]%*%p + n[i, ]%*%(p*c) ) )

       }
    }
    

    # vector de residuos de la población generada
    e <- k - ( p%*%t(n) + t(c*p)%*%t(n) )

    W<-t(((p)*t(n)/k))
 
    # vector con los rendimientos de la población generada
    Rp<-(t(E_R*p)%*%t(n))/k + (e*rf)/k  - (((t*t(E_R*p)+p)*c2)%*%t(n)/k )/t
    
    # vector con las varianzas de la población generada    
    Sp<-diag(W%*%S%*%t(as.matrix(W)))

    # vector con los valores de adaptación de la población generada
    F<-lambda*Rp-(1-lambda)*Sp  

    x<-order(F)

    # ordenamos F, W, n de peor a mejor adaptado
    F<-as.matrix((F)[x])
    W<-W[(1:N)[x],]
    n<-n[(1:N)[x],]

    # guardamos el elemento mejor adaptado de esta población
    Fg[1]<-F[N,]


    # función de corrección de la población, para cruce y mutación
    nfix<-function(nn) {

       pp<-(p*nn)/(p%*%nn)

       y<-order(pp, decreasing=TRUE)
       idx<-((1:nAssets)[y])[pp[y]>0]

       nc<-rep(0, nAssets)

       ke<-k

       for(l in idx) nc[l]<-trunc( (pp[l]*k)/( p[l] * (1+c[l]) ) )


       return(nc);
    }  



    # generamos las m-1 generaciones faltantes
    for(j in 2:m) {

       # calculamos la probabilidad de cruzamiento, según adaptación
       pb<-(F/sum(F))
       z<-rep(0, N)
       for(i in 1:N) z[i]<-sum(pb[1:i])

       # obtenemos elementos a cruzar
       crossIdx<-rep(0, N)
       for(i in 1:N) {
          crossIdx[i]<-((1:N)[runif(1)<z])[1]
       }

       # proceso de cruzamiento
       for(i in seq(1,N,2)) {

          # soluciones a cruzar
          n1<-n[crossIdx[i],]
          n2<-n[crossIdx[i+1],]

          # mutamos, la tasa de mutación aumenta con las generaciones 
          if(runif(1)<(mr+j/m*.3)) 
             n1[sample(1:nAssets, 1)]<-sample(1:(2*limn), 1)

          # punto de cruce aleatorio
          nx<-sample(1:(nAssets-1), 1)

          # obtenemos los nuevos elementos y los guardamos en una población 
          # temporal
          ntmp[i,]<-c(n1[1:nx], n2[(nx+1):nAssets])
          ntmp[i+1,]<-c(n2[1:nx], n1[(nx+1):nAssets])

          # corrección de las soluciones
          ntmp[i,]<-nfix(ntmp[i,])
          ntmp[i+1,]<-nfix(ntmp[i+1,])

       }

       # residuos de la nueva población
       e2 <- k - ( p%*%t(ntmp) + t(c*p)%*%t(ntmp) )

       # proporciones
       W2<-t(((p)*t(ntmp)/k))
 
       # rendimiento y varianza de la nueva población
       Rp2<-t(E_R*p)%*%t(ntmp)/k + (e2*rf)/k  
                                 - (((t*t(E_R*p)+p)*c2)%*%t(ntmp)/k)/t
       Sp2<-as.numeric(diag(W2%*%S%*%t(as.matrix(W2))))

       # adaptación de la nueva población
       F2<-lambda*Rp2-(1-lambda)*Sp2
    
       x<-order(F2)

       F2<-as.matrix((F2)[x])
       W2<-W2[(1:N)[x],]
       ntmp<-ntmp[(1:N)[x],]

       #ordenamos del peor al mejor adaptado
       Ftmp<-rbind(F, F2)
       x<-order(Ftmp)

       # mantenemos en la población sólo a los N mejor adaptados
       # de la población anterior y la nueva
       F<-as.matrix(((Ftmp)[x])[(N+1):(2*N)])
       W<-(rbind(W,W2)[ (1:(2*N))[x] ,])[(N+1):(2*N), ]
       n<-(rbind(n,ntmp)[ (1:(2*N))[x] ,])[(N+1):(2*N), ]

       # guardamos al mejor adaptado de esta generación
       Fg[j]<-F[N,]


    }


    # graficamos a los elementos mejor adaptados de cada generación
    if(plot) plot(Fg, type="l", xlab="Generación", ylab="Adaptación")


    # obtenemos a al mejor adaptado de todas las generaciones
    # y calculamos su rendimiento y desviación estándar
    w<-W[N,]
    n0<-n[N,]
    e0 <- k - ( sum(p*n0) + sum(c*p*n0) )
    F0<- F[N,]

    er<-(E_R*p)%*%n0/k + (e0*rf)/k - (sum(((t*E_R*p)+p)*n0*c2)/k )/t
    sd<-sqrt(w%*%S%*%w)

    weights<-round(w, 4)

    portfolio <- weights[ weights > 0 ]
    names(portfolio) <- names( pObject$r[ weights > 0 ] )

    nportfolio <- n0[ n0 > 0 ]
    names(nportfolio) <- names( pObject$r[ n0 > 0 ] )

    # presentamos el resultado
    results <- list("er" = er, "sd" = sd, "e" = e0, "F" = F0, 
          "n" = n0, "portfolio" = portfolio, "nportfolio" = nportfolio)
    return( results )


}




# obtenemos la frontera eficiente variando lambda
GAlambdaNFrontier <- function( pObject, m=500, plot=TRUE, prices, cb=0, 
                                                   cs=0, cap, rf=0, t=1) {

    # número de valores a graficar
    n <- 40

    retM <- rep(0, n)
    sdM <- rep(0, n)

    E_R <- pObject$r

    range <- seq( 0, 1, ( 1 ) / ( n-1 ) )

    # valores de lambda
    range <- round( range, 6 )

    # obtenemos el rendimiento y desviación estándar para cada lambda
    for( i in 1:n) {

       p <- GAlambdaNPortfolio( pObject, lambda=range[i], plot=FALSE, 
                   m=m, prices=prices, cb=cb, cs=cs, cap=cap, rf=rf, t=t)
       retM[i] <- p$er
       sdM[i] <- p$sd

    }

    # devolvemos los pares desviación estándar-rendimiento para graficar
    return( cbind(sdM, retM ) )

}
\end{verbatim}

\bibliographystyle{plain}                                                                                                                                                                             
\bibliography{portfolio_arxiv}

@Book{mark2_2,
  author =	"{Harry M.} Markowitz",
  title =	"Portfolio selection: efficient diversification of investments",
  year = 	"1991",
  edition =	"2nd",
  publisher =	"Basil Blackwell",
  address =	"New York"
}

@book{mark3,
 author = {{Harry M.} Markowitz},
 title = {Mean-Variance analysis in portfolio choice and capital markets},
 year = {1987},
 publisher = {Basil Blackwell}
}

@Book{mark2,
  author =	"{Harry M.} Markowitz",
  title =	"Portfolio selection: efficient diversification of investments",
  year = 	"1959",
  publisher =	"John Wiley \& Sons",
  address =	"New York"
}

@Article{mark1,
  author =	"{Harry M.} Markowitz",
  title =	"Portfolio selection",
  journal =	"Journal of Finance",
  volume =	"7",
  number =      "1",
  year = 	"1952",
  pages =	"77--91"
}

@Book{Scherer:2005:IMP,
  author =	"Bernd Michael Scherer and R. Douglas Martin",
  title =	"Introduction to modern portfolio optimization with {NuOPT} and {S-PLUS}",
  publisher =	"Spring{\-}er-Ver{\-}lag",
  year = 	"2005",
}

@book{1204793,
 author = {Maringer, Dietmar},
 title = {Portfolio Management with Heuristic Optimization (Advances in Computational Management Science)},
 year = {2005},
 isbn = {0387258523},
 publisher = {Springer-Verlag},
 }

@book{citeulike:1884757,
    author = {Vince, Ralph},
    day = {25},
    howpublished = {Hardcover},
    isbn = {0471757683},
    publisher = {John Wiley \& Sons},
    title = {The Handbook of Portfolio Mathematics: Formulas for Optimal Allocation \& Leverage},
    year = {2007}
}

@book{citeulike:2550034,
    author = {Jorion, Philippe and Garp},
    day = {15},
    howpublished = {Paperback},
    isbn = {0470126302},
    month = {may},
    publisher = {John Wiley \& Sons},
    title = {Financial Risk Manager Handbook (Wiley Finance)},
    year = {2003}
}

@Book{ nla.cat-vn2525693,
author = { Satchell, S. and Scowcroft, Alan. },
title = { Advances in portfolio construction and implementation },
edition = { 1st ed. },
publisher = { Butterworth-Heinemann, Amsterdam; Oxford},
pages = { xvi, 365 p. ; },
year = { 2003 },
type = { Book },
language = { English }
}

@Book{Pastor00portfolioselection,
	author = {Shouyang Wang and Yusen Xia},
	title = {Portfolio selection and asset pricing},
        publisher = {Springer-Verlag},
	journal = {Journal of Finance},
	year = {2002}
}

@ARTICLE{Yoshimoto96themean-variance,
	author = {Atsushi Yoshimoto},
	title = {The Mean-Variance Approach to Portfolio Optimization Subject to Transaction Costs},
	journal = {Journal of the Operations Research Society of Japan},
	year = {1996},
	number = {1},
        volume = {39},
	pages = {99--117}
}

@book{mark4,
 author = {John JR. Guerard},
 title = {Handbook of Portfolio Construction: Contemporary Applications of Markowitz Techniques},
 year = {2009},
 publisher = {Springer-Verlag}
 }

@book{invest,
 author = {Frank J. Travers},
 title = {Investment Manager Analysis: A Comprehensive Guide to Portfolio Selection, Monitoring and Optimization},
 year = {2004},
 publisher = {John Wiley \& Sons}
 }

@mastersthesis{gen1,
 author = {F\'{e}lix Roudier},
 title = {Portfolio Optimization and Genetic Algorithms},
 year = {2007},
 School = {Department of Management, Technology and Economics-Swiss Federal Institute of Technology Zurich}
 }

@book{ag2,
 author = {Natyhelem Gil Londono},
 title = {Algoritmos gen\'{e}ticos},
 year = {2006},
 publisher = {Escuela de Estad\'{i}stica-Universidad Nacional de Colombia}
 }

@book{mitchell,
    address = {Cambridge, MA, USA},
    author = {Mitchell, Melanie},
    publisher = {The MIT Press},
    title = {An Introduction to Genetic Algorithms},
    year = {1999}
}

@incollection {gentran,
   author = {Lin, Dan and Li, Xiaoming and Li, Minqiang},
   affiliation = {School of Sciences, Tianjin University, Tianjin 300072 P.R. China},
   title = {A Genetic Algorithm for Solving Portfolio Optimization Problems with Transaction Costs and Minimum Transaction Lots},
   booktitle = {Advances in Natural Computation},
   series = {Lecture Notes in Computer Science},
   publisher = {Springer Berlin / Heidelberg},
   pages = {808-811},
   volume = {3612},
   year = {2005}
}

@book{portmo,
 author = {Philip J. McDonnell},
 title = {Optimal Portfolio Modeling},
 year = {2008},
 publisher = {John Wiley \& Sons}
 }

@book{Levy84portfolioand,
	author = {H. Levy and M. Sarnat},
	title = {Portfolio and Investment Selection: Theory and Practice},
	booktitle = {Prentice-Hall Intentional, Inc},
	year = {1984},
	publisher = {Prentice Hall}
}

@techreport{JEM_transcosts,
    AUTHOR  =  {J. E. Mitchell and S. Braun},
    TITLE   =  {Rebalancing an Investment Portfolio in the Presence of Transaction Costs},
    YEAR    =  2003,
    MONTH   =  {December},
    INSTITUTION={Mathematical Sciences, Rensselaer Polytechnic Institute},
    URL={http://www.rpi.edu/~mitchj/papers/transcosts.html}  
}

@article{semiv,
  type={Working Paper Series},
  title={{Mean-Semivariance Optimization: A Heuristic Approach}},
  author={Estrada, Javier},
  journal={SSRN eLibrary},
  year={2007},
  publisher={SSRN},
  location={http://ssrn.com/paper=1028206},
  language={English}
}

@MISC{Finance00yahoofinance,
	author = {Yahoo Finance},
	title = {http://finance.yahoo.com},
	year = {2010}
}

@article{andreevichproblems,
  title={PROBLEMS OF CHOOSING MODELS FOR THE FORMATION AND MANAGEMENT OF INVESTMENT PORTFOLIOS},
  author={Andreevich, Gorskiy Mark},
  journal={SCIENCE EDUCATION PRACTICE},
  pages={40}, 
  year={2020}
}

@Article{erwin2023metaheuristics,
  author  = {Erwin, Kyle and Engelbrecht, Andries},
  title   = {Meta-heuristics for portfolio optimization},
  journal = {Soft Computing},
  volume  = {27},
  pages   = {19045--19073},
  year    = {2023}
}
\nocite{*} \addcontentsline{toc}{chapter}{Bibliografía}


\end{document}